\documentclass[a4,12pt]{revtex4-1}

\usepackage{graphicx}
\usepackage{url}

\begin{document}

\title{Perspective: How good is DFT for water?}

\author{M. J. Gillan}
\email[Author to whom correspondence should be addressed. Electronic mail: ]{gillan.mike@gmail.com}
\affiliation{London Centre for Nanotechnology, Gordon St., London WC1H 0AH, UK}
\affiliation{Thomas Young Centre, University College London, London WC1H 0AH, UK}
\affiliation{Dept. of Physics and Astronomy, University College London, London WC1E 6BT, UK}

\author{D. Alf\`{e}}
\affiliation{Dept. of Earth Sciences, University College London, London WC1E 6BT, UK}
\affiliation{London Centre for Nanotechnology, Gordon St., London WC1H 0AH, UK}
\affiliation{Thomas Young Centre, University College London, London WC1H 0AH, UK}
\affiliation{Dept. of Physics and Astronomy, University College London, London WC1E 6BT, UK}

\author{A. Michaelides}
\affiliation{Dept. of Chemistry, University College London, London WC1E 6BT, UK}
\affiliation{London Centre for Nanotechnology, Gordon St., London WC1H 0AH, UK}
\affiliation{Thomas Young Centre, University College London, London WC1H 0AH, UK}

\begin{abstract}
Kohn-Sham density functional theory (DFT) has become established as an indispensable tool for investigating
aqueous systems of all kinds, including those important in chemistry, surface science, biology and
the earth sciences. Nevertheless, many widely used approximations for the exchange-correlation (XC)
functional describe the properties of pure water systems with an accuracy that is not fully satisfactory.
The explicit inclusion of dispersion interactions generally improves the description, but there remain
large disagreements between the predictions of different dispersion-inclusive methods. We present here
a  review of DFT work on water clusters, ice structures and liquid water, with the aim of elucidating
how the strengths and weaknesses of different XC approximations manifest themselves across this variety
of water systems. Our review highlights the crucial role of dispersion in describing the delicate balance
between compact and extended structures of many different water systems, including the liquid.
By referring to a wide range of published work, we argue that the correct description of
exchange-overlap interactions is also extremely important, so that the choice of semi-local or hybrid
functional employed in dispersion-inclusive methods is crucial. The origins and consequences
of beyond-2-body errors of approximate XC functionals are noted, and we also discuss the substantial
differences between different representations of dispersion. We propose a simple numerical scoring system
that rates the performance of different XC functionals in describing water systems, and we suggest
possible future developments. 
\end{abstract}

\maketitle

\section{Introduction}
\label{sec:intro}

Water is an endlessly fascinating substance with many anomalous properties, of which its expansion
on freezing and its density maximum at $4$~$^\circ$C are just two of the most famous. The fascination is
only deepened by the apparent simplicity of the H$_2$O molecule itself. Because of its importance for
life, for the Earth's geology and climate, and for innumerable domestic and industrial processes, 
water in all its forms has been one of the most widely studied of all substances. From the theoretical
viewpoint, it offers unrivaled opportunities to deepen our understanding of hydrogen bonding (H-bonding).
A little over $20$ years ago, the first attempts were reported to derive and interpret the properties
of liquid water from electronic-structure calculations based on density functional theory 
(DFT)~\cite{laasonen1993a,tuckerman1994a,sprik1996a}. This idea has proven
immensely productive, and has been developed by many research groups, but the search for a fully satisfactory
DFT description of water systems has been unexpectedly arduous, and is not yet complete. Our aim
here is to review what has been learnt so far, and to assess the challenges that remain. An important feature
of the review is that we aim to cover DFT work not only on the liquid, but also on clusters and ice structures.

The first ever DFT simulations of liquid water, pioneered by Parrinello, Car and 
co-workers~\cite{laasonen1993a,tuckerman1994a,sprik1996a,silvestrelli1997a,silvestrelli1999a}, followed a long history
of water modeling based on force 
fields (see e.g. Refs.~\cite{bernal1933a,rowlinson1951a,barker1969a,rahman1971a,berendsen1981a,jorgensen1983a}). 
It was recognized over $80$ years ago that the bent shape of the H$_2$O
molecule and the electronegativity of oxygen make electrostatic forces very important~\cite{bernal1933a}. The earliest
force fields represented the Coulomb interactions in terms of point charges, with overlap repulsion and dispersion
modeled by simple potentials, the molecules being assumed rigid and unpolarizable. Such elementary models
can be remarkably successful for the ambient liquid~\cite{guillot2002a,abascal2005a}, but their transferability 
is poor. The dipole moment of the H$_2$O
molecule is known to increase by $40 - 50$~\% from the gas phase to the ice and
liquid phases~\cite{coulson1966a,batista1998a,silvestrelli1999b,badyal2000a}, so that the neglect of polarizability
is a serious limitation. A large research effort has been devoted to the development of accurate models
that treat the molecules as polarizable and flexible~\cite{dang1997a,burnham2002a,ren2003a,habershon2011a,wikfeldt2013a}
and include the weak intermolecular covalency that has often been thought
significant~\cite{isaacs1999a,nilsson2005a,glendening2005a,cobar2012a}. The most sophisticated of these models have the declared aim of describing all water systems,
from clusters through ice structures to the liquid (see e.g. Refs.~\cite{fanourgakis2008a,babin2012a}). Reviews of
the many force fields that have been proposed can be found elsewhere, e.g. Refs.~\cite{guillot2002a,vega2009a,vega2011a}.

By definition, close approximations to the true exchange-correlation (XC) functional of DFT would automatically deliver
everything offered by force-fields and more. The search for such approximations for pure water systems is important
for several reasons. DFT gives direct access to electronic charge distributions, which are important for the
interpretation of experimental observables such as infra-red spectra~\cite{sharma2005a,zhang2011c}, 
dielectric properties~\cite{sharma2007a}, x-ray scattering intensities~\cite{krack2002a}, surface potentials~\cite{kathmann2011a}, etc.
It also provides a way of investigating water systems completely independent of force-fields,
so that it gives the possibility of fruitful dialog between the approaches. 
Although our review is restricted to pure water systems, the development of improved XC approximations reviewed
here is highly relevant to the DFT description of more general aqueous systems, including solutions and acid-base
systems~\cite{laasonen1994a,tuckerman1995a,tuckerman1995b,white2000a,naor2003a,iftimie2006a,iftimie2008a,marx2006a,marx2010a,hassanali2013a,hassanali2014a},
as well as confined water~\cite{coudert2006a,cicero2008a,munoz-santiburcio2013a} and water adsorbed on 
surfaces~\cite{lindan1998a,carrasco2011a,carrasco2012a,sulpizi2012a,cheng2010a,wang2012b,wood2013a}. 
The crucial role of H-bonding in the cohesion of water systems~\cite{pauling1935a,stillinger1980a}
makes them an outstanding paradigm of this bonding mechanism,
which is so widespread in many other molecular systems, including those important in biology~\cite{jeffrey1997a}. This means that the challenges
to DFT to be described here for pure water systems will have a wide relevance.

The XC functionals known as generalized gradient approximations (GGAs) are among the most popular
and successful for a great variety of condensed-matter systems, and they were used for the first DFT simulations
of liquid water. Appropriately chosen GGAs were found to give quite satisfactory binding energies for the water
dimer~\cite{sim1992a,laasonen1992a,laasonen1993b,santra2007a} 
and the common form of ice~\cite{hamann1997a,feibelman2008a}, and when used in molecular dynamics (MD) simulations 
gave a reasonable structure for the liquid~\cite{laasonen1993a,sprik1996a,silvestrelli1999a}. The early 
successes prompted a surge of interest in the use of GGA-based simulations
to explore a wide range of important questions concerning water itself, such as H-bond dynamics, electronic properties, and the structure
and diffusion of hydronium and hydroxyl ions. Simulations of this kind have also been very widely used to probe
the solvation shells around a variety of ions and other neutral solutes in water. A review of DFT-based MD work both on pure water
and on a wide range of aqueous systems can be found in Ref.~\cite{hassanali2014a}. The widespread use of
DFT for simulating interfaces involving aqueous systems and water adsorbed on surfaces
is also noteworthy (see e.g. Refs.~\cite{lindan1998a,michaelides2006a,carrasco2011a,carrasco2012a,skulason2007a,sulpizi2012a}). 
The new insights gained in these investigations
would in many cases have been difficult or impossible to achieve with force-field methods, and the enormous value
of DFT-based simulations of aqueous systems is beyond dispute.

Nevertheless, it became clear over $10$ years ago that the description of liquid water given by GGAs was not
completely satisfactory~\cite{asthagiri2003a,grossman2004a,schwegler2004a,fernandez-serra2004a,sit2005a}. 
Fortuitous cancelation of errors in the earlier work had made the approximations seem more
accurate than they really were~\cite{grossman2004a,kuo2004b,vandevondele2005a}. It was 
also discovered that GGA predictions of energy differences
between extended and compact structures of some water systems, including larger clusters and ice, are
qualitatively wrong~\cite{santra2008a,shields2008a,dahlke2008a,santra2011a}. These discoveries stimulated 
a re-examination of XC approximations for pure
water systems that continues to this day, and we shall try to decribe what has been learnt from this.
An important outcome will be that dispersion is crucially important, and that some of the errors of GGAs
come from their failure to describe dispersion correctly. It has been recognized for many years that H-bonding 
is the dominant mechanism of cohesion in water systems~\cite{pauling1935a,stillinger1980a}. However, H-bonding
is a complex phenomenon, which can be analyzed
into electrostatic attraction, polarization, dispersion and partial covalency~\cite{arunan2011a}, though the relative contributions
of these components in water remain controversial, depending significantly on definition~\cite{cobar2012a}. The contribution 
of partial covalency (often termed charge transfer), for example, has been particularly 
contentious~\cite{isaacs1999a,ghanty2000a,romero2001a,nilsson2005a}. In addition, good XC functionals must
correctly describe exchange-repulsion and monomer deformation. Our point of view here will be
that all these energy components can be in error, and our review of the research will try to assess where
the main errors lie. The evidence will indicate that dispersion is far from being the only culprit.

We will start by reviewing DFT work on the water monomer and water clusters. The DFT description of the monomer (Sec.~\ref{sec:monomer})
is important for the electrostatic, polarization and monomer deformation parts of the energy, while the dimer (Sec.~\ref{sec:dimer})
provides tests of H-bonding, where exchange-repulsion, dispersion and weak covalency also play a role.
Energies of the dimer in non-H-bonding geometries may help in separating the covalency contribution. DFT work on
clusters from the trimer to the pentamer (Sec.~\ref{sec:coop}) gives further information about polarizability, while
the isomers of the hexamer (Sec.~\ref{sec:hexamer}) and larger clusters (Sec.~\ref{sec:larger_clusters}) help to
separate dispersion and exchange repulsion. We shall see that the energetics of ice structures (Sec.~\ref{sec:ice}) is vital in assessing the
roles of these energy components. The lessons learnt up to this point form the background to our review of DFT work on 
the liquid (Sec.~\ref{sec:liquid}). In the final Sec.~\ref{sec:discussion}, we draw together the evidence from all the
water systems to assess the ability of current XC functionals to account for all the components of the energy. We summarize
by proposing a simple scoring scheme, which assigns a numerical score to any given XC functional, based on the quality of
its predictions for clusters and ice structures. The scheme may help to gauge the likely performance
of the functional on the liquid. We should note at the outset that water is a vast subject, so that our review will inevitably be
incomplete, as well as reflecting our own personal perspective. We provide in the Appendix a brief survey of the main XC approximations 
that will be relevant.


\section{The water monomer}
\label{sec:monomer}

Since electrostatic interactions are very important in all water systems, we need to know that available XC functionals
reproduce the charge distribution of the free H$_2$O monomer, or at least its leading multipole moments.
This will ensure the correctness of the so-called first-order electrostatic energy, i.e. the Coulomb interaction energy of an assembly
of molecules when the monomer charge distributions are taken to be those of free monomers. In reality, the electric fields
of the monomers distort each other's charge distributions, so it is important that XC functionals reproduce the polarizabilities
of the free monomers, and ideally this should mean the response of the dipole and higher multipole moments to dipolar
and higher multipolar fields.

Almost all published DFT work on the charge distribution of the free H$_2$O monomer reports only the dipole
moment $\mu$, though some information is available for the quadrupole moments. Only a limited number of GGAs appear to have been
studied, but the BLYP~\cite{becke1988a,lee1988a} and BP86~\cite{becke1988a,perdew1986a} functionals reproduce the 
benchmark value of $\mu$ to within $\sim 3$~\%, and the hybrid functionals
B3LYP~\cite{becke1993a,stephens1994a} and PBE0~\cite{perdew1996b,adamo1999a} 
are even better than this~\cite{adamo1999a,calaminici1998a,fantin2007a}. The rather sparse
results for the quadrupole moments indicate that these are also correct to within a few percent~\cite{cohen1999a}. So far as 
we are aware, DFT calculations have been reported only for the components of the dipolar polarizability, and there is general
agreement that GGAs always overestimate them 
by $\sim 10$~\%~\cite{mcdowell1995a,tozer1998a,calaminici1998a,adamo1999a,cohen1999a,vancaillie2000a}.
(The overestimation of molecular polarizabilities by GGAs is a general phenomenon, which arises from the
underestimation of the energies of virtual Kohn-Sham states relative to those of occupied states, which
in turn is related to the incorrect behavior of the Kohn-Sham potential in the asymptotic region far from the 
molecule~\cite{tozer1998a,tozer1998b}.) 
Hybrid functionals do much better, with the
average polarizability from B3LYP and PBE0 being in error by $\sim 2$~\% and less than $1$~\% 
respectively~\cite{tozer1998a,adamo1999a,vancaillie2000a,hammond2009a}.

The water monomer is flexible, and a correct description of its deformation energetics is likely to be important,
for two reasons. One is that, in water clusters and condensed phases, formation of a H-bond weakens and
lengthens the OH bond of the donor, and these effects help to determine the strength of the H-bond. The other reason
is that the spectrum of intramolecular vibrations is an important experimental diagnostic of H-bond formation, which it is
desirable to reproduce in simulations.

\begin{figure}
\centering
\includegraphics[width=0.9\linewidth]{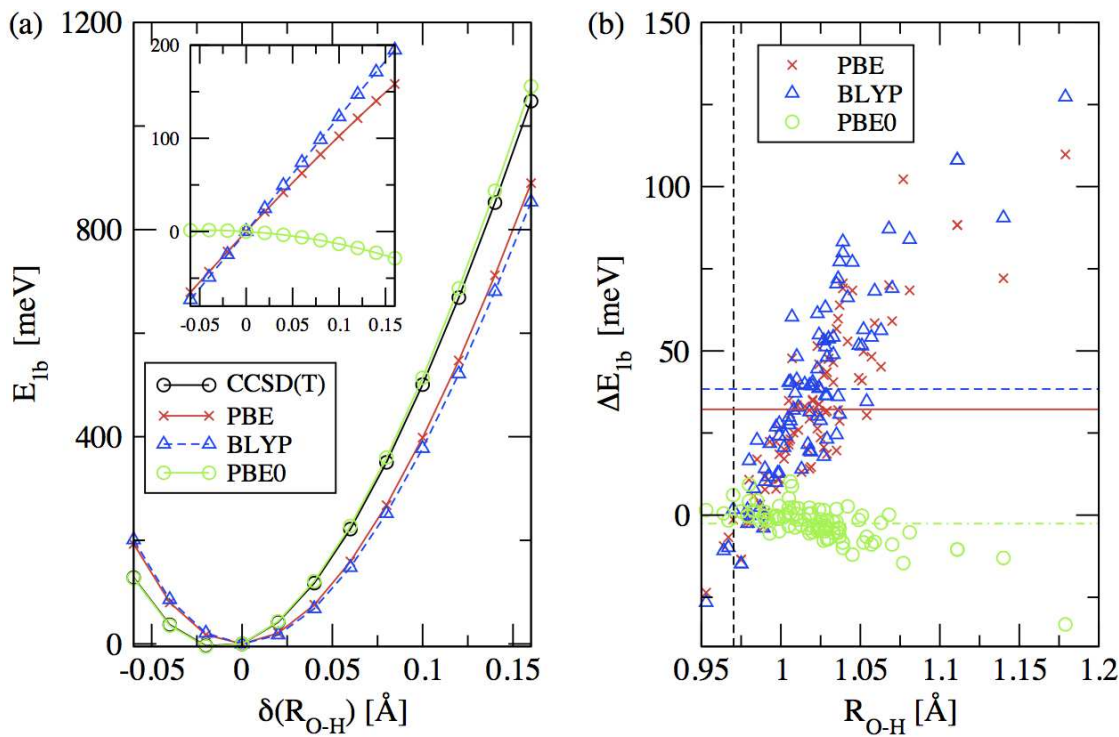}
\caption{a)~Variation of the monomer deformation energy $E_{\rm 1 b}$ with deviation $\delta R_{\rm O-H}$ of the
O-H bond length of an H$_2$O molecule from its equilibrium value in the symmetric stretch mode, calculated
with CCSD(T), PBE, BLYP and PBE0. Inset shows the errors $\Delta E_{\rm 1 b}$ of the deformation energy
with the three XC functionals relative to CCSD(T) (benchmark minus DFT value of $E_{\rm 1 b}$). (Note that 
$\delta R_{\rm O-H}$ is computed relative to the equilibrium
value of the O-H bond length given by PBE, which is slightly greater than that
calculated with PBE0 and CCSD(T).)
(b)~Errors $\Delta E_{\rm 1 b}$ for deformed monomers drawn from a simulation of liquid water as a function
of the longest O-H bond of each monomer. The vertical dashed line indicates the gas-phase equiibrium O-H bond length
($0.97$~\AA) of a monomer (optimized with PBE) and the horizontal solid, dashed and dash-dotted lines represent the
average errors of PBE, BLYP and PBE0 respectively. Note that a positive error $\Delta E_{\rm 1 b}$ indicates that
it is too easy to stretch O-H bonds of the monomers with a given XC functional compared with CCSD(T).
Reproduced with permission from J. Chem. Phys. {\bf 131}, 124509 (2009). Copyright 2009, American Institute
of Physics.}
\label{fig:monomer_deformation}
\end{figure}

The ability of GGA and hybrid functionals to describe monomer deformation was investigated by Santra \emph{et al.}~\cite{santra2009a},
who showed that the energy cost of stretching the O-H bonds of the monomer is significantly underestimated
by PBE and BLYP but is very accurately given by PBE0. This is shown in panel~(a) of Fig.~\ref{fig:monomer_deformation}
which plots the deformation energy $E_{\rm 1 b}$ in the symmetric mode as a function of the departure 
$\delta R_{\rm O-H}$ of the O-H bond length
from its equilibrium value, $E_{\rm 1 b}$ being computed with the PBE, BLYP and PBE0 approximations and with the benchmark
CCSD(T) technique (coupled-cluster with single and double excitations and a perturbative treatment of triples~\cite{helgaker2000a}). 
(We note that the small offsets of the minima of the plots of $E_{\rm 1 b}$ \emph{vs} $\delta R_{\rm O-H}$ are due to the fact
that $\delta R_{\rm O-H}$ is computed in all cases relative to the equilibrium bond length given by PBE, which is slightly
greater than the bond lengths given by PBE0 and CCSD(T).)
The deviations of the GGA values of $E_{\rm 1 b}$ from the benchmark values (inset
of panel~(a)) are \emph{ca.}~100~meV for a bond stretch $\delta R_{\rm O-H}$ of $0.1$~\AA, but the errors of PBE0 are much smaller. 
The authors examined the consequences of this for the liquid by drawing a large sample of monomers from an MD simulation of liquid water
performed with flexible monomers. They found that bond stretches of up to \emph{ca.}~0.1~\AA\ are very common, and they
confirmed the accuracy of PBE0 and the substantial underestimates of the deformation energy given by PBE and BLYP (see
panel~(b) of Fig.~\ref{fig:monomer_deformation}). This underestimate by GGAs, also 
noted by other authors~\cite{xu2004b,zhang2011b,gillan2012a},
correlates with an erroneous softening of the intramolecular OH stretch frequencies, which are underestimated by $\sim 3$~\%
 and $\sim 4.2$~\% with PBE and BLYP respectively, but are reproduced almost exactly by PBE0~\cite{santra2009a}.

The comparisons for the monomer thus help us to assess the accuracy of XC functionals for three important parts of
the energy in general water systems, namely first-order electrostatics, polarization and monomer deformation. For all
three, GGAs appear to give reasonable, but far from perfect accuracy, while the accuracy of hybrid functionals is
considerably better.


\section{Testing hydrogen-bonding: the dimer}
\label{sec:dimer}

\begin{figure}
\centering
\includegraphics[width=0.7\linewidth]{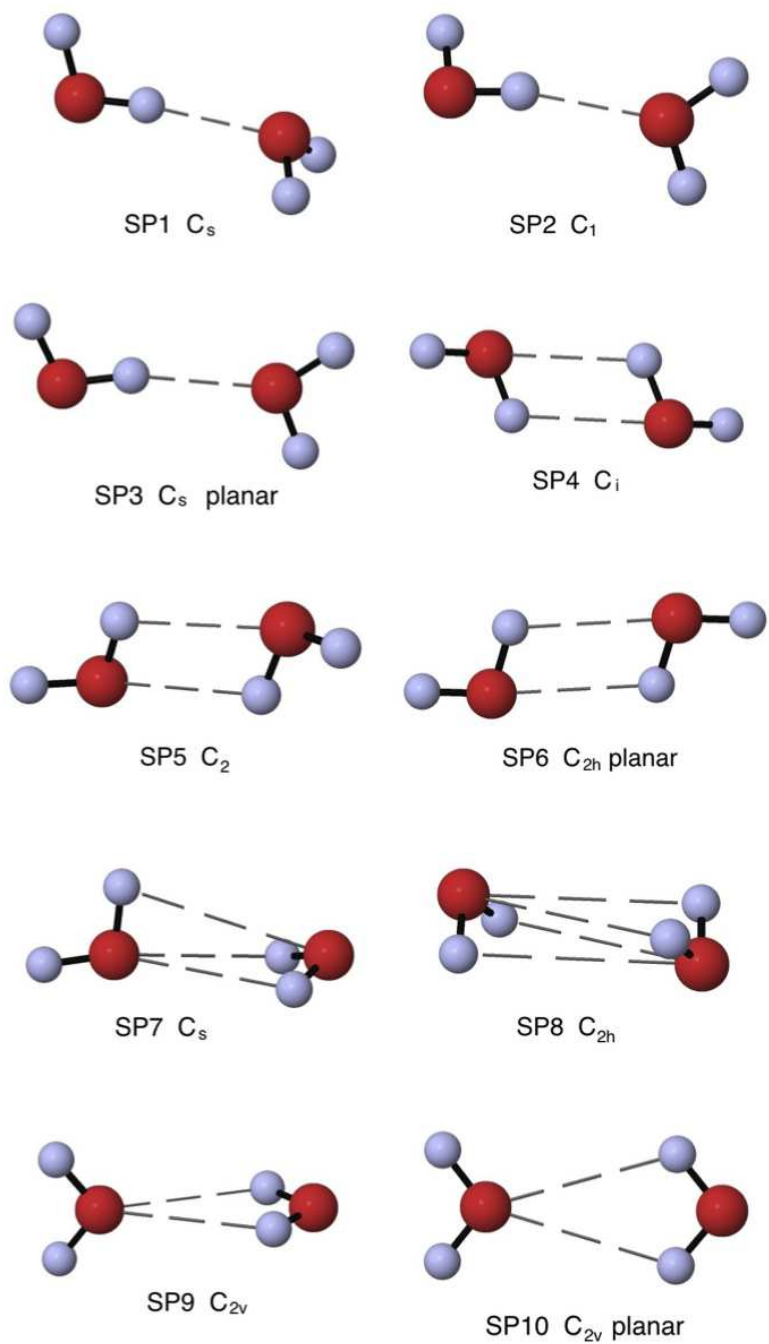}
\caption{Geometries of the 10 Smith stationary points SP$n$ ($n = 1, 2, \ldots 10$) of the water dimer, with their
point-group symmetries.}
\label{fig:Smith_SP}
\end{figure}

The dimer is the simplest water system we can use to test the accuracy of XC functionals for the energy of interaction
between H$_2$O monomers in H-bonding and other geometries. The (H$_2$O)$_2$ system
has been thoroughly studied by very accurate CCSD(T) calculations, which are believed to give the interaction energy 
in any geometry with errors relative to the exact value of no more
than $\sim 5$~meV~\cite{klopper2000a,tschumper2002a}. We know from such calculations that the geometry
having the global minimum energy is the configuration labeled SP1 in Fig.~\ref{fig:Smith_SP}. This is a typical
H-bonding geometry, with the OH bond of the donor directed towards the O atom of the acceptor. 
We define the dimer binding energy $E_{\rm b}^{\rm dim}$ as twice the energy of an isolated 
equilibrium monomer minus the energy of the dimer in its global-minimum geometry. According to
CCSD(T), $E_{\rm b}^{\rm dim}$ is $217.6 \pm 2.0$~meV, the equilibrium O-O distance $R_{\rm O O}^{\rm dim}$ being
$2.909$~\AA~\cite{tschumper2002a}; these values are consistent with the somewhat uncertain 
experimental values~\cite{curtiss1979a,odutola1980a,mas2000a}.

Acceptable DFT approximations must reproduce benchmarks for the binding energy $E_{\rm b}^{\rm dim}$ and the geometry of the 
global minimum, and in fact most DFT work on the water dimer has focused exclusively on this configuration. However, this is not enough,
because both in the liquid and in compressed ice phases water monomers approach each other closely in non-H-bonded
geometries, and the energetics of such geometries is very important. A simple way of going beyond the global minimum is to study the
set of configurations on the energy surface of the dimer known as the Smith stationary points (Fig.~\ref{fig:Smith_SP})~\cite{smith1990a}, some of
which resemble geometries found in dense ice structures. We also review here assessments of XC functionals made using large 
statistical samples of dimer geometries designed to be relevant to condensed phases.

We discuss first local and semi-local XC functionals, including the local density approximation 
(LDA~\cite{kohn1965a,ceperley1980a,perdew1981a}), GGAs of different kinds, 
and hybrids, confining ourselves initially to the global minimum geometry. (See the Appendix for information about
the relevant XC approximations.) Several extensive surveys have been published on the predictions  of semi-local
functionals for the water dimer~\cite{santra2007a,xu2004b,dahlke2005a}.
In some work on the dimer, full basis-set convergence was not achieved, so that the accuracy of the functionals
themselves was not completely clear. However, Ref.~\cite{santra2007a} reported calculations 
very close to the complete basis set (CBS) limit
for 16 semi-local functionals applied to the dimer and other small water clusters. Table~\ref{tab:dimer_gm} reproduces some of the
$E_{\rm b}^{\rm dim}$ and $R_{\rm O O}^{\rm dim}$ values from that work, supplemented with results from Ref.~\cite{zhang2011a}
and from our own calculations performed in the present work. We performed our own 
calculations using the {\sc molpro}~\cite{molpro} and {\sc vasp}~\cite{vasp}
codes, following the procedures described elsewhere~\cite{alfe2014a}. The wide spread of predicted binding energies
is striking. The LDA is clearly unacceptable, since it overbinds the dimer by nearly a factor of 2~\cite{lee1994a,lee1995a},
and there are other functionals, such as PBEsol~\cite{perdew2008a}, which also overbind significantly. At the other extreme, functionals such
as BLYP and revPBE~\cite{zhang1998a} underbind quite seriously. Among 
the best functionals for $E_{\rm b}^{\rm dim}$ are PBE and its hybrid version PBE0.
It will become clear below that functionals predicting a good $E_{\rm b}^{\rm dim}$ can still give a poor description of larger
clusters, ice structures and liquid water. The wide variability of $E_{\rm b}^{\rm dim}$ with different
semi-local functionals will turn out to be crucial for the understanding of extended water systems. 

The reason why different semi-local functionals give such different binding energies in dimers of small molecules such as water is
well known. It was recognized long ago~\cite{harris1985a} that the gross overbinding given by LDA arises from a spurious
exchange attraction. In GGAs, this spurious attraction is suppressed by the exchange-enhancement factor $F_{\rm X} ( s )$,
which depends on the reduced gradient $x = | \nabla \rho | / \rho^{4/3}$ of the electron density $\rho$
through the quantity $s = x / ( 2 ( 3 \pi^2 )^{1/3} )$.
Roughly speaking, the exchange-overlap interactions of BLYP and revPBE are strongly repulsive and those of
PBE and PW91 are weakly repulsive because of the very different behavior of their $F_{\rm X} ( s )$ factors
in the region of large $s$ where the tails of the monomer densities 
overlap~\cite{lacks1993a,zhang1997a,wu2001a,kannemann2009a,murray2009a,kanai2009a}. 
This important difference between GGAs will be referred to again several times.

Dispersion plays a vital role in the energetics of water systems, as we shall see later, but is not
correctly described by the semi-local functionals discussed above. The past 20 years have seen the introduction
of several different ways of accounting for dispersion (see the Appendix, and recent reviews~\cite{klimes2012a,dilabio2014a}). 
One approach consists of the addition of potentials of various kinds to existing functionals, first
explored nearly 20 years ago (see e.g. Refs.~\cite{meijer1996a,gianturco2000a,wu2001a})
and then extensively developed by Grimme~\cite{grimme2004a,grimme2006a,grimme2010a},
Tkatchenko and Scheffler (TS)~\cite{tkatchenko2009a} and others. An alternative is the
incorporation of explicitly non-local correlation functionals, pioneered by Lundqvist, Langreth and 
others~\cite{dion2004a,klimes2010a,vydrov2010a}. The DCACP method of Rothlisberger and 
co-workers~\cite{vonlilienfeld2004a} and the closely related DCP method of Ref.~\cite{torres2012a}
 are also noteworthy. In all these methods, a representation
of non-local correlation energy is added to a chosen semi-local functional. 
The naming of these different approaches is not completely uniform in the literature, so we summarize
briefly the nomenclature used throughout this review. The 
approach now generally known as the Grimme method comes in three versions,
which we denote by func-D1, func-D2 and func-D3, where ``func'' is the name of the semi-local or hybrid
functional to which dispersion is added. Similarly, we denote TS methods by func-TS, and methods based on the non-local functional
of Ref.~\cite{dion2004a} by func-DRSLL (the acronym DRSLL stands for the authors of Ref.~\cite{dion2004a}).
We denote by rPW86-DF2 the method of Ref.~\cite{lee2010a} (sometimes known as LMKLL after the authors of this Ref.), 
which employs a modified form of the
DRSLL non-local correlation functional added to a revised version of the PW86 semi-local functional~\cite{perdew1986a}.

We summarize in Table~\ref{tab:dimer_gm} the dimer binding energies $E_{\rm b}^{\rm dim}$ and equilibrium 
O-O distances predicted for the global minimum geometry by some of these schemes.
This shows that the addition of dispersion to a semi-local functional always increases $E_{\rm b}^{\rm dim}$, as
expected. For BLYP and revPBE, which significantly underestimate $E_{\rm b}^{\rm dim}$,
the dispersion-inclusive versions BLYP-D3 and revPBE-DRSLL give improved values of $E_{\rm b}^{\rm dim}$,
though the latter functional is still significantly underbound. By contrast, the addition of TS dispersion
to PBE and PBE0, which already gave accurate values of $E_{\rm b}^{\rm dim}$, inevitably worsens
the predictions. The DRSLL-type functionals are particularly instructive in this regard. Their original form~\cite{dion2004a}, based on
revPBE, generally underbinds molecular dimers, so the underestimate of $E_{\rm b}^{\rm dim}$
 for the H$_2$O dimer by revPBE-DRSLL comes as no surprise. It was pointed out ~\cite{gulans2009a,klimes2010a} that better binding
 energies are obtained if less repulsive semi-local functionals are used in place of revPBE. It turns out that
 PBE is too weakly repulsive, so that PBE-DRSLL generally overbinds molecular dimers, including (H$_2$O)$_2$.
 However, if the exchange functional is appropriately tuned, much better approximations can be obtained.
 This is illustrated in the Table by the optPBE-DRSLL and optB88-DRSLL approximations, which are based on tuned forms
 of the PBE and B88 functionals. Similar arguments underlie the rPW86-DF2 non-local functional~\cite{lee2010a}. The comparisons
 shown in the Table indicate that the addition of dispersion increases $E_{\rm b}^{\rm dim}$ by up to $\sim 35$~meV.
 This is comparable with the variation of $E_{\rm b}^{\rm dim}$ resulting from different choices of semi-local or hybrid
 functional. We shall see in Sec.~\ref{sec:liquid} that addition of dispersion to a semi-local functional can bring
 large changes in the structure and equilibrium density of the liquid, so that errors as large as $35$~meV
 in $E_{\rm b}^{\rm dim}$ are important.

We noted earlier the importance of accuracy for configurations other than the global minimum,
and many authors have drawn attention to the role of non-H-bonded pair configurations in condensed phases
of water~\cite{wu2001a,dahlke2005a,ireta2004a,lin2009a,wang2011a,zhang2011a}. The Smith stationary points (Fig.~\ref{fig:Smith_SP})
provide some relevant geometries, but the only systematic studies of DFT errors in these configurations
appear to be those of Refs.~\cite{gillan2012a,anderson2006a}. Anderson and 
Tschumper~\cite{anderson2006a} studied 10 different GGA and hybrid
methods. All approximations gave the correct energy ordering, but most of them overestimated 
the energies relative to that of the global minimum, particularly for
configurations SP4, SP5 and SP6, which closely resemble configurations in ice VIII. 
Relative energies for some semi-local and hybrid functionals are summarized in Table~\ref{tab:Smith_energies}, which shows
that PBE and BLYP both overestimate the relative energy of configuration~SP6 by over $30$~meV.
Hybrid approximations are appreciably better, but still overestimate the relative
energies. Calculations of the energies relative to the global minimum with dispersion-inclusive
methods do not appear to have been published, so we have made our own (see Table~\ref{tab:Smith_energies}). We find that
optPBE-DRSLL and rPW86-DF2 are both quite satisfactory, their errors in the relative energies all being less than
$15$~meV. However, both PBE-TS and PBE0-TS are less satisfactory, giving relative energies of configurations
4, 5 and 6 in error by $\sim 30$~meV.

A characterization of XC errors for the global minimum and some special geometries is illuminating and useful,
but a full characterization should cover all relevant O-O distances and monomer orientations. One way of doing this
is to analyze the errors for large samples of dimers drawn from a MD simulation of the liquid, as has been done
by Santra \emph{et al.}~\cite{santra2009a} and Gillan \emph{et al.}~\cite{gillan2012a}. For this purpose, we must separate
1-body and 2-body errors in the sense of the many-body expansion
(MBE)~\cite{hankins1970a,xantheas1994a,pedulla1998a}, which will also be important
later. In general, the total energy $E ( 1, 2, \ldots N )$ of a system of $N$ monomers can be exactly expressed as:
\begin{equation}
E ( 1, 2, \ldots N ) = \sum_i E^{(1)} ( i ) + \frac{1}{2} {\sum_{i, j}}^\prime E^{(2)} ( i, j ) +
\frac{1}{6} {\sum_{i, j, k}}^\prime E^{(3)} ( i, j, k ) + \ldots \; .
\label{eqn:MBE}
\end{equation}
Here, $E^{(1)} ( i )$ is the 1-body energy of monomer $i$ in free space, with the argument
$i$ being short-hand for the set of coordinates specifying its geometry. Similarly, $E^{(2)} ( i, j )$ is the
2-body energy of dimer $( i, j )$, i.e. its total energy minus the 1-body energies of monomers $i$ and $j$, with
the arguments $i$ and $j$ being short-hand for the geometries of the monomers (the prime on the summation
indicates the omission of terms $i = j$). The 3-body and higher terms needed for systems larger than the dimer
 are defined analogously. The zero of energy is conveniently taken as $N$ times the energy of an isolated equilibrium 
 monomer. To analyze the errors of a chosen XC functional, the total energy of each dimer in the sample and the 1-body 
 energies of its monomers are computed, and the 2-body energy is obtained by subtracting from the total dimer energy
the sum of the two monomer energies. The 1- and 2-body errors are then obtained by subtracting benchmark values of
the 1- and 2-body energies, typically computed with CCSD(T).

\begin{figure}
\centering
\begin{tabular}{cc}
\includegraphics[width=0.5\linewidth]{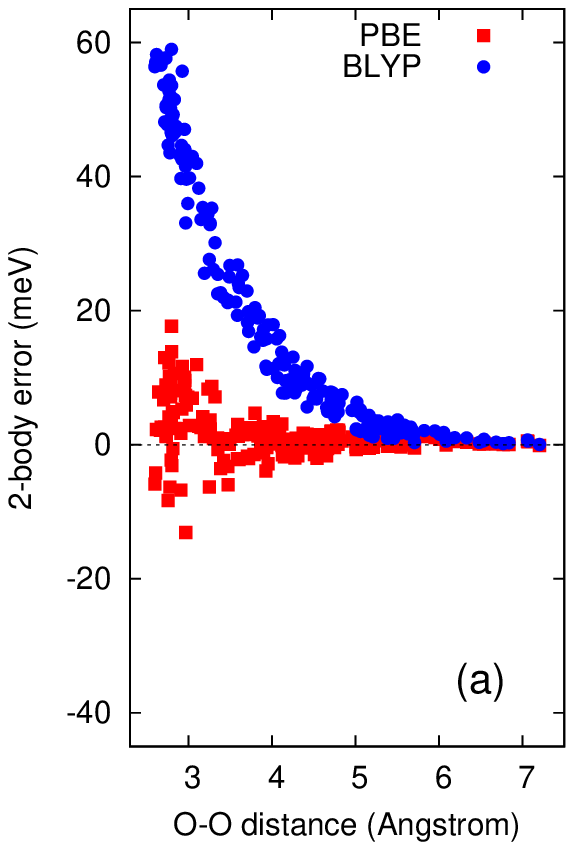} &
\includegraphics[width=0.5\linewidth]{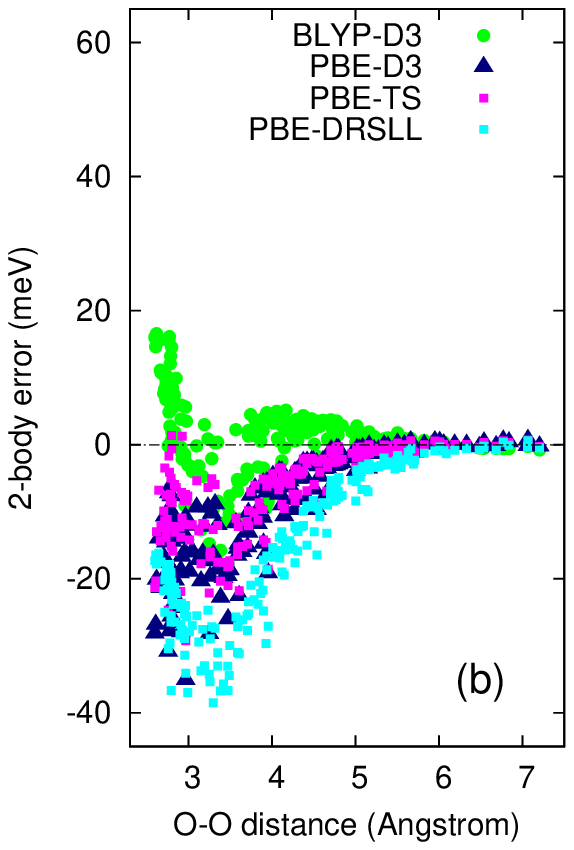}
\end{tabular}
\caption{Errors of (a) semi-local functionals PBE and BLYP and (b) dispersion-inclusive 
functionals BLYP-D3, PBE-D3, PBE-TS and PBE-DRSLL for the 2-body energies of a sample of dimer geometries drawn 
from a MD simulation of liquid water~\cite{gillan2012a}.}
\label{fig:dimer_2body_errors}
\end{figure}

Santra \emph{et al.}~\cite{santra2009a} analyzed the errors of the BLYP, PBE and PBE0 functionals for dimers drawn
from the liquid. BLYP values of the 2-body energies were underbound by an average of $\sim 40$~meV,
while PBE and PBE0 values were overbound by $\sim 10$~meV. Interestingly, the errors 
of all three functionals showed a scatter of around $\pm 15$~meV about their averages
at typical O-O nearest-neighbor distances of $\sim 2.8$~\AA. There may be a link here with the errors in the
relative energies of the Smith stationary points. An interesting finding from the same work~\cite{santra2009a}
was that the known enhancement of H-bond energy by elongation of the donor OH bond is significantly
exaggerated by both GGA functionals. The work of Ref.~\cite{gillan2012a} examined the 2-body errors computed
with semi-local functionals for a thermal sample of dimers covering a range of O-O distances from $2.5$
to over $7.0$~\AA. This revealed the expected systematic underbinding of BLYP
over this range, and the much smaller errors of PBE, as illustrated 
in Fig.~\ref{fig:dimer_2body_errors}. We also present in this Figure calculations on the same thermal sample 
performed with the BLYP-D3, PBE-D3,  PBE-TS and PBE-DRSLL
functionals. We see that the excessive repulsion of BLYP is largely eliminated by BLYP-D3, which becomes
slightly overbound between $3.0$ and $4.0$~\AA. However, the PBE-based functionals
are all overbound, with the overbinding errors of PBE-DRSLL being particularly strong in this
range. Interestingly, the typical difference of 2-body energy between the BLYP and PBE
functionals is comparable with the shifts due to dispersion, so that addition of D3 dispersion to BLYP
brings its 2-body energy rather close to that of uncorrected PBE. The typical energy difference
between the dispersion-corrected functionals BLYP-D3 and PBE-DRSLL is on the same scale
as the energy shifts due to dispersion, so that the robustness of the methods clearly needs discussion.
It is notable that the substantive differences between the various PBE-based methods in the region $3.0 - 4.0$~\AA\
exist despite the close agreement between their global-minimum binding energies (Table~\ref{tab:dimer_gm}).

It has been shown by Bart\'{o}k \emph{et al.}~\cite{bartok2013a} that machine-learning
techniques operating on very large thermal samples of dimers can be used to create very accurate,
but rapidly computable representations of the 1- and 2-body errors of chosen XC functionals.
These techniques, based on the GAP (Gaussian approximation potential) method of machine
learning~\cite{bartok2010a}, give a way of compensating almost exactly for the
1- and 2-body errors of any chosen XC functional. We review below (Secs.~\ref{sec:larger_clusters} and 
\ref{sec:ice}) work on large water clusters and ice based on XC functionals corrected in this way.

In summary, we have seen that both semi-local and dispersion-inclusive functionals vary quite 
widely in their predictions of the H-bond energy. The variability affects the  
2-body energy in the important region of O-O separation extending from $2.5$ to at least $4.0$~\AA.
We have noted the importance of the exchange-enhancement factor as one cause of this variability.
The variation between different functionals is comparable with the increase of binding energy resulting
from the addition of dispersion. The energies of the Smith stationary points relative to the global minimum 
are significantly overestimated by both semi-local and dispersion-inclusive functionals.


\section{Cooperative hydrogen-bonding in small clusters}
\label{sec:coop}

It has long been known that hydrogen bonding in water and in many other molecular systems
is a cooperative effect: if molecule A donates a H-bond to molecule B, the propensity of molecule B
to donate a H-bond to molecule C is thereby 
enhanced~\cite{frank1957a,elrod1994a,jeffrey1997a,karpfen2002a,xu2002a}. 
The cooperativity manifests itself in the non-additivity of H-bond energies. In suitable geometries, an assembly
of water molecules is stabilized by the mutual enhancement of 
H-bonds~\cite{hankins1970a,white1990a,xantheas1994a,xantheas2000a}, so that the overall binding energy
is greater than the sum of dimer binding energies. This non-additivity of binding energies can be understood
as resulting from molecular polarizability: the electron cloud on each monomer is distorted by the
electric fields of its neighbors, and this changes its electrostatic interaction with other monomers. 
We noted in the Introduction that this is a strong effect, since molecular polarizability
enhances the dipole moment of water monomers in ice and liquid water by
$40 - 50$~\% above that of monomers in free space~\cite{coulson1966a,batista1998a,silvestrelli1999a,badyal2000a}.
It was suggested by Frank and Wen~\cite{frank1957a} that the cooperativity
would play an important role in the dynamical making and breaking of H-bonds in liquid water, and subsequent
calculations and experiments have fully confirmed the importance of H-bond cooperativity 
in water systems of all kinds, including ice and clusters (see e.g. Ref.~\cite{luck1998a}).
It is clearly important to know whether DFT approximations reproduce these cooperative
effects. 

Fortunately, H-bond cooperativity is already important in small water clusters, where it has 
been extensively studied both experimentally and 
theoretically~\cite{xantheas1993a,saykally1993a,xantheas1994a,liu1996b,gregory1996a,xantheas2000a,keutsch2003a,glendening2005a,santra2007a,bryantsev2009a,cobar2012a,neela2010a,guevara-vela2013a}. 
These systems provide a simple way of testing the accuracy
of DFT approximations in describing non-additivity, since very accurate benchmarks are readily available. We 
discuss here the trimer, tetramer and pentamer, leaving till Sec.~\ref{sec:hexamer} the hexamer, which raises 
issues beyond H-bond cooperativity. Experiments and accurate quantum chemistry calculations show that
the most stable configurations of the (H$_2$O)$_n$ clusters with $n = 3$, $4$ and $5$ are quasi-planar and 
cyclic~\cite{xantheas1993a,xantheas1994a,liu1996b,xantheas2000a}.
An important indication of the progressive strengthening of the H-bonds
is that the O-O distance shortens from $\sim 2.91$~\AA\ in the dimer to $2.72$~\AA\ in the pentamer~\cite{liu1996b,santra2007a}. Another 
commonly used measure for the strength of a H-bond is the red-shift of the intramolecular stretching frequency of the donor
O-H bond. Theory and experiment both find an increasing red-shift on passing from the trimer to the pentamer~\cite{buck2000a}.

For reference data on the non-additivity of the binding energies, we rely on benchmark MP2 and CCSD(T)
calculations near the CBS limit, since accurate experimental data is not available. (MP2 is the second-order
M\o{}llet-Plesset approximation~\cite{helgaker2000a}, which is often nearly as accurate as CCSD(T) for water systems.) The non-additivity
can be quantified using the many-body expansion (MBE) introduced above in eqn~(\ref{eqn:MBE}). The non-additive
parts of the energy are characterized by the 3-body and higher-body terms $E^{(n)}$. For a small cluster (H$_2$O)$_n$
in a given geometry, it is straightforward to compute benchmark total energies of all the monomers, dimer, trimers, etc.
that can be formed from the cluster, and from these the terms $E^{(1)} ( i )$, $E^{(2)} ( i, j )$, $E^{(3)} ( i, j, k )$ etc. of the MBE
can be extracted. It has been shown by Xantheas~\cite{xantheas2000a} that the 3-body and higher components of the energy play a vital role
in determining the relative energies of different conformations of the water trimer, tetramer and pentamer.

The performance of a wide variety of semi-local and hybrid XC functionals on the binding energies of the water trimer, tetramer and pentamer
in their most stable geometries has been assessed by Santra \emph{et al.}~\cite{santra2007a}. MP2  energies near the CBS
limit were used as reference, the evidence being that these energies should differ from CCSD(T) values by no more than
a few meV per H-bond. The functionals studied included the popular GGAs PBE, PW91 and BLYP and some less
common ones such as XLYP, PBE1W, mPWLYP and BP86, the hybrid functionals PBE0, B3LYP and X3LYP, and
the meta-GGA functional TPSS (references to the definitions of these functionals can be found in Ref.~\cite{santra2007a}). 
As expected from their performance on the dimer, BLYP was found to be always quite strongly
underbound, and PBE and PW91 always overbound. The B3LYP functional was a considerable improvement on BLYP, while the
hybrids PBE0 and X3LYP gave almost perfect binding energies. The TPSS meta-GGA turned out to be somewhat worse
than its parent functional PBE. 

For present purposes, the most important finding of Ref.~\cite{santra2007a} is that all the functionals
reproduce semi-quantitatively the cooperative enhancement of the H-bond energies given by the MP2
benchmarks. According to the benchmarks, the binding energy per H-bond is enhanced by a modest $6$~\%
in the trimer, increasing to a more impressive $46$~\% in the pentamer. With very few exceptions, all the
functionals give enhancements between $4$ and $8$~\% in the trimer and between $45$ and $55$~\% in the pentamer.
Notably, the enhancement is overestimated by almost all the functionals in the tetramer and pentamer, perhaps because
most functionals overestimate the polarizability of the water monomer (see Sec.~\ref{sec:monomer}). It is also noted in Ref.~\cite{santra2007a}
that for all the functionals the error in the binding energy per H-bond is almost independent of cluster size, though the
error becomes more positive (more strongly bound) for some functionals, including PBE. This suggests that for a given
DFT functional its error in the H-bond energy of the dimer is likely to be a good guide to its error in the H-bond energy
in larger water aggregates. The authors also note that the known shortening of the equilibrium O-O distance with increasing
cluster size is also semi-quantitatively reproduced by all the functionals.

To summarize, the work on small clusters up to the pentamer shows that H-bond cooperativity becomes a strong effect
as we go to larger aggregates, but most XC functionals appear to describe the non-additivity of energies fairly accurately.
Semi-local functionals generally overestimate the enhancement of H-bond strength, but hybrid functionals do better than GGAs.


\section{Compact versus extended geometries: the hexamer}
\label{sec:hexamer}

\begin{figure}
\centering
\begin{tabular}{cc}
\includegraphics[width=0.5\linewidth]{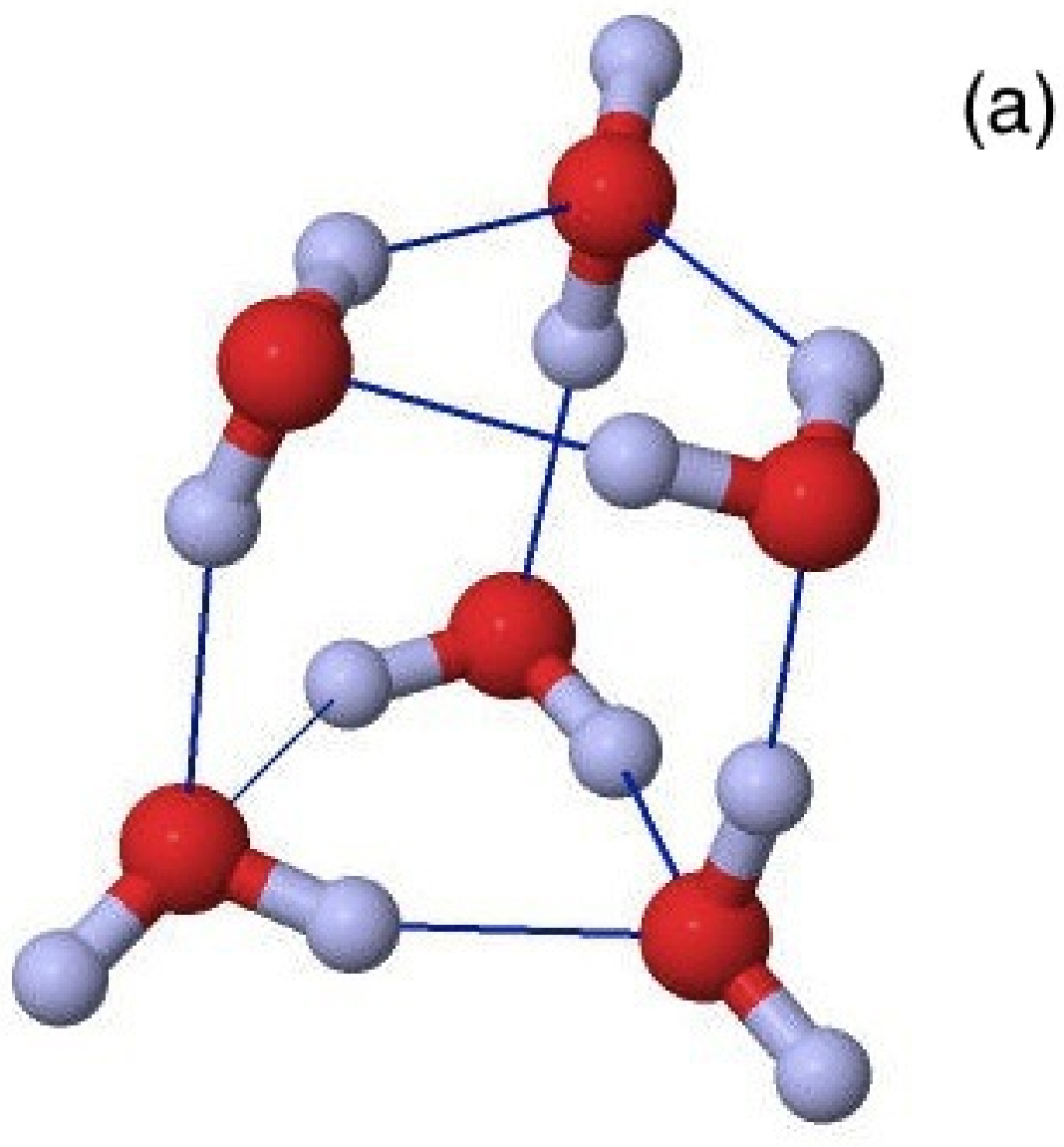} &
\includegraphics[width=0.5\linewidth]{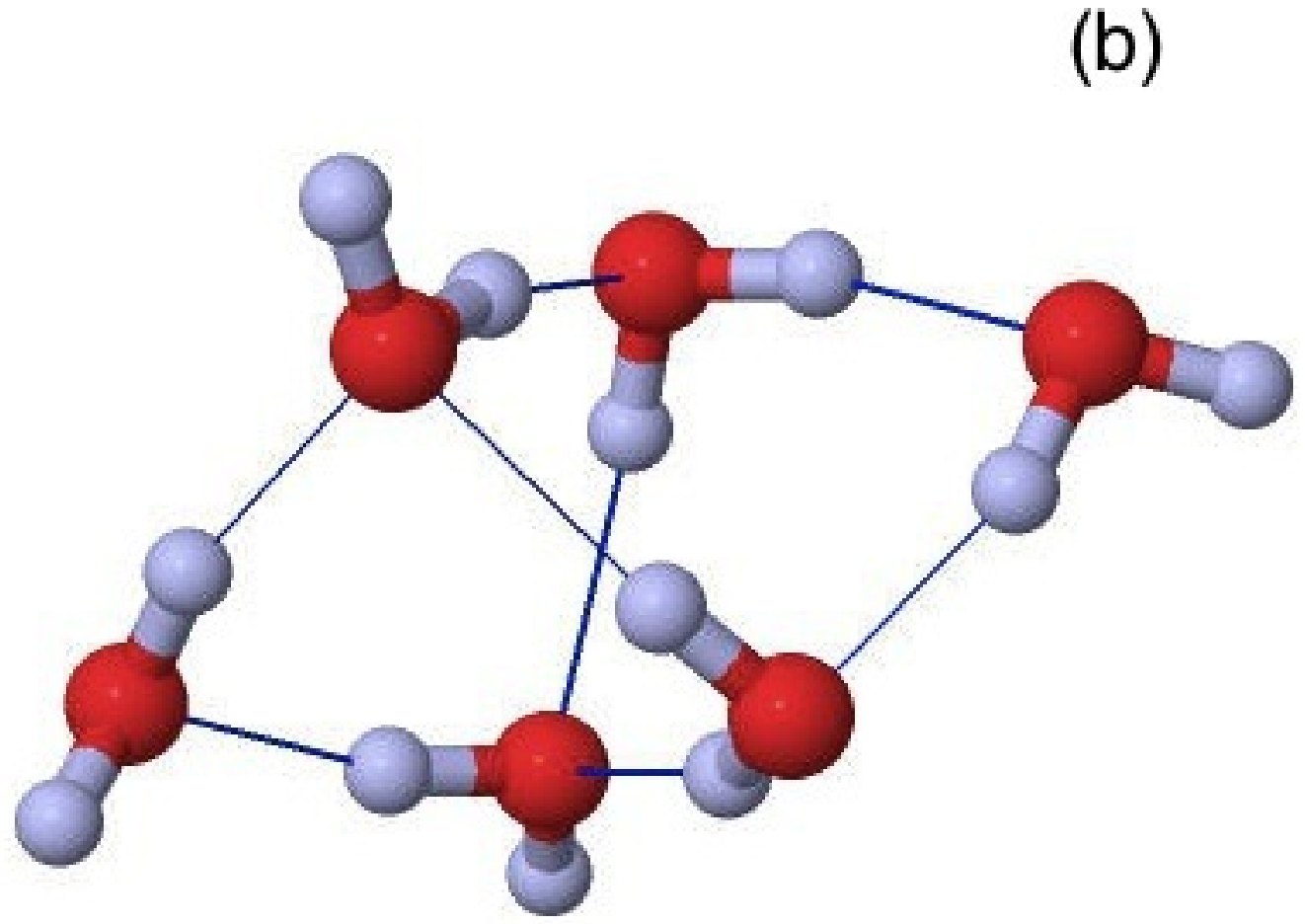} \\
\includegraphics[width=0.5\linewidth]{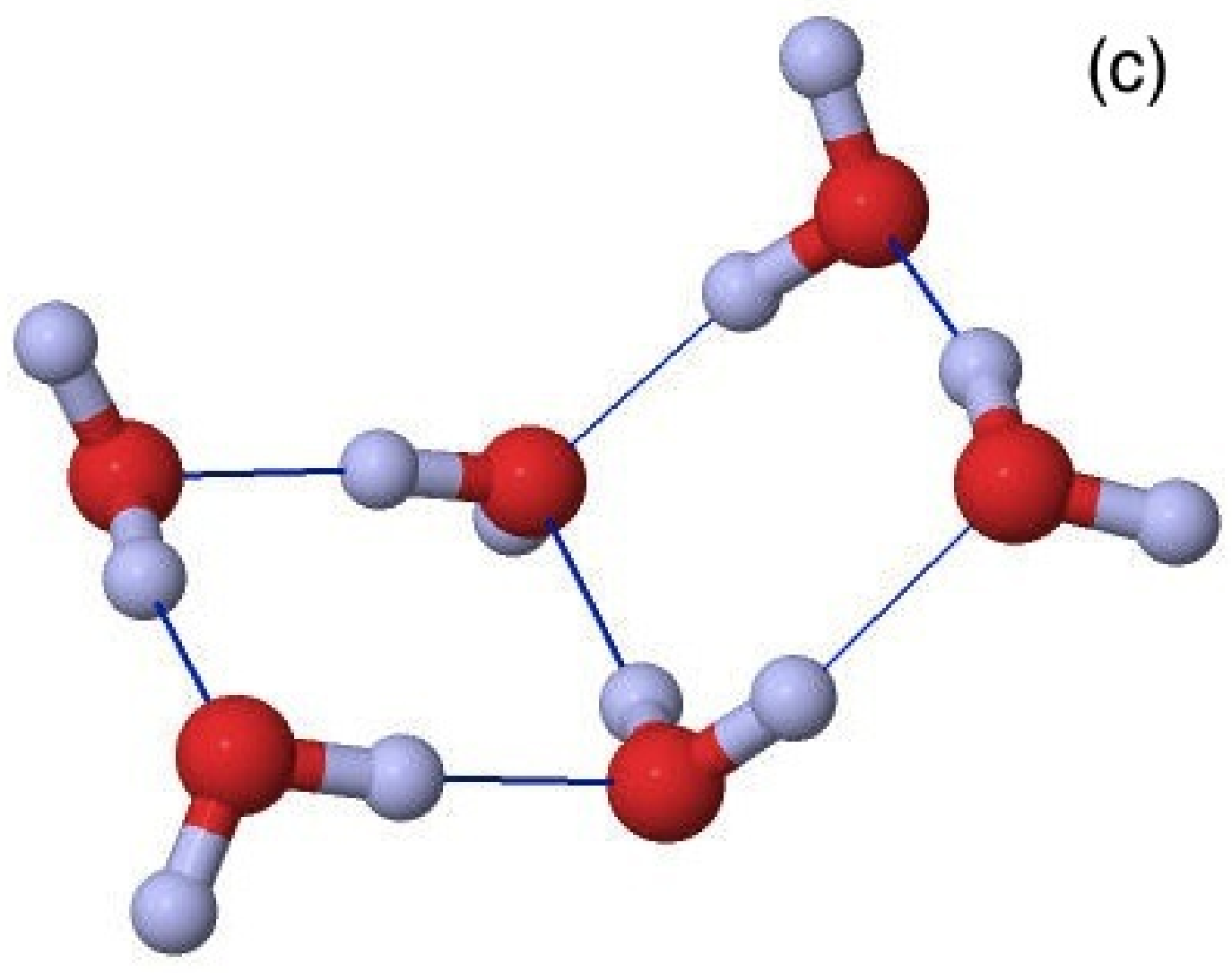} &
\includegraphics[width=0.5\linewidth]{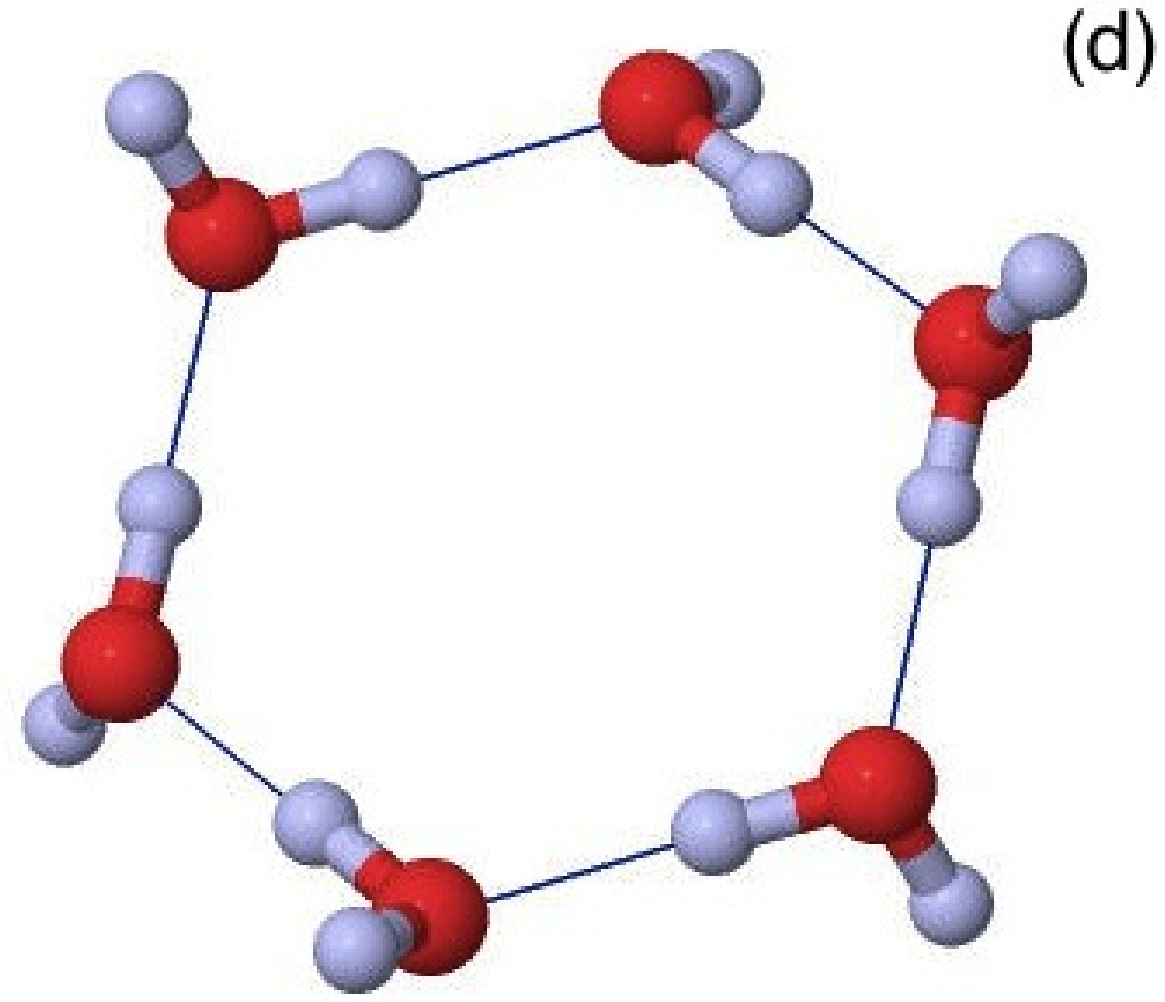}
\end{tabular}
\caption{Isomers of the water hexamer: (a)~prism, (b)~cage, (c)~book and (d)~ring.}
\label{fig:hexamers}
\end{figure}

The hexamer occupies a special place in water studies, because it is the smallest cluster for which three-dimensional
structures compete energetically with the two-dimensional cyclic structures just discussed for the trimer, tetramer and pentamer. There 
are many local minima on its complex energy surface~\cite{tsai1993a,kryachko1999a,hincapie2010a},
but here we pay particular attention to four of them, known as the prism, the cage, the book and the ring (Fig.~\ref{fig:hexamers}).
In the ring, each monomer is H-bonded to two neighbors, and all six H-bonds are of canonical form, with the
donor O-H bond pointing directly at the acceptor O atom. In the prism, by contrast, each monomer is 3-fold
coordinated, but the nine H-bonds are strongly distorted. In their coordination and H-bond count, the cage resembles the prism and
the book resembles the ring. Are the extended ring and book structures with fewer but
stronger H-bonds more or less stable than the compact cage and prism with more but weaker bonds? This is an important
question, because the competition between compact and extended structures is central to the energetics
of solid and liquid water phases.

In fact, CCSD(T) calculations close to the CBS limit leave no doubt that the energy ordering from lowest to highest
is prism~$<$ cage~$<$ book~$<$ ring~\cite{gillan2012a,olson2007a,dahlke2008a,bates2009a,wang2010a}. 
This ordering is confirmed by diffusion Monte Carlo (DMC) calculations~\cite{foulkes2001a,needs2010a},
which give total binding energies
relative to free monomers and total energy differences between the isomers in accord with CCSD(T) to within
$\sim 5$~meV/monomer~\cite{santra2008a}. The fact that the compact prism and cage are more stable than extended structures such as the ring was
already suggested by early MP2 calculations~\cite{tsai1993a,pedulla1998a,xantheas2002a}. It is now established that MP2 in the CBS limit gives the same stability
ordering as CCSD(T) for the prism, cage, book and ring, though MP2 underestimates the difference of total
binding energy of the ring and prism by $\sim 25$~meV~\cite{olson2007a,santra2008a,dahlke2008a,bates2009a}.
The prism and cage isomers are very close in energy, but CCSD(T) makes
the prism more stable by $\sim 10$~meV~\cite{bates2009a}. (We note that this statement refers to energy-minimized structures;
in the real world, zero-point and thermal vibrational energies appear to be large enough to reverse the stabilities
of these two isomers~\cite{liu1996a,wang2012a}.)

Several detailed studies have investigated the accuracy of DFT approximations for the relative energies of the different
isomers and also for their overall binding energies with respect to free monomers. Between them, these studies cover a 
wide variety of methods. The works of Dahlke \emph{et al.}~\cite{dahlke2008a} and 
Santra \emph{et al.}~\cite{santra2008a} investigated respectively 11 and 12 different
functionals, including GGAs, meta-GGAs, hybrids and hybrid-meta-GGAs. Later 
studies~\cite{gillan2012a,wang2010a,gillan2014a,pruitt2013a}
covered a range of GGAs and hybrids, and Ref.~\cite{pruitt2013a} studied a number of dispersion-inclusive methods.
The remarkable outcome of these studies is that almost all the semi-local approximations
erroneously make the ring or book more stable than the prism and cage. The only exceptions reported in these papers are the
three Minnesota functionals M06-L, M05-2X and M06-2X~\cite{dahlke2008a}. On the positive side, a number of commonly
used functionals (e.g. PBE, PBE0) give quite accurate values for the average binding energies of the four isomers,
though others (e.g. BLYP, revPBE) are seriously underbound. The problems of semi-local XC functionals 
are illustrated in Table~\ref{tab:hexamer_binding}, where we compare their predicted binding
energies with the CCSD(T) and DMC benchmarks.

The semi-local functionals just referred to do not explicitly describe dispersion, and it is natural to assume that this
is the main cause of the erroneous stability ordering of the hexamers. This makes physical sense, because compact
structures will be more strongly stabilized than extended structures by a pairwise attraction. The importance of
dispersion has been confirmed by several studies of the hexamer~\cite{santra2008a, kelkkanen2009a,silvestrelli2009a},
showing that different dispersion-inclusive DFT methods all correct the wrong stability
ordering and bring the relative energies of the isomers into respectable agreement with the benchmarks,
as can be seen from the illustrative examples given in Table~\ref{tab:hexamer_binding}.

In spite of this compelling evidence, there are clear indications that dispersion is not the only source of 
errors in the XC functionals. These indications come from the many-body analyses reported in
several papers. We saw in previous Sections how the binding energies of clusters can
be separated into their 1-body, 2-body, and beyond-2-body components.
The same approach can be applied to the errors of XC functionals,
i.e. the deviations of approximate DFT energies from benchmarks. If the errors of XC approximations
for the hexamer were entirely due to poor dispersion, we would expect them to be mainly 2-body
errors, because beyond-2-body dispersion is generally much less significant than 2-body dispersion in water
systems. (See e.g. Ref.~\cite{vonlilienfeld2010a}, which indicates that 3-body dispersion contributes \emph{ca.}~100
times less than 2-body dispersion to the cohesive energy of ice Ih.)
However, the reality is more complex. For the energy differences between the isomers, the errors of
the BLYP and revPBE approximations are indeed mainly 2-body errors, but it turns out that the errors of PW91, PBE
and PBE0 are largely beyond-2-body errors~\cite{wang2010a,gillan2012a,gillan2013a,gillan2014a}. 
The importance of beyond-2-body errors in the energetics of the hexamer has been noted in Ref.~\cite{pruitt2013a}.
On the other hand, the average binding energy of the four isomers
is quite accurately given by PBE and PBE0 (see above), which reproduce both its 2-body and
beyond-2-body components. (By ``average binding energy'' we mean the sum of the binding energies
of the four isomers divided by 4.) However, the large error in the average binding energy given by BLYP
arises from its excessive 2-body repulsion together with a smaller but still significant beyond-2-body
overbinding~\cite{gillan2012a,gillan2013a}. These facts indicate that the errors cannot be explained by incorrect dispersion alone.

It has been proposed recently~\cite{gillan2014a} that the many-body errors of GGAs for water  (and 
other molecular systems) are closely linked
to the choice of exchange-enhancement factor $F_{\rm X}$, which we know has an important influence
on the 2-body interaction energy (see Sec.~\ref{sec:dimer}). It appears that an $F_{\rm X}$ that gives excessive
exchange-overlap repulsion in the dimer also tends to produce an overly attractive 3-body
interaction, while an $F_{\rm X}$ whose 2-body exchange-overlap repulsion is too weak
produces a spurious 3-body repulsion. The suggestion is that the accurate dimer binding
energy given by PBE results from an overly weak exchange-repulsion mimicking
the missing dispersion, the indirect consequence being the erroneous stability ordering
in the hexamer due to the exaggerated 3-body repulsion. Conversely, the excessive
2-body repulsion of BLYP, to which dispersion should be added, is linked to the unduly attractive
3-body interaction noted above. The idea of an inverse correlation between the 2-body and beyond-2-body
errors of semi-local functionals is confirmed by recent work on a wide range of molecular trimers~\cite{rezac2015a}.

The extensive work on the hexamer teaches us several lessons. First, good accuracy for the dimer
(and other clusters smaller than the hexamer) is no guarantee
of even qualitative correctness for the relative stability of compact and extended conformations. Second,
many-body errors may be at least as important as 2-body errors, but the relative importance of the
two kinds of error depends strongly on the exchange-correlation functional. Third, dispersion is crucial,
and demands to be correctly described. Fourth, dispersion is not the only source of errors, and there is evidence
that the many-body errors are associated with exchange. We shall see in the following how these lessons are
reinforced by DFT work on larger clusters, ice structures and the liquid.


\section{Larger clusters}
\label{sec:larger_clusters}
For many years, quantum-chemistry benchmark calculations were feasible only on rather small clusters, with the 
hexamer representing the practical limit for basis-set converged CCSD(T). However, technical innovations
are now helping to extend the calculations to much larger clusters without significant loss of accuracy.
In fact, well converged MP2 calculations were possible on clusters of 20 - 30 monomers
several years ago, but with the development of linear-scaling methods, they can now be 
applied to even larger systems~\cite{werner2003a,doser2009a}.
Corrections for the difference between CCSD(T) and MP2 can then be made using the many-body
expansion for this difference~\cite{gora2011a}. In addition, there is growing evidence that the accuracy of
quantum Monte Carlo methods~\cite{foulkes2001a,needs2010a} rivals that of CCSD(T) for non-covalent 
interactions~\cite{gurtubay2007a,ma2009a,santra2011a,gillan2012a,dubecky2013a,ambrosetti2014a}, and the mild scaling of these methods
with system size and their high efficiency on parallel computers make it practical to compute benchmarks for water systems
containing $30$ or more monomers~\cite{alfe2013a,morales2014a}.

The availability of very accurate benchmark energies for large clusters opens the interesting possibility
of tracing out the development of DFT errors as a function of cluster size. The notion here is that as we go
to ever larger clusters the errors should come to resemble ever more closely those seen in ice structures and
the liquid. Furthermore, an analysis of the errors into their 2-body and beyond-2-body components, similar
to that discussed above for the hexamer, allows us to deepen our understanding of the continuous connections
between these errors in clusters and in condensed phases.

These ideas have been explored~\cite{alfe2013a} for thermal samples of water clusters containing up to $27$
monomers, the benchmark energies being computed with the DMC technique. The sets of
configurations for each water cluster were drawn from MD simulations performed with a realistic
force field that treats the monomers as flexible and polarizable. For large enough clusters, these configurations
should be roughly typical of those found in small water droplets. The time-varying radius of a cluster of $N$ monomers
can be characterized by the quantity $R_{\rm gyr} ( t )$ defined by:
\begin{equation}
R_{\rm gyr} ( t )^2 = \frac{1}{N} \sum_{i = 1}^N \left| {\bf r}_i ( t ) - \bar{\bf r} ( t ) \right|^2 \; ,
\end{equation}
where ${\bf r}_i ( t )$ is the position of the O atom of monomer $i$ at time $t$, and
$\bar{\bf r} ( t )$ is the centroid of these O positions. This ``radius of gyration'' $R_{\rm gyr} ( t )$ fluctuates
in time as the cluster spontaneously breathes in and out, exploring compact and extended configurations.
We saw in Sec.~\ref{sec:hexamer} that the errors of GGAs for the hexamer grow more positive as we pass from extended to
compact isomers, and one might expect a similar dependence on $R_{\rm gyr}$ for thermal clusters.

\begin{figure}
\centering
\includegraphics[width=0.75\linewidth]{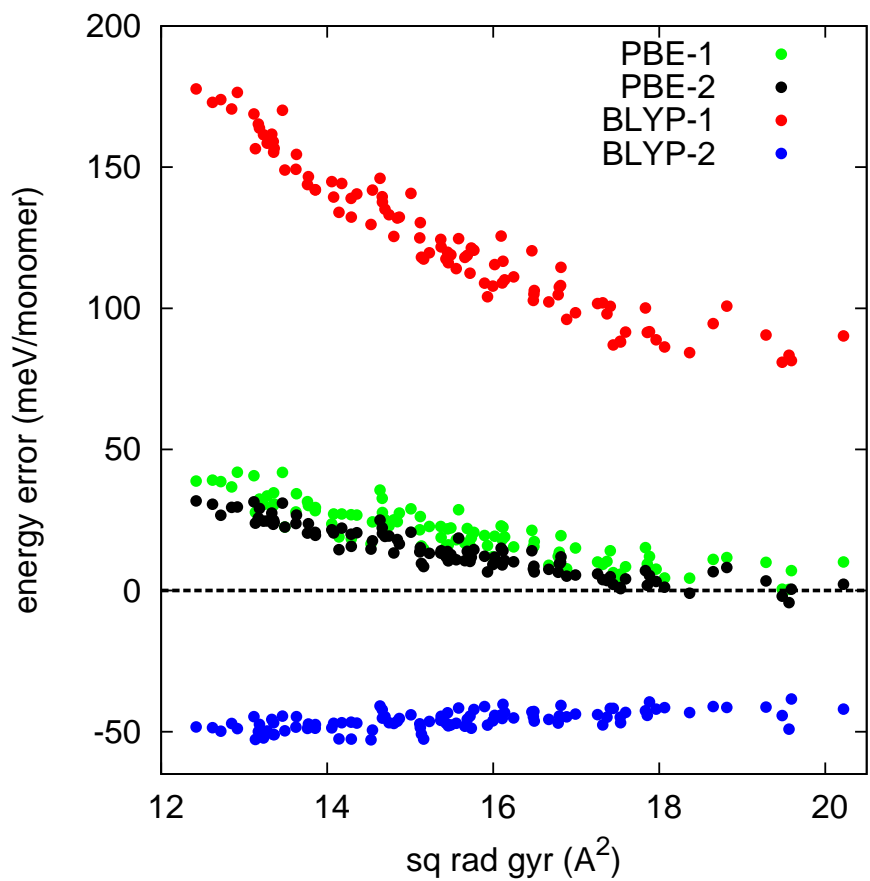}
\caption{Errors in computed binding energy per monomer (meV units) of the water 27-mer as a function of radius of gyration (see text).
Binding energies are computed with the PBE and BLYP functionals, corrected for 1- and 2-body errors,
with benchmark values from diffusion Monte Carlo (DMC) calculations. Adapted with permission from
J. Chem. Phys. {\bf 141}, 014104 (2014) .}
\label{fig:nanodrop27}
\end{figure}

This expectation is amply fulfilled for the GGAs examined so far, namely BLYP and PBE. The analysis of
the errors of these approximations has been reported~\cite{alfe2013a,alfe2014a} 
for thermal clusters containing $N = 6$, $9$, $15$ and $27$
monomers, with the GAP techniques mentioned above (Sec.~\ref{sec:dimer}) used to correct almost exactly for 1- and 2-body errors. If only
1-body errors are corrected (the resulting approximations are called BLYP-1 and PBE-1), it is found for all the clusters
that BLYP-1 has large positive errors, while PBE-1 has much smaller errors, the errors in both
cases growing more positive with decreasing $R_{\rm gyr}$ (see Fig.~\ref{fig:nanodrop27}). After correction for both 1- and 2-body errors (approximations
BLYP-2 and PBE-2), BLYP-2 has negative errors showing a weak downward trend with decreasing $R_{\rm gyr}$, while
PBE-2 has almost the same errors as PBE-1, trending upwards with decreasing 
$R_{\rm gyr}$ (Fig.~\ref{fig:nanodrop27}). All these trends are
very much the same as for the isomers of the hexamer, though the magnitude of the errors increases markedly with
cluster size. Exactly as for the hexamers, the erroneous destabilization of compact relative to extended configurations
is mainly a 2-body effect with BLYP, but mainly a beyond-2-body effect for PBE. We shall see exactly the same patterns
of erroneous energetics in the ice structures.


\section{Ice structures}
\label{sec:ice}

Water ice exhibits a rich and complex phase diagram. Besides the hexagonal Ih structure, familiar
as the ice and snow found in colder
regions of the Earth's surface, and the closely related cubic ice Ic, there are $15$ other experimentally
known structures~\cite{petrenko1999a}. Ice Ih and some of the other phases are ``proton disordered'', meaning that the molecular
orientations show a degree of randomness, with a corresponding orientational entropy. 
However, at low temperatures the proton-disordered phases all undergo  transitions to proton-ordered structures;
for example, ice Ih transforms to proton-ordered ice XI at \emph{ca.}~$72$~K~\cite{kawada1972a,tajima1982a}.
Quite moderate pressures of up to a few kbar are enough to
stabilize a series of well characterized structures. The energy differences between the structures are
remarkably small, being considerably less than ``chemical accuracy'' of  $1$~kcal/mol ($43.4$~meV). This means that ice energetics
provides an exquisitely delicate test of DFT methods. We will concentrate here mainly on the sublimation
energies and equilibrium volumes predicted by DFT approximations, with some
comments also on proton order-disorder energetics and the relative energies of the Ih and Ic phases.
When we refer to a computed ``sublimation energy'' $E_{\rm sub}$ here, we mean the energy of an isolated,
relaxed, static water monomer minus the energy per monomer of a relaxed, static ice structure, with no account
taken of zero-point vibrational energy. (A greater $E_{\rm sub}$ signifies a more strongly bound
ice structure.) When we compare a computed $E_{\rm sub}$ with experiment, a calculated or estimated
value of the zero-point vibrational contribution must first be removed from the experimental value.

The performance of LDA and GGAs for the sublimation energy and equilibrium volume of ice Ih was
first investigated by Hamann~\cite{hamann1997a}. To deal with the proton-disorder, Hamann followed Bernal
and Fowler~\cite{bernal1933a} in representing the structure approximately by a 12-molecule repeating cell, a procedure
that is known to incur only very small errors (see e.g. ref. ~\cite{pan2010a}). 
His calculations showed that LDA overbinds ice Ih
by $\sim 70$~\% and underestimates its equilibrium volume by $\sim 20$~\%, these very large
errors being expected from its poor treatment of the water dimer. The GGAs studied by Hamann
performed better, though substantial over- or under-binding was found in some cases.
More accurate DFT calculations on ice Ih were reported later by Feibelman~\cite{feibelman2008a}, who 
studied a rather broad set of GGAs, finding that the predicted sublimation energies come in the order
revPBE~$<$ RPBE~$<$ BLYP~$<$ PBE~$<$ AM05~$<$ PW91~$<$ PBEsol~$<$
LDA. (We gave references for most of these functionals earlier, but we note here the references for RPBE~\cite{hammer1999a},
AM05~\cite{armiento2005a} and PW91~\cite{perdew1992a,perdew1993a}.) 
Of the GGAs considered, PBE predicts a sublimation energy of \emph{ca.}~$640$~meV/H$_2$O, 
which is only \emph{ca.}~$30$~meV/H$_2$O larger 
than the experimental value of $610$~meV/H$_2$O \cite{whalley1984a}. 
This experimental value, which excludes zero-point contributions, was reported many years ago
by Whalley~\cite{whalley1984a}, but has since been corroborated by two kinds of high-level
electronic structure calculations. These consist of DMC calculations~\cite{santra2011a},
which gave $E_{\rm sub} = 605 \pm 5$~meV/H$_2$O, and CCSD(T) 
calculations~\cite{gillan2013a} implemented with an embedded many-body
expansion, which gave $601$~meV/H$_2$O. The substantial spread of $E_{\rm sub}$ values found in the GGA calculations
of Feibelman has been confirmed by other, more recent studies~\cite{brandenburg2015a,fang2013a}, as we show
in Table~\ref{tab:ice_energies}, where we summarize the $E_{\rm sub}$ values obtained from a range of
semi-local functionals. (The Table also reports dispersion-inclusive predictions, which will be discussed later
in this Section.)

\begin{figure}
\centering
\includegraphics[width=1.0\linewidth]{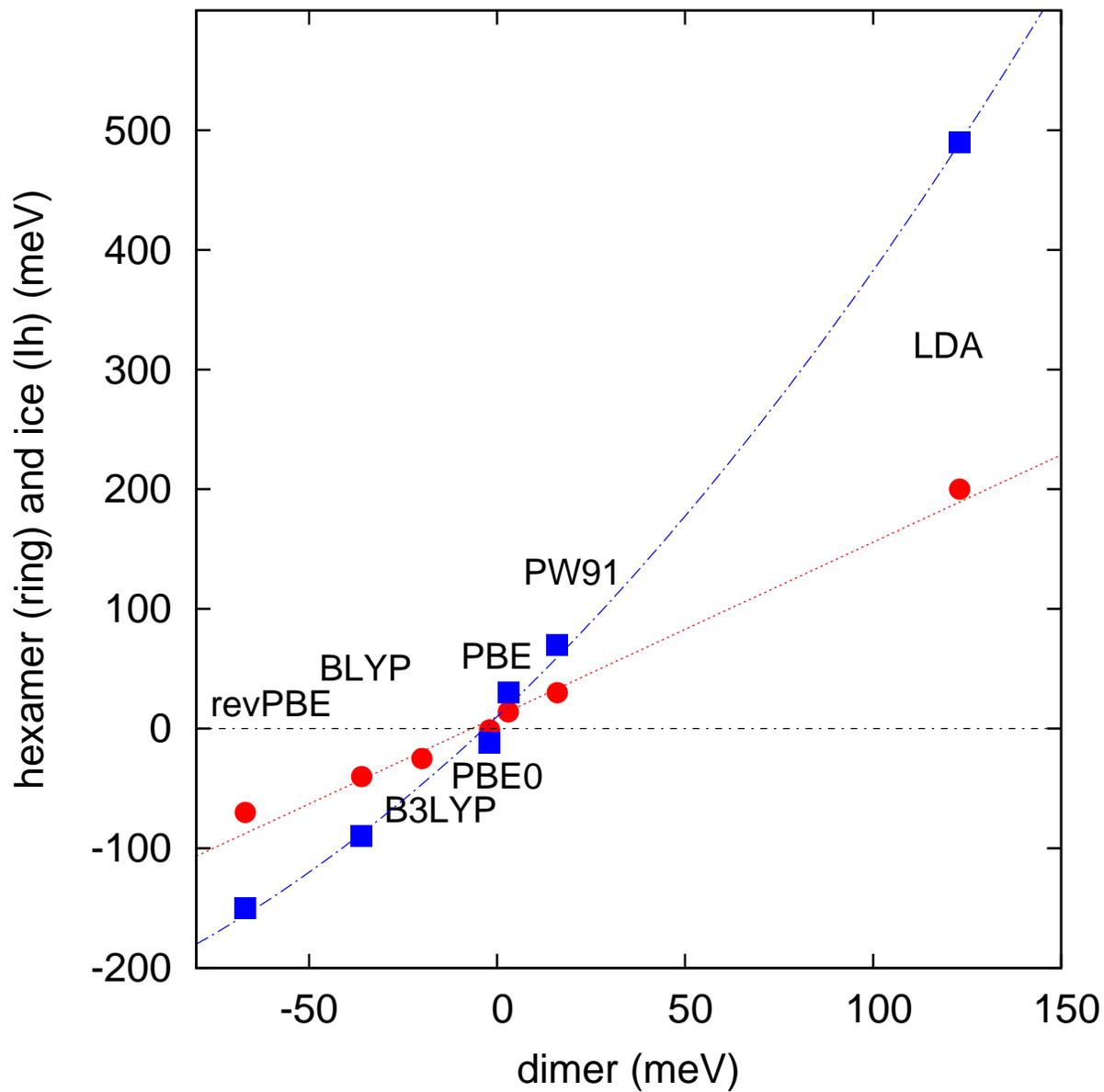}
\caption{
Errors of sublimation energy per monomer of ice Ih crystal (blue squares) and binding energy per monomer 
of ring-hexamer (red circles) plotted against error of dimer binding
energy for chosen semi-local and hybrid exchange-correlation functionals.
}
\label{fig:H-bond_energy}
\end{figure}

The ordering of sublimation energies of ice~Ih given by the GGAs in Table~\ref{tab:ice_energies}
is reminiscent of the ordering of GGA dimer energies (Table~\ref{tab:dimer_gm})
and ring-hexamer energies (Table~\ref{tab:hexamer_binding}). To bring out the close relationship between these different manifestations
of H-bond energy in water, we show in Fig.~\ref{fig:H-bond_energy} plots of the GGA errors of the ice Ih sublimation energy and the ring-hexamer
binding energy \emph{vs} the corresponding error in the binding energy of the dimer in its global-minimum geometry. 
The close relationship between the three kinds of error is immediately apparent, and the
smoothness of the curves indicates that for GGAs a knowledge of the error in the dimer energy
suffices to predict the errors in the hexamer and ice Ih energies. It is noteworthy that the plots
in Fig.~\ref{fig:H-bond_energy} pass almost exactly through zero, something that would presumably not happen if GGA errors
of polarizability caused a serious mis-description of the cooperative enhancement of H-bonding
(see also Sec.~\ref{sec:coop}). Hybrid approximations give more accurate polarizabilities
than GGAs (Sec.~\ref{sec:monomer}), so it is interesting to compare hybrids with their parent GGAs for the sublimation energy of ice Ih.
Published information on this is very sparse, but we include in Fig.~\ref{fig:H-bond_energy} the sublimation
energy from PBE0, which is smaller than from PBE by \emph{ca.}~$40$~meV. This difference is certainly not negligible
but only a part of this appears to be due to errors of cooperative enhancement. 

We turn now to the energetics of compressed ice structures. Experiment tells us that as the pressure increases
from atmospheric to $\sim 3$~GPa, the most stable low-temperature structures of ice pass through the series
known for historical reasons as Ih, IX, II, XIII, XIV, XV, VIII. Fortunately, only a few key features of these structures 
need concern us here. The first is that the H$_2$O monomers retain their integrity
in all the structures, though there are small changes in the intramolecular geometries. A second feature is that the
number of H-bonds per monomer does not change, each monomer donating two H-bonds and accepting two from
four of its neighbors. Even the O-O distance in each H-bond changes only a little through the series, the
surprising fact being that it is slightly longer in the more compressed structures. Nonetheless, the volume per
monomer decreases strongly from Ih to VIII, the volume in ice VIII 
being about two thirds that of ice Ih at zero pressure~\cite{whalley1984a}.
This dramatic compression is entirely due to the ever closer approach of monomers that are not H-bonded
to each other as we progress through the series. The coordination number is four in ice Ih but is eight in ice VIII, the
O-O nearest-neighbor distance for the four non-H-bonded neighbors in ice VIII being slightly shorter than for the
four H-bonded neighbors. Remarkably, in spite of the enormous compression, extrapolation of experimental data shows
that the energies per monomer in the Ih and VIII structures, when both are at zero pressure, differ by a mere $33$~meV.
This very small experimental energy difference is 
corroborated by both DMC and CCSD(T) calculations, which concur in predicting an energy difference 
between Ih and VIII of \emph{ca.}~$30$~meV/H$_2$O \cite{santra2011a,gillan2013a}.

\begin{figure}
\centering
\includegraphics[width=1.0\linewidth]{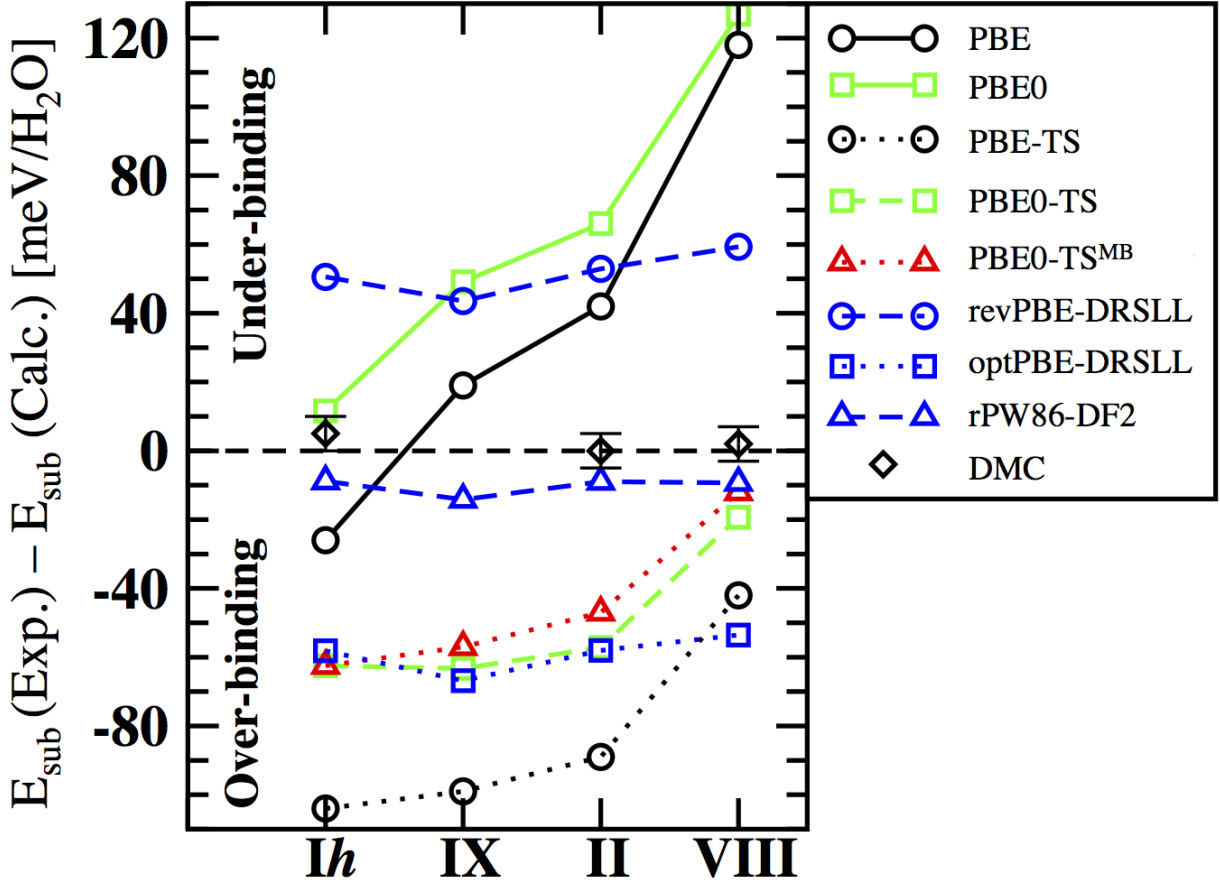}
\caption{
Differences $E_{\rm sub} ( {\rm Exp} ) - E_{\rm sub} ( {\rm Calc} )$ between experimental and calculated values
of the zero-pressure sublimation energies of the ice Ih, IX, II and VIII structures. Calculated values employ the
semi-local functionals PBE and PBE0, a selection of dispersion-inclusive functionals, and diffusion Monte Carlo (DMC).
Zero-point vibrational contributions to the experimental values have been subtracted. Adapted with permission
from J. Chem. Phys. {\bf 139}, 154702 (2013).
}
\label{fig:santra_ice_2013}
\end{figure}

Semi-local XC functionals completely fail to reproduce the small energy differences between compressed ice structures and ice Ih,
as can be seen from the calculated sublimation energies of ice~VIII at zero pressure summarized in Table~\ref{tab:ice_energies}.
With PBE and BLYP, the energy differences per monomer between zero-pressure ice VIII and Ih are calculated
to be $177$ and $208$~meV respectively, which are $\sim 6$ times the experimental value of $33$~meV.
We show in Fig.~\ref{fig:santra_ice_2013} plots of the errors of selected functionals for the sublimation energies
of the sequence of increasingly compressed ice structures Ih, IX, II and VIII, reproduced from Ref.~\cite{santra2013a}.
The rapidly growing errors of PBE along the sequence are striking, and we note that the hybrid functional PBE0
shows a very similar trend. Since PBE0 gives much better polarizabilities than PBE,
the large semi-local errors clearly cannot stem mainly from errors of polarizability. The energies
of compressed ice structures relative to ice Ih have been computed with a very wide range of semi-local
functionals in Refs.~\cite{kambara2012a,brandenburg2015a}. The functionals studied include the GGAs
PW91, PBE, PBEsol, BLYP, RPBE and revPBE, the hybrid functionals PBE0 and B3LYP, the range-separated
hybrid HSE06 and the meta-GGAs TPSS and M06L. In almost every case, the energy difference betwen ice Ih and VIII was
found to be grossly overestimated. This indicates that essentially all semi-local and hybrid functionals suffer from the same
kind of problem.

Why do semi-local and hybrid approximations incorrectly destabilize the compact, high-pressure ice structures 
relative to the more extended
low-pressure structures? There appears to be a connection here with their behavior for the
compact and extended configurations of the hexamer (Sec.~\ref{sec:hexamer}) and the 
larger water clusters (Sec.~\ref{sec:larger_clusters}).
For the hexamer, we saw the clear evidence that lack of dispersion is one of the main reasons for the wrong
compact-extended balance given by semi-local approximations, and we will see the same for the ice structures. However,
before reviewing dispersion-inclusive approximations for ice, we recall the many-body evidence
that dispersion is not the only cause of trouble for the compact-extended balance in the hexamer. A similar
many-body analysis has been reported for the errors of BLYP and PBE for the relative energies
of high- and low-pressure ice structures~\cite{gillan2013a}. The outcome was that for BLYP the enormous overestimate
of the energy difference between VIII and Ih is mainly a 2-body error, and that good relative energies
are obtained once this error is corrected. However, the same is not true of PBE, where there appears
to be a large beyond-2-body contribution to the error in relative energies. 

A variety of dispersion-inclusive DFT approaches have been used to study the energetics
of ambient and compressed ice phases, including Grimme DFT-D 
methods~\cite{kambara2012a,fang2013a,brandenburg2015a},
TS dispersion paired with PBE and PBE0~\cite{santra2011a,santra2013a}, and DRSLL-type
approximations~\cite{hamada2010a,kolb2011a,murray2012a,fang2013a,santra2013a}.
Most of these have paid particular attention to the energy differences between compressed structures
and ice Ih, and in some cases have studied the equilibrium volumes of the structures and the transition pressures
between them. All the different ways of including dispersion give a large improvement in the relative energies
of the extended and compact structures. However, not all the methods are equally good, because in some
cases the improvement in the relative energies is accompanied by a worsening of the sublimation energy of the ice Ih structure.

The range of results that can be given by different dispersion-inclusive methods is illustrated by the
work of Ref.~\cite{santra2013a}, where the performance of different versions of the TS and DRSLL-type schemes
was compared (see Fig.~\ref{fig:santra_ice_2013}). 
The work showed that the original version of DRSLL based on revPBE~\cite{dion2004a}, referred to here
as revPBE-DRSLL, gives very satisfactory energies of the structures IX, II and VIII relative to ice Ih, but that
all the structures are underbound compared with experiment by $\sim 50$~meV. This is expected, because
the excessive exchange-repulsion of revPBE-DRSLL generally gives underbinding in H-bonded systems, as we
saw for the water dimer and ring-hexamer in Secs.~\ref{sec:dimer} and \ref{sec:hexamer}. The optPBE-DRSLL
approximation has weaker exchange-repulsion, and performs better for H-bonding energies, giving much 
better dimer and ring-hexamer energies. It also gives accurate energies of the IX, II and VIII structures relative
to Ih, but all these structures are now overbound. However, the revised version of DRSLL due to 
Lee \emph{et al.}~\cite{lee2010a}, which we refer to as rPW86-DF2,
performs very well for both the relative energies and the sublimation energies of the the ice structures,
(Fig.~\ref{fig:santra_ice_2013}), as was also found by Murray and Galli~\cite{murray2012a}. 
Not reported in Fig.~\ref{fig:santra_ice_2013} but given in Table~\ref{tab:ice_energies} are results for optB88-DRSLL. 
As with optPBE-DRSLL, it describes the energy difference between ice I and VIII well but overbinds both phases. 

Predictions from the scheme of Tkatchenko and Scheffler~\cite{tkatchenko2009a} in which dispersion is added to PBE or PBE0 
provide an instructive contrast. Since ice Ih is already somewhat overbound with PBE, it is no surprise
that PBE-TS overbinds this structure by over $100$~meV/monomer. Nevertheless, PBE-TS gives a considerable
improvement over PBE itself for the relative energies, though it is not as good as any of the DRSLL-type
methods. The PBE0-TS approximation also overbinds
ice Ih, but only by $\sim 60$~meV, and the difference between the energies of the VIII and Ih structures is also slightly
better than with PBE-TS. Also included in Fig.~\ref{fig:santra_ice_2013} are the predictions obtained
by adding many-body dispersion~\cite{tkatchenko2012a} to PBE0. These differ only slightly from PBE0-TS, so that the effects of 
beyond-2-body dispersion appear to be very small for these ice structures, as might be expected from previous
work on the contribution of 3-body dispersion to the energetics of ice~\cite{vonlilienfeld2010a}. 

Insight into the performance of the DFT-D methods of Grimme \emph{et al.} for the energetics of
ice structures can be gained from Refs.~\cite{brandenburg2015a,kambara2012a}, both of which
demonstrate the major improvements brought by the inclusion of dispersion. In the first of these
papers, the binding energies of $10$ ice structures were computed with Grimme D3 dispersion added
to a variety of semi-local functionals. These DFT-D3 approximations give reasonably good binding energies
of ice Ih, except for PBE-D3, which is overbound by an unacceptable $\sim 140$~meV. The PBE-D3 approximation
also gives a greatly overestimated value of $\sim 130$~meV for the energy difference between the ice VIII
and Ih structures, which is also overestimated by the other DFT-D3 approximations, though less seriously. 

Turning now to the equilibrium volumes, we report in Table~\ref{tab:ice_volumes} results for ice I and 
VIII from a selection of GGAs, PBE0, and several dispersion inclusive functionals.
By and large, the trends found for sublimation energies are mirrored in the predictions
of equilibrium volumes. For example, at the GGA level the sublimation 
energies decrease from PBE to BLYP to revPBE, while the 
equilibrium volumes show the opposite trend, with revPBE~$>$ BLYP~$>$ PBE. 
The errors in the volumes for ice Ih predicted by GGAs are less than $4$~\%,
but they are much larger for ice VIII, in the range $9$~- $28$~\%, as might be expected from the
substantial under-binding of ice~VIII predicted by GGAs.
It has been shown that in ice~VIII zero-point effects increase the equilibrium volume
by \emph{ca.}~$5$~\%~\cite{murray2012a,santra2013a,brandenburg2015a}, so in assessing the errors in the predicted volumes we compare with
the experimental volume reduced by this amount. Zero-point corrections to the equilibrium volume
are much smaller for ice~Ih~\cite{murray2012a,santra2013a,brandenburg2015a}, so 
for this phase we simply compare with the uncorrected experimental value. As shown in Table~\ref{tab:ice_volumes},
accounting for exact exchange by going from PBE to PBE0 does little to reduce the errors in the volumes. 
By contrast, dispersion has a significant impact on the volumes, generally decreasing them, as one would expect. 
However, the results are very sensitive to the particular choice of dispersion inclusive functional. 
Of the functionals reported, BLYP-D3, revPBE-D3, optPBE-DRSLL do reasonably well for the two phases, while
revPBE-DRSLL stands out as offering the worst performance (volume of ice~VIII overestimated by 20\%).
Despite performing very well in terms of sublimation energies, the volumes predicted by rPW86-DF2
are rather disappointing. Overall, it is clear that for predictions of equilibrium volumes in the ice phases 
there is still considerable room for improvement. 

We noted at the start of this Section that proton-disordered ice phases such as Ih transform to proton-ordered
structures at low temperatures. This is a subtle phenomenon in both experiment and theory. The very slow kinetics of
molecular reorientation at low temperatures makes it difficult to measure the true thermodynamic transition
temperatures accurately, and unambiguous identification of the symmetry of the proton-ordered phase
has sometimes been controversial. Widely used force fields sometimes yield completely erroneous
predictions for these transitions~\cite{buch1998a}, so that there has been considerable interest in DFT 
treatments~\cite{kuo2004c,singer2005a,knight2006a,tribello2006a,tribello2006b,fan2010a,pan2010a,singer2011a,delben2014a}.

The transformation of ice Ih to a low-temperature ordered phase was conjectured over 60 years ago~\cite{bjerrum1952a}, but the experimental
evidence for a transition near $72$~K appeared only more recently~\cite{kawada1972a,tajima1982a}. Simple electrostatic
arguments~\cite{bjerrum1952a,davidson1984a,buch1998a} suggest that the low-temperature phase ice XI should
be antiferroelectric, but this expectation is contradicted by diffraction and thermal depolarization experiments~\cite{jackson1995a,jackson1997a},
which indicate that it is actually ferroelectric, though the interpretation of the experiments has been challenged~\cite{iedema1998a}. DFT calculations
based on the BLYP functional~\cite{singer2005a}, in conjunction with graph-theoretic methods used to enumerate H-bonding
topologies~\cite{kuo2001a,kuo2003a}, support the ferroelectric assignment. The parameterized models produced in the course of
this computational approach, when used in statistical-mechanical calculations, yield a transition temperature of $98$~K~\cite{singer2005a}
in respectable agreement with the experimental value. It was shown later~\cite{knight2006a,tribello2006b,fan2010a} that this outcome is not
significantly altered if other XC functionals are used in place of BLYP. Remarkably, even the unsophisticated LDA functional
yields essentially the same result~\cite{tribello2006b}. It has been argued from this that the electrostatic part of the energy dominates the energetics
of proton ordering, as originally proposed by Bjerrum~\cite{bjerrum1952a}, and that the inability of common force fields
to describe the energetics correctly arises from their failure to reproduce the high multipole moments of the 
charge distribution of the H$_2$O monomer~\cite{tribello2006b,fan2010a}.
Another approach to the transformation between ice Ih and XI was taken by Ref.~\cite{schoenherr2014a}. There, MC sampling based on the
dispersion-inclusive functionals PBE-D2 and BLYP-D2, and also their hybrid counterparts PBE0-D2 and 
B3LYP-D2 was used to compute the
static dielectric constant $\epsilon$ and the ice Ih/XI transition temperature. It was found that PBE0-D2 at $T = 273$~K gives $\epsilon$
roughly $20$~\% greater than the experimental value of $95$, while the value with PBE-D2 is roughly $50$~\% greater, so that the over-polarizability
of the H$_2$O molecule with PBE (see Sec.~\ref{sec:monomer}) is clearly significant. The transition temperature computed with PBE0-D2 in this work was in the
range $70 - 80$~K, in satisfactory agreement with experiment. 

Calculations combining DFT with graph-theoretic methods have also been successful in treating the transformation
of other proton-disordered phases to their low-temperature ordered counterparts, including 
ice VII to VIII~\cite{knight2006a,tribello2006b}, XII to XIV~\cite{tribello2006a} and V to XIII~\cite{knight2008a}.
Where comparison with experiment is possible, the symmetries of the ordered phases are correctly predicted, and the transition
temperatures are approximately correct. However, there appears to be one exception, namely the transition from ice VI to XV,
where experiment shows the ordered phase to be antiferroelectric~\cite{salzmann2009a}, while DFT 
calculations consistently make the most stable
structure ferroelectric~\cite{knight2005a,delben2014a}. Hybrid and dispersion-inclusive 
functionals give the same result, as do calculations based on MP2
and the random phase approximation (RPA)~\cite{delben2014a}, which are expected to be even more accurate. 
In discussion of the possible origins of this paradox, it was suggested in Ref.~\cite{delben2014a} that the so-called
``tin-foil'' boundary conditions implicitly used in conventional electronic-structure calculations on periodic systems
may not be appropriate for the ice VI-XV transition. It was shown that if instead the boundary conditions are allowed to
reflect the electrostatic environment in which ice XV grows, then the ferroelectric phase may be sufficiently disfavored
for the antiferroelectric phase to become more stable. The experimental work of Ref.~\cite{shephard2015a} appears to
offer support for this idea.

Although hexagonal ice Ih is the naturally occurring form under ambient conditions, there is evidence
that the cubic variant known as ice Ic can form in the upper atmosphere~\cite{zhang2006a,murray2006a},
though it is not yet clear whether pure ice Ic or only a disordered mixture of ice Ic and Ih is formed~\cite{malkin2012a,kuhs2012a}. 
In either case, the question of the energy difference between the 
two forms of ice is of some importance for environmental science. 
Raza \emph{et al.}~\cite{raza2011a} tackled this question by performing DFT calculations on the lowest-energy
proton-ordered forms of the two crystal structures, using the XC functionals PBE, PBE0, BLYP-D3 and optPBE-DRSLL;
accurate reference calculations with DMC were also reported. The conclusion from their work was that the
two structures are isoenergetic within the technical tolerances of the calculations, the indication being that
the energy difference is less than $1$~meV/monomer. This conclusion is supported by later work from
Geiger \emph{et al.}~\cite{geiger2014a}. It has been shown very recently~\cite{engel2015a} that the clear preference for ice Ih
over ice Ic observed in nature may be due to the difference of anharmonic vibrational energy between the two phases.

We conclude this Section by commenting briefly on the influence of isotope effects on the properties of ice.
In almost solids, the replacement of a heavier isotope by a lighter one causes an expansion of the equilibrium volume.
The reason is that expansion normally reduces the vibrational frequencies and hence the zero-point energy,
the reduction being greater for lighter isotopes. It therefore comes as a surprise that the equilibrium volume of 
the H$_2$O form of ice Ih is less than the D$_2$O form~\cite{roettger1994a}. It was shown recently~\cite{pamuk2012a} that this
anomalous isotope effect occurs because expansion of ice weakens the H-bonds, which in turn strengthens
the intramolecular O-H bonds, thus increasing their frequency. However, the consequences of this increase
are partially canceled by the softening of other vibrational modes. Some of the common force-fields
for water wrongly predict a normal isotope effect, but DFT calculations generally give the observed
anomalous isotope effect, at least with the XC functionals examined so far. It has been noted recently~\cite{pamuk2015a}
that the same mechanism leads to an interesting isotope effect on the temperature of the transition from ice Ih to XI.


\section{DFT simulation of liquid water}
\label{sec:liquid}

Our review of DFT work on liquid water will be concerned with the structure of the liquid, its thermodynamic
properties, and the dynamics of the molecules. We shall pay particular attention to the three radial
distribution functions (rdfs) $g_{\alpha \beta} ( r )$, the density $\rho_{\rm eq}$ of the liquid at near-ambient pressure,
and the self-diffusion coefficient $D$, all of which are available from experiment. Some aspects of the DFT
simulation of the liquid have recently been reviewed by Khaliullin and K\"{u}hne~\cite{khaliullin2013a}.

As initial orientation, we refer to panel~(a) of Fig.~\ref{fig:gab_GGA_expt}, showing the O-O rdf $g_{\rm O O} ( r )$
from recent high-energy x-ray diffraction measurements~\cite{skinner2013a} performed
at temperatures close to $297$~K. (The simulation results shown in the Figure will be discussed later.)
It is worth commenting here that there have been many experimental measurements of
$g_{\rm O O} ( r )$ over the years (see literature cited in Ref.~\cite{skinner2013a}), and there has been considerable controversy about the height of the
first peak, but the measurements we compare with here are generally accepted to supersede earlier work.
The first peak at O-O separation $R_{\rm O O} = 2.80$~\AA\ and
the second peak at $R_{\rm O O} = 4.5$~\AA\ correspond rather closely to the first- and second-neighbor
O-O distances in ice Ih. The first-neighbor coordination number $N_1$ in the liquid is not uniquely
defined, but is conventionally taken to be the integral $N_1 = 4 \pi n \int_0^{r_{\rm min}} dr \; r^2 g_{\rm O O} ( r )$
under the first peak ($n$ is the bulk number density) up to the radius $r_{\rm O O}^{\rm min}$ of the first minimum.
The diffraction experiments all give a coordination number $N_1$ in the region of $4.3$, which is consistent with
roughly fourfold tetrahedral bonding. However, there is clearly considerable disorder, since the experimental $g_{\rm O O} ( r )$
is quite close to unity over the range $3.0 < R_{\rm O O} < 4.0$~\AA, throughout which there are no neighbors
in ice Ih; all diffraction measurements agree that the value of $g_{\rm O O} ( r )$ at its first minimum
is $g_{\rm O O}^{\rm min} = 0.84 \pm 0.02$. This means that there is substantial penetration of molecules
from the second shell into the shell of H-bonded first neighbors.

The phenomenon of cross-shell penetration is crucially important in liquid water, and is closely linked to the density
increase on melting. The amount of penetration is sensitive to pressure, and diffraction 
experiments~\cite{straessle2006a,weck2009a,katayama2010a} show that
a pressure of only $\sim 1$~GPa ($10$~kbar) is enough to increase the O-O coordination number from \emph{ca.}~$4$ to
 \emph{ca.}~$8$, an effect that occurs by collapse of the second shell into the region of the first shell, without 
 significant breaking of H-bonds~\cite{straessle2006a}. This 
 implies that the penetration in the liquid is closely related to changes of ice structure with increase of pressure,
 exemplified by the presence of non-H-bonded first neighbors at approximately the same distance as the H-bonded
 neighbors in ice VIII. Cross-shell penetration also appears to be intimately linked to diffusion, which
 requires molecules to cross the region $3.0 < R_{\rm O O} < 4.0$~\AA; since the lifetime of H-bonds
 under ambient conditions is estimated to be in the region of $1$~ps~\cite{luzar1996a, bankura2014a}, this crossing must be frequent.
 
 We saw in Sec.~\ref{sec:ice} that GGAs grossly exaggerate the energy difference betwen
 the extended ice Ih structure and compact structures such as ice VIII. They make it too difficult for
 a molecule to approach another molecule that is already H-bonded to four others, and the problem
 is cured by accounting for dispersion. We can anticipate that the same mechanisms will operate
 in the liquid, so that GGAs will generally hinder penetration, making the liquid over-structured and
 under-diffusive, with a high pressure needed to maintain the experimental density. Accordingly, a leading
 theme of our survey of DFT work on the liquid will be the concerted efforts of the past few years to address
 the difficulties of liquid structure, thermodynamics and dynamics
 with dispersion-inclusive methods. We shall see that these efforts have enjoyed considerable
 success, but it will also become apparent that current dispersion-inclusive methods are still immature,
 and that over-correction can be as much of a peril as under-correction. We shall also see that purely technical
 issues have sometimes made it difficult to judge the true capabilities of any given DFT method.

\subsection{Technical issues}
\label{sec:liq_tech}

It is not trivial to assess how well an XC functional performs for the liquid, because the first-principles simulation of the liquid raises
many tricky technical issues that are completely absent from the study of clusters or static ice structures. Thermodynamic, structural
and dynamical quantities such as the pressure $P$, the radial distribution functions $g_{\alpha \beta} ( r )$ and the self-diffusion
coefficient $D$ are all calculated as time averages over the duration of the simulation, which must be long enough
to yield useful statistical accuracy. Furthermore, collection of the data to be averaged can only begin after the system has been
`equilibrated', i.e. simulated for long enough to ensure that memory of the initial conditions has been lost. The problems
of equilibration and time averaging are exacerbated in water by the wide separation of timescales between intramolecular
and intermolecular motions, which hinders energy transfer between these degrees of freedom. Memory times can be reduced
by the use of thermostats, but dynamical quantities may then be falsified. In MD simulations of water based on force fields, the equilibration and
production phases commonly have durations of $100$~ps or more (see e.g. Refs.~\cite{fernandez2006a,fanourgakis2008a,habershon2009a}), 
but such durations have been impractical in DFT simulations until
recently~\cite{kuehne2009a,khaliullin2013a}. (In the earliest first-principles simulations of 
water~\cite{laasonen1993a,tuckerman1994a,sprik1996a,silvestrelli1999a,silvestrelli1999b}, these durations were 
between $2$ and $10$~ps.)

First-principles simulations can be performed by Born-Oppenheimer MD (BOMD), in which the Kohn-Sham orbitals
are relaxed to the self-consistent ground state at every time step, or by Car-Parrinello MD (CPMD), in which the orbitals
are treated as dynamical degrees of freedom, their dynamics being governed by a fictitious mass $\mu$. 
The method known as second-generation CPMD~\cite{kuehne2007a,kuehne2009a,kuehne2013a} combines
the strengths of these two techniques. In addition,
first-principles Monte Carlo simulation is also feasible~\cite{kuo2004b}. If rigorously implemented,
all these techniques should be equivalent, but practical compromises in the setting of technical parameters can cause differences.
It took several years to discover that the close agreement with experiment reported in some early 
papers~\cite{silvestrelli1999a,silvestrelli1999b,izvekov2002a} was partly due to
simulation errors caused by $\mu$ being set too 
high~\cite{asthagiri2003a,grossman2004a,schwegler2004a,kuo2004b,fernandez-serra2004a,vandevondele2005a,sit2005a}.
Significant discrepancies may also arise from the use of different statistical-mechanical
ensembles, which are expected to yield identical results in the thermodynamic limit, but not in finite systems. Inadequate basis sets too can be 
a source of inaccuracy. System-size errors present yet another hazard: ideally, the size of the periodically repeated
simulation cell should be systematically increased at constant density until the observables of interest settle to converged values,
but in practice the cell size is constrained by the computer budget. 

These and other technical difficulties plagued much of the pioneering DFT work on water, but the problems started to be clearly
identified and addressed around 10 years 
ago~\cite{asthagiri2003a,grossman2004a,schwegler2004a,fernandez-serra2004a,kuo2004b,sit2005a} 
and it is now understood in principle how to bring most of them under firm control. The conditions necessary
for adequate equilibration, statistical averaging and thermostating were analyzed in Ref.~\cite{vandevondele2005a}. 
The equivalence of Car-Parrinello
and Born-Oppenheimer dynamics and the range of fictitious mass needed to ensure this equivalence were 
established in Refs.~\cite{schwegler2004a,kuo2006a}.
System-size errors have turned out to be only a minor concern for many thermodynamic and structural quantities, provided systems
of 64 molecules or more are used~\cite{vandevondele2005a,kuehne2009a}, 
although substantial (but well understood) size corrections are needed for the diffusivity~\cite{kuehne2009a}.
Increasing computer power has made it much less necessary to compromise on basis-set completeness. Nevertheless, technical
uncertainties continue to be troublesome even now, as will become clear below.

\subsection{Quantum nuclear effects}
\label{sec:QNE}

In addition to the technical issues just outlined, there is also the physical phenomenon of quantum nuclear effects (QNE).
Almost all first-principles simulations of water have treated
the nuclei classically, but it is well known from diffraction experiments~\cite{soper2008a} and 
force-field simulations~\cite{fanourgakis2008a,habershon2009a} that QNE due to the light mass
of hydrogen are not negligible. Path-integral simulation~\cite{chen2003a,morrone2008a,ceriotti2013a,wang2014a} 
allows QNE to be included almost exactly, at least for
thermodyamics and structure, but the computational cost then escalates still further. (We note, however, that
recent algorithmic developments~\cite{ceriotti2014a} mitigate the additional cost.) If QNE are not included in 
simulations of the liquid, then perfect agreement with experimental data cannot be expected, but it is not straightforward
to separate errors due to neglect of QNE from those due to inaccuracy of the XC functional. We will note the possible influence
of QNE where appropriate in the following, but we do not aim to treat this important matter in depth in this review, since QNE
in water have recently been reviewed by Ceriotti \emph{et al.}~\cite{ceriotti2015a}.

\subsection{GGAs and hybrids}
\label{sec:liq_GGA_hybrid}

We start our survey of DFT methods for liquid water by focusing on GGAs, and since BLYP and PBE have been the most widely
used GGAs we discuss these first. After this, we summarize the rather few simulations that have been performed with
other GGAs and with hybrid functionals. Most simulations have been performed at constant volume, corresponding to an experimental
density close to $1$~g/cm$^3$, and these conditions apply unless we mention otherwise. A summary of the main features of
the simulations reviewed here is provided in Tables~\ref{tab:liq_BLYP} and \ref{tab:liq_PBE}.

\begin{figure}
\centering
\begin{tabular}{c}
\includegraphics[width=0.6\linewidth]{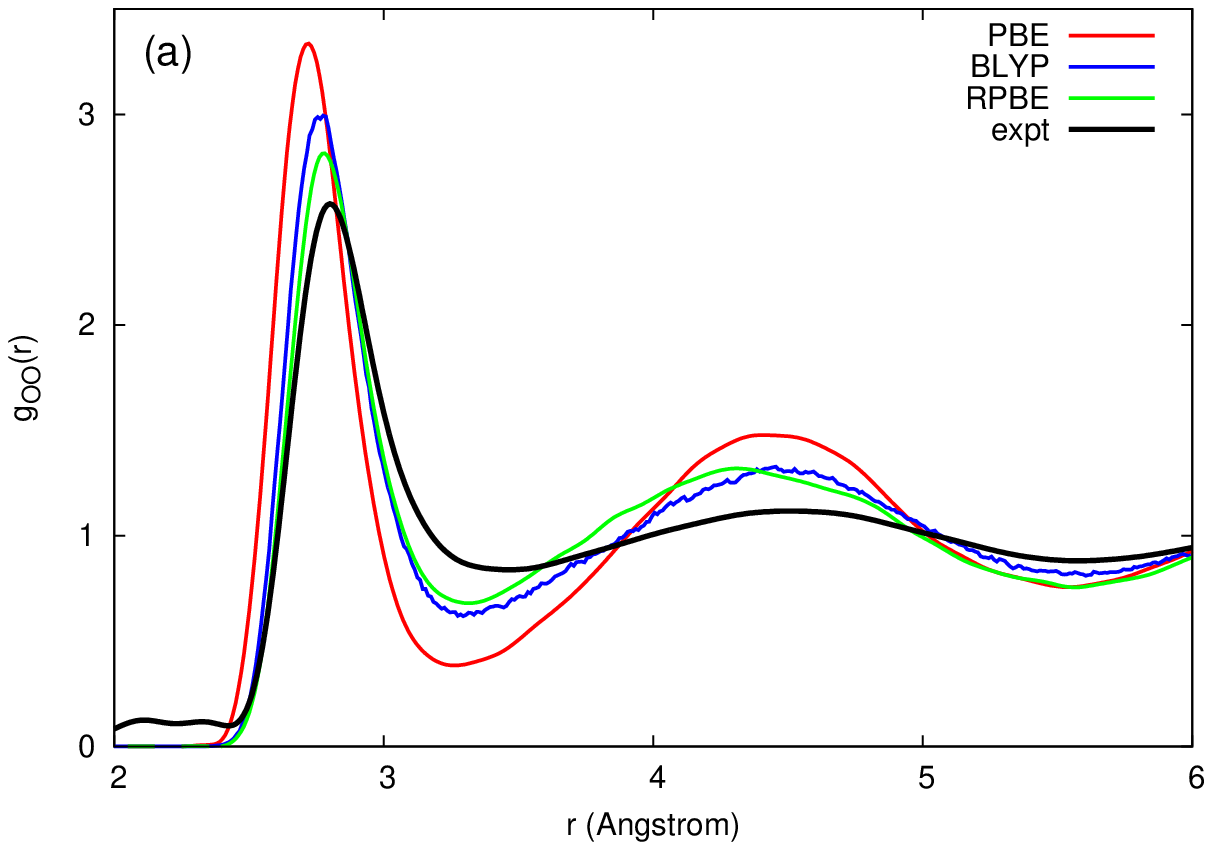} \\
\includegraphics[width=0.6\linewidth]{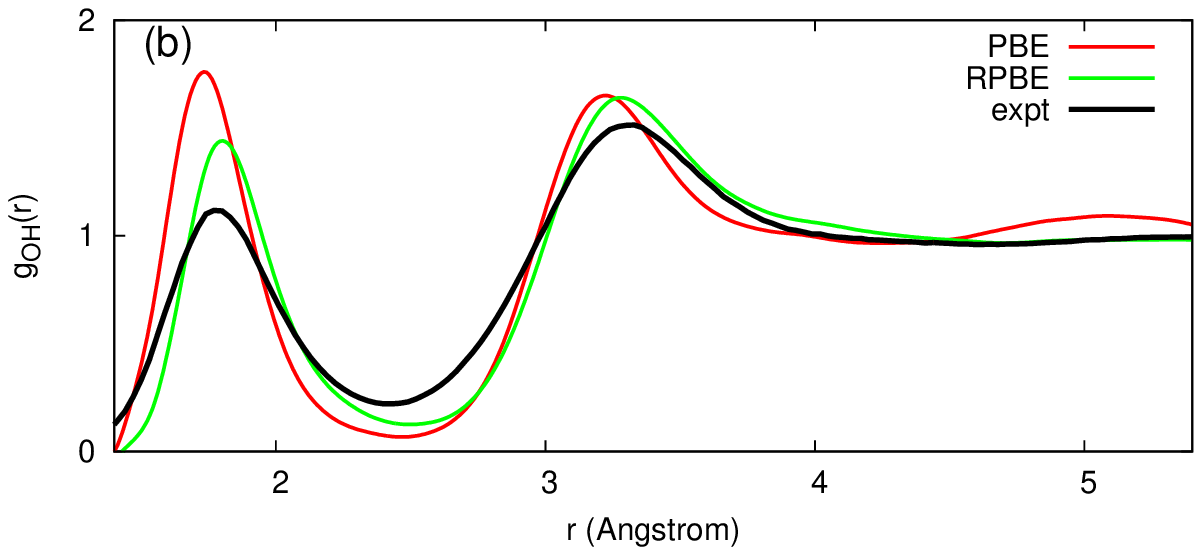} \\
\includegraphics[width=0.6\linewidth]{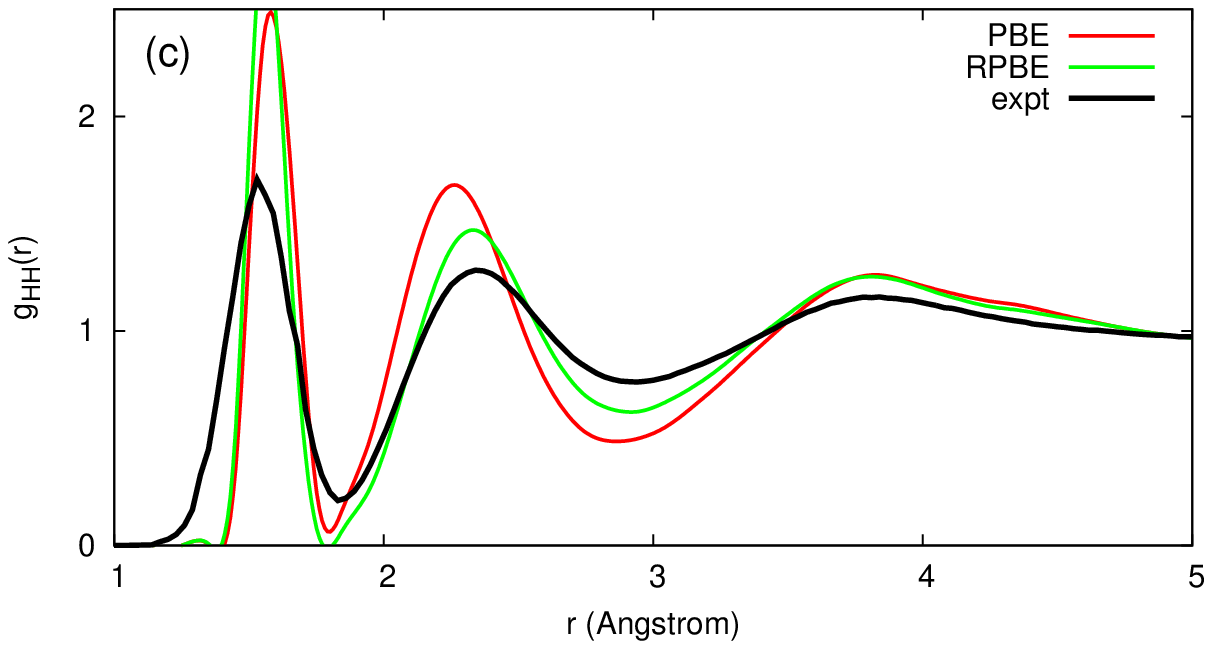}
\end{tabular}
\caption{
Panels (a), (b) and (c) show O-O, O-H and H-H radial distribution functions (rdfs) of liquid
water at ambient temperature and experimental density from experiment and
from MD simulations based on the PBE and RPBE approximations (results from BLYP
are also shown for $g_{\rm O O} ( r )$). Simulation rdfs with PBE and RPBE are
from Ref.~\cite{forster-tonigold2014a} and results with BLYP from Ref.~\cite{bankura2014a}.
Experimental rdf $g_{\rm O O} ( r )$ is from high-energy x-ray diffraction
measurements~\cite{skinner2013a}, and rdfs $g_{\rm OH} ( r )$ and $g_{\rm H H} ( r )$ are from 
joint refinement of neutron and x-ray diffraction measurements~\cite{soper2007a}.
}
\label{fig:gab_GGA_expt}
\end{figure}

Well over $20$ substantial simulations of the near-ambient liquid have been reported with the BLYP 
functional since the resolution of the technical difficulties mentioned 
above (references in Table~\ref{tab:liq_BLYP}).  Most of these simulations appear to be free of
major technical limitations, and there is now a reasonable consensus about the properties predicted by BLYP. Comparison of the O-O rdf
$g_{\rm O O} ( r )$ with neutron and x-ray diffraction data is generally regarded as the cleanest way of testing the liquid structure,
since this rdf is much less affected by QNE than $g_{\rm O H} ( r )$ and $g_{\rm H H} ( r )$~\cite{fanourgakis2008a,habershon2009a}. 
An example of the comparison between BLYP and experiment is shown in panel~(a) of Fig.~\ref{fig:gab_GGA_expt}.
As in most BLYP-based simulations published over the past $10$ years, the liquid
is somewhat over-structured, the height of the first peak and the depth of the first minimum being somewhat exaggerated, but the radial
positions of all the main features agree well with experiment (see Table~\ref{tab:liq_BLYP}). 
Many of the BLYP-based simulations have reported values of the self-diffusion
coefficient $D$, for which accurate experimental values are available over a range of temperatures for both light and heavy water. As
might be expected from the over-structuring, the values of $D$ given by BLYP (see Table~\ref{tab:liq_BLYP}) are lower 
than the experimental values, usually by a factor of $\sim 3$.

Some BLYP-based studies have also reported the O-H and H-H rdfs $g_{\rm O H} ( r )$ and $g_{\rm H H} ( r )$. The
O-H rdf is instructive, because its peak at $r \simeq 1.75$~\AA\ directly probes the H-bond between the donor H and the acceptor O.
All the BLYP-based simulations that reported $g_{\rm O H} ( r )$ 
(see e.g. Refs.~\cite{grossman2004a,lee2006a,morrone2008a,bartok2013a,bankura2014a}) found the height of this H-bond peak
to be overestimated by $\sim 30$~\%. Both force-field and DFT-based simulations that have accounted for QNE by the
path-integral technique~\cite{fanourgakis2008a,morrone2008a} suggest that the quantum nature of the proton 
might explain as much as half of this overstructuring. However, there is little
doubt that the other half is due to errors of the BLYP functional. The overstructuring also affects 
$g_{\rm H H} ( r )$~\cite{grossman2004a,lee2006a,morrone2008a,bartok2013a}.

The disagreements with experiment are more severe with the PBE functional. All the PBE-based simulations reported over the past 10
years (references in Table~\ref{tab:liq_PBE}) give a substantially over-structured liquid, the height of the first peak 
and the first minimum in $g_{\rm O O} ( r )$ being typically $g_{\rm O O}^{\rm max} \simeq 3.4$ and $g_{\rm O O}^{\rm min} \simeq 0.4$ 
respectively, compared with the experimental values of $2.57$ and $0.84$ 
(Fig.~\ref{fig:gab_GGA_expt} and Table~\ref{tab:liq_PBE}). The very deep first minimum is particularly significant, since it indicates
a strong suppression of close approaches by non-H-bonded monomers. The overstructuring is even more striking in $g_{\rm O H} ( r )$
and $g_{\rm H H} ( r )$~\cite{schwegler2004a,forster-tonigold2014a}, with the height of the H-bond peak in $g_{\rm O H} ( r )$
at $r \simeq 1.75$~\AA\ being overestimated by $\sim 50$~\% and similarly large errors in $g_{\rm H H} ( r )$
(see Fig.~\ref{fig:gab_GGA_expt}). Correspondingly, the self-diffusion coefficient is grossly underestimated, being too low compared 
with experiment by about a factor of $10$. The only other GGAs that have seen more than occasional use in water simulations are revPBE 
and the closely related RPBE (references in Table~\ref{tab:liq_PBE}). Almost all the simulations
reported with these functionals agree that they yield a liquid with much less structure
(see Fig.~\ref{fig:gab_GGA_expt}) and a higher diffusivity than BLYP and PBE (but see Ref.~\cite{kuehne2009a}).

\begin{figure}
\centering
\includegraphics[width=1.0\linewidth]{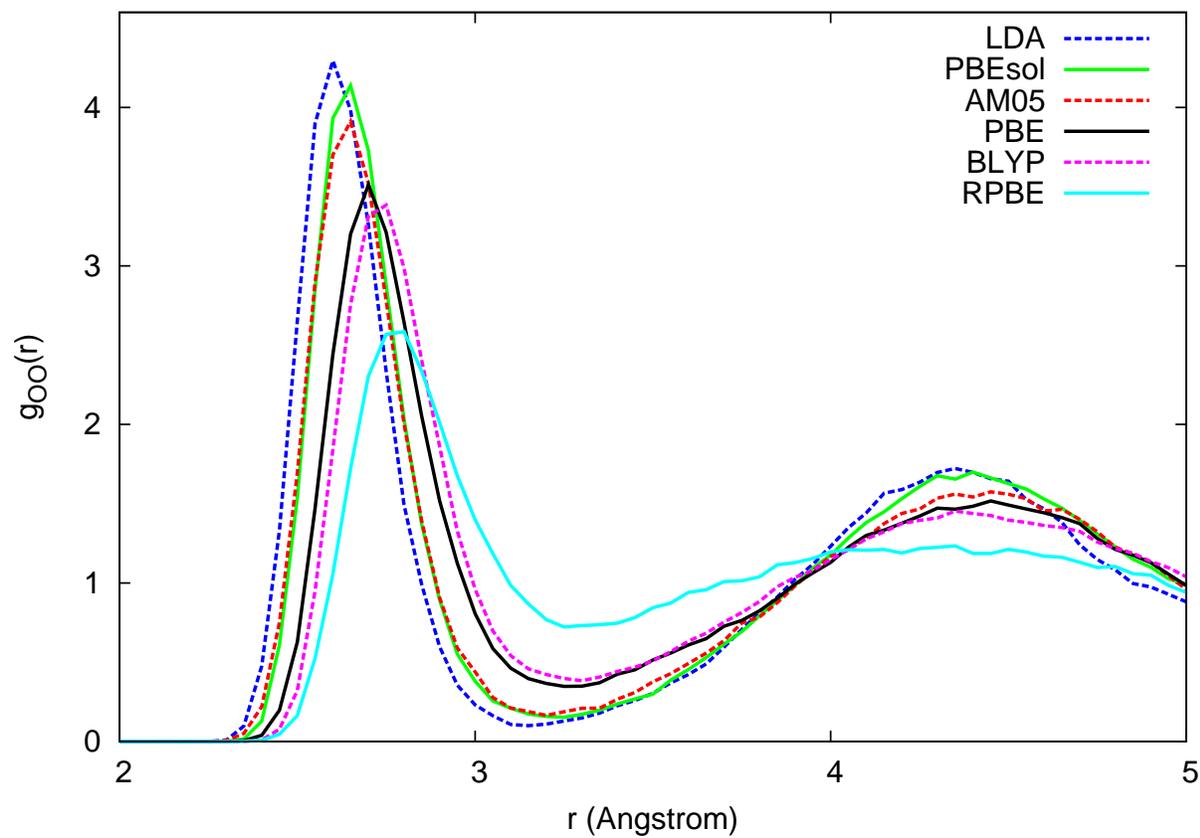} 
\caption{
The O-O rdf of liquid water at experimental density and $T = 300$~K computed with a series
of semi-local functionals. Adapted from Ref.~\cite{mattsson2009a}, with permission.
}
\label{fig:Mattsson_rdf}
\end{figure}

The foregoing comparisons show that the structure and diffusivity of the liquid differ substantially for different GGAs.
It would be surprising if this were not so, since we saw in Secs.~\ref{sec:dimer}, \ref{sec:hexamer} and \ref{sec:ice}
that the H-bond energy depends strongly on the choice of GGA, and Fig.~\ref{fig:H-bond_energy}
demonstrates the close relationship between the errors in GGA binding energies of clusters and ice. The
properties of the liquid at specified density and temperature presumably depend on the H-bond energy, and it is natural 
to ask if the trends in predicted structure and diffusivity correspond in some way to the trends illustrated in Fig.~\ref{fig:H-bond_energy}.
The paper of Mattsson and Mattsson~\cite{mattsson2009a} appears to be the only one that has systematically addressed this
important question, and we reproduce in Fig.~\ref{fig:Mattsson_rdf} their O-O rdfs computed at $T = 300$~K
with the series of semi-local functionals RPBE, BLYP, PBE, AM05, PBEsol, LDA. The functionals that yield highly overstructured
water are likely to give a glassy state at $300$~K, but the trend of increasing
structural ordering along this series is nonetheless clear, as is the steady leftward shift of the first-peak position $r_{\rm O O}^{\rm max}$. The order
of this trend is also the order of increasing dimer binding energy (see Table~\ref{tab:dimer_gm}), with the possible exception
of the AM05 approximation, which appears to give a slightly lower dimer binding energy than PBE~\cite{mattsson2009a}.
We return later to the close correlation between structural ordering in the liquid and H-bond energy.

As noted in Secs.~\ref{sec:monomer} and  \ref{sec:coop}, GGAs generally overestimate the polarizability of 
the water molecule  by $\sim 10$~\%, so that they might
be expected to exaggerate the cooperative enhancement of H-bonding in condensed phases. Furthermore, they underestimate the stretching
frequencies of the water monomer by $3 - 5$~\%, and the excessive flexibility of the intramolecular OH bonds may further
strengthen H-bonding. These are reasons for expecting hybrids to soften the structure of the liquid, since they give better polarizabilities
and monomer frequencies. The evidence on this point is not entirely unanimous. The early simulations of Todorova
\emph{et al.}~\cite{todorova2006a}, performed at $T = 350$~K, showed a substantial softening 
when the PBE functional is replaced by PBE0, with $g_{\rm O O}^{\rm max}$ decreasing from
$2.99$ to $2.58$ and $g_{\rm O O}^{\rm min}$ increasing from $0.47$ to $0.73$.
Subsequent work of Zhang \emph{et al.}~\cite{zhang2011a,zhang2011b} and
DiStasio \emph{et al.}~\cite{distasio2014a} also found that PBE0 produces a softer structure than PBE, but the
softening found in the latter work is fairly weak, with $g_{\rm O O}^{\rm max}$ decreasing from $3.22$ to $2.98$ and
$g_{\rm O O}^{\rm min}$ increasing from $0.38$ to $0.52$.  Simulations reported by 
Guidon \emph{et al.}~\cite{guidon2008a,guidon2010a} found very little change in structure from PBE to PBE0 even when
the fraction of exact exchange in PBE0 is increased from its normal value of $25$~\% all the way to $100$~\%. 
However, there is very recent evidence~\cite{delben2015a} that basis-set errors in the ADM (Auxiliary Density Matrix)
technique employed in Ref.~\cite{guidon2010a} may have had the effect of slightly hardening the structure,
so that the small softening due to exact exchange was perhaps masked. 

All the comparisons we have presented so far for GGA and hybrid functionals have been for fixed densities very close to
the experimental value. However, a major failing of GGAs is that they give very inaccurate pressure-volume
relations for the liquid. One manifestation of this is that if simulations are performed at zero pressure, for example by
working with the ($N P T$) ensemble, then the average density $\rho_{\rm eq}$ deviates markedly from the experimental value.
Equivalently, in simulations at fixed volume, the average pressure is poorly predicted. Since pressure is routinely
calculated (though not always reported) in constant-volume simulations, these errors are straightforward to detect.

The simulations of Kuo \emph{et al.}~\cite{kuo2004a}, in which BLYP-based simulations were used to study 
the density profile in a water slab
with free surfaces, suggested that BLYP underestimates $\rho_{\rm eq}$ by $\sim 15$~\%,
giving a value of $\sim 0.85$~g/cm$^3$. An underestimate of this order was confirmed soon afterwards by simulations
performed in the ($ N, P, T $) ensemble using MC sampling~\cite{mcgrath2005a}, which also employed the BLYP
functional. These indicated an equilibrium density at $T = 298$~K of \emph{ca.}~$0.80$~g/cm$^3$. This substantial
underestimate has since been confirmed by many
studies~\cite{schmidt2009a,baer2011a,ma2012a,lin2012a,bartok2013a,alfe2013a,delben2013a}, 
though the numerical values have been surprisingly variable, with $\rho_{\rm eq}$ ranging from $0.74$~g/cm$^3$~\cite{schmidt2009a}
to $0.92$~g/cm$^3$~\cite{lin2012a,ma2012a}. Related work has shown that revPBE gives even 
lower densities~\cite{wang2011a,lin2012a}, with $\rho_{\rm eq}$
ranging from $0.69$ to $0.85$~g/cm$^3$. PBE also underestimates the 
density~\cite{schmidt2009a,wang2011a,lin2012a,corsetti2013a,miceli2015a,gaiduk2015a}, but 
more modestly, with $\rho_{\rm eq}$ varying from $0.86$ to $0.96$~g/cm$^3$
in different studies. One of the few DFT studies
of the moderately compressed liquid is that of Ref.~\cite{alfe2014a}, which reported BLYP and 
PBE simulations at a density of $1.245$~g/cm$^3$,
and $T = 420$~K, where the experimental pressure is $15$~kbar. As would be expected from the predictions of these GGAs
near ambient conditions, BLYP was found to give a greatly overestimated pressure of $32$~kbar, while the PBE value of
$24$~kbar was less in error. Values of the equilibrium density from GGA simulations are summarized in 
Tables~\ref{tab:liq_BLYP} and \ref{tab:liq_PBE}, and we note that
all studies that compare values of $\rho_{\rm eq}$ predicted by different functionals
agree that the estimated densities come in the order revPBE~$<$ BLYP~$<$ PBE~$<$ expt. 
This trend is consistent with the natural supposition that $\rho_{\rm eq}$ computed with GGAs will tend
to increase with increasing H-bond energy. 

The large body of work based on semi-local and hybrid functionals thus indicates that increasing the H-bond energy
enhances the  ordering of the liquid and reduces the diffusivity, and also (slightly more tentatively) increases
the equilibrium density. However, functionals that give a more realistic degree of ordering and diffusivity
also severely underestimate the equilibrium density. This comes as no surprise, since the work on
clusters and ice structures taught us that to get both good H-bond energies and a correct extended-compact
balance, dispersion is needed.

\subsection{Dispersion-inclusive approximations}
\label{sec:disp_incl}

With these ideas in mind, we now turn to simulations of the liquid performed with dispersion-inclusive DFT approximations,
which have been reported using various representations of dispersion added to a number of different GGAs and hybrids.
The main features of the simulations reviewed here are summarized in Tables~\ref{tab:liq_disp_BLYP_PBE} and \ref{tab:liq_disp_revPBE_etc}.
We start by discussing the very popular approach in which correction potentials are added to the BLYP 
functional~\cite{schmidt2009a,jonchiere2011a,mcgrath2011a,baer2011a,yoo2011a,ma2012a,delben2013a,lin2009a,lin2012a,bankura2014a}.
As noted in the Sections on clusters and ice structures and in the Appendix, there are a number of methods of this
type, including the different versions of BLYP-D due to the Grimme group~\cite{grimme2004a,grimme2006a,grimme2010a}, 
and DCACP-corrected BLYP as developed by Rothlisberger and co-workers~\cite{vonlilienfeld2004a}. 
It is convenient to include here simulations based on BLYP+GAP~\cite{bartok2013a,alfe2013a}, which corrects
almost exactly for the 1- and 2-body errors of BLYP and therefore accounts for dispersion, which is expected to be
mainly a 2-body interaction in water systems~\cite{vonlilienfeld2010a}. We already know 
that all these methods give a much better description
of clusters and ice structures than BLYP itself, so there is good reason to hope for improvements in the liquid as well.

\begin{figure}
\centering
\begin{tabular}{c}
\includegraphics[width=0.8\linewidth]{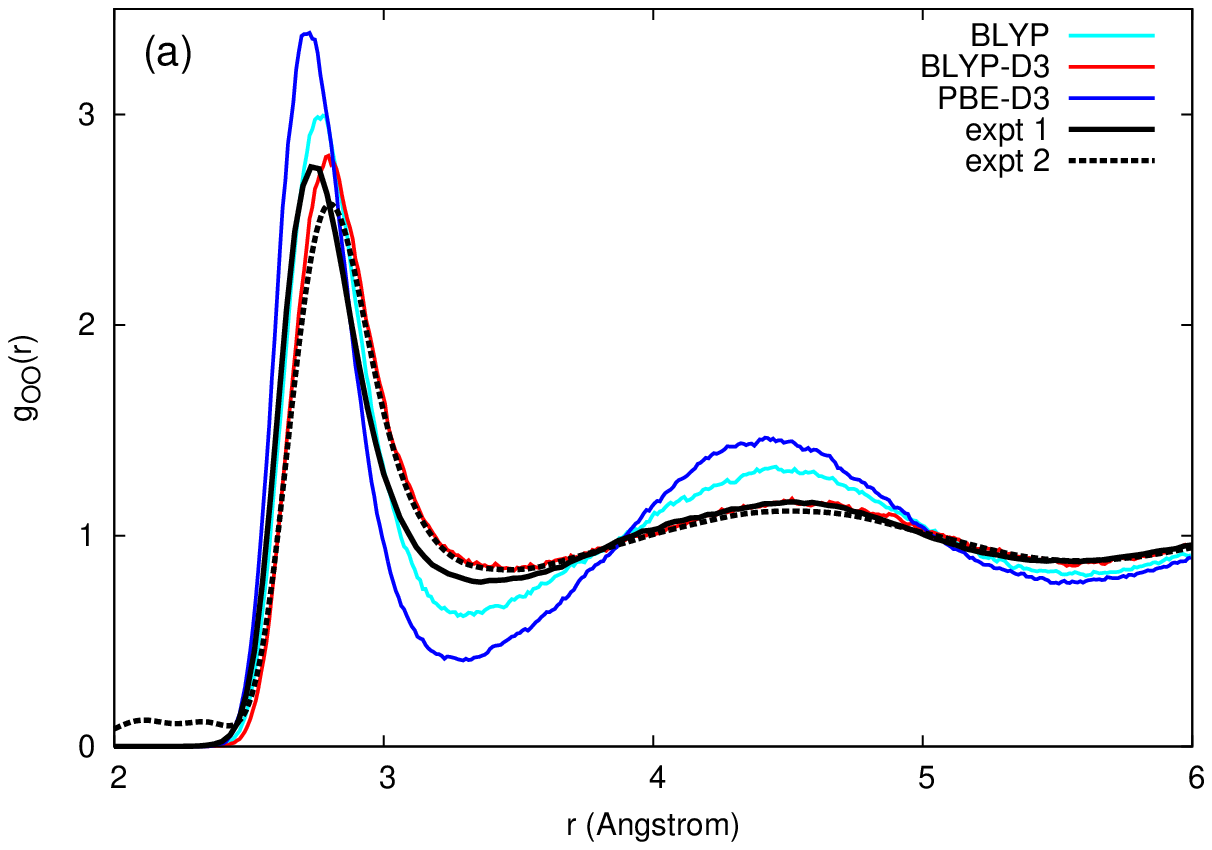} \\
\includegraphics[width=0.8\linewidth]{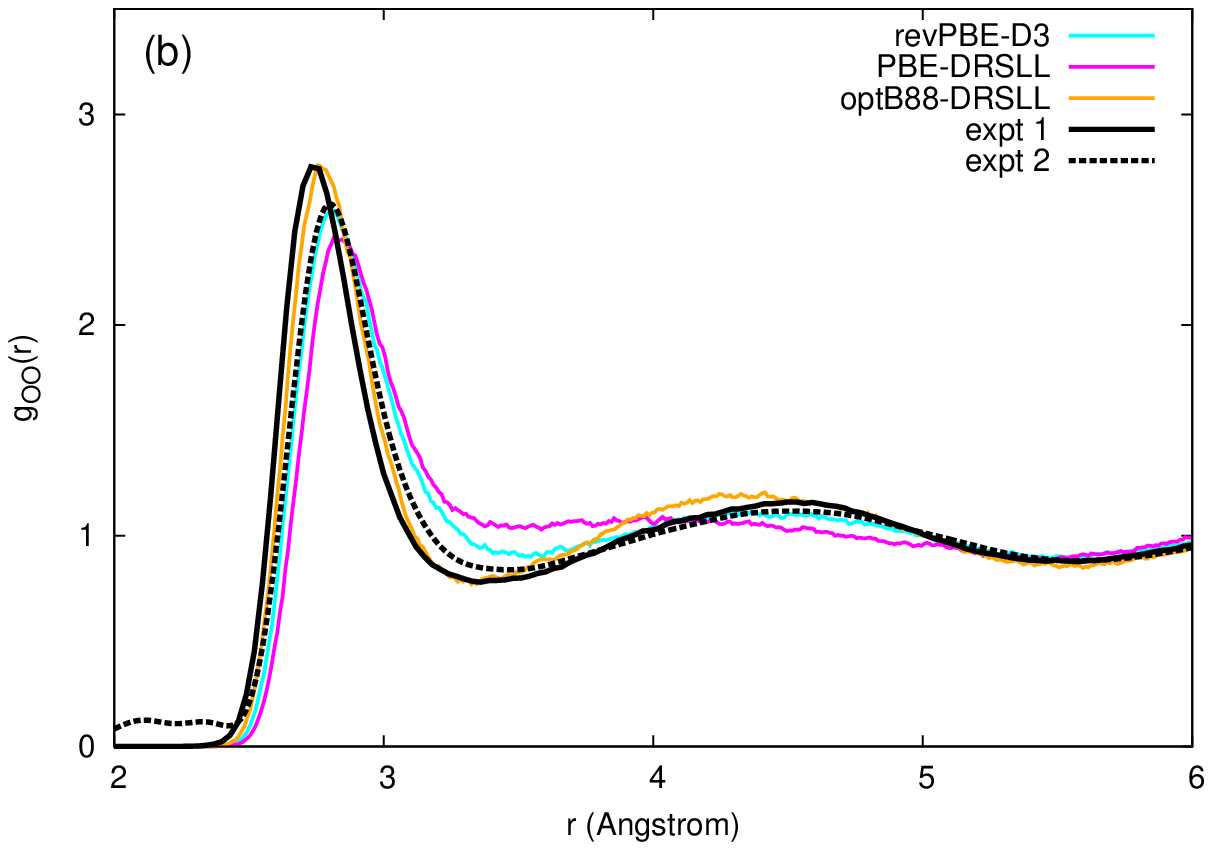}
\end{tabular}
\caption{
Oxygen-oxygen rdf $g_{\rm O O} ( r )$ of liquid water from BLYP and dispersion-inclusive DFT approximations,
compared with experiment. Simulation results are from Ref.~\cite{bankura2014a}. Experimental results
are from joint refinement of neutron and x-ray measurements~\cite{soper2007a} (expt~1: solid black curve) and from
high-energy x-ray diffraction measurements~\cite{skinner2013a} (expt~2: dashed black curve).
}
\label{fig:gOO_disp_incl_expt}
\end{figure}

Almost all the simulations based on corrected BLYP agree that the addition of dispersion substantially increases the equilibrium
density $\rho_{\rm eq}$, softens the liquid structure, and increases the diffusivity. Since uncorrected BLYP considerably
underestimates $\rho_{\rm eq}$, somewhat overstructures the liquid, and suppresses diffusion, this means that the changes
brought by dispersion are all in the right direction. The large differences
in the liquid properties given by different simulations based on uncorrected BLYP (see above) carry over to
dispersion-corrected BLYP, as can be seen in Table~\ref{tab:liq_disp_BLYP_PBE}. Focusing on the areas of
agreement first, we note from the Table that all the correction methods increase $\rho_{\rm eq}$
by between $0.10$ and $0.25$~g/cm$^3$~\cite{schmidt2009a,baer2011a,lin2012a,ma2012a,delben2013a,alfe2013a}. 
Taken  together, the calculations suggest that with dispersion-corrected BLYP
$\rho_{\rm eq}$ is slightly overestimated, perhaps by $\sim 5$~\%. This is not unexpected, since we saw that
BLYP corrected for 1- and 2-body errors using GAP overbinds both water clusters and ice structures, and tends
to compress both. The liquid structure is softened in similar ways by the different forms of dispersion-corrected BLYP,
the value of $g_{\rm O O}^{\rm max}$ being lowered by $\sim 0.30$ and the value of
$g_{\rm O O}^{\rm min}$ being raised by $\sim 0.20$. Perhaps most important is that most of 
the simulations give $g_{\rm O O}^{\rm min} \simeq 0.78$, which is  close to the well established experimental value of $0.84$
and is very much better than the value of $g_{\rm O O}^{\rm min} \simeq 0.60$ given by uncorrected BLYP.
An example of the softening effect of adding dispersion to BLYP is shown in Fig.~\ref{fig:gOO_disp_incl_expt}.
(We include in this Figure $g_{\rm O O} ( r )$ obtained from the recent high-energy x-ray data of Ref.~\cite{skinner2013a}, which 
is expected to be more accurate than earlier experimental measurements of $g_{\rm O O} ( r )$.)
As expected from the structural softening, all the simulations agree that correcting BLYP for dispersion
increases the diffusion coefficient $D$ by a factor of $2 - 3$, thus bringing it into fair agreement with experiment
(correction for system-size errors is essential in making this comparison, as emphasized above). The general
consensus is that BLYP corrected for dispersion, whether in the form of BLYP-D, BLYP-DCACP or BLYP+GAP
is a considerable improvement over BLYP itself, as might be hoped from the reasonable performance of these
approximations for clusters and ice structures.

Two caveats should be noted, however. First, the BLYP-D2 simulation of Ma \emph{et al.}~\cite{ma2012a}, which
gave a high $\rho_{\rm eq}$ of $1.07$~g/cm$^3$, predicted a significant under-structuring of the liquid.
The authors of Ref.~\cite{ma2012a} stress that their simulations are performed very close to the CBS limit, and
they suggest that the basis sets used to generate Grimme's D2 representation of dispersion may not have
been large enough to guarantee compatibility with their own basis sets. This suggestion appears to deserve
further investigation. Second, dispersion-corrected BLYP cannot be regarded as
fully satisfactory, since we know that it suffers from beyond-2-body over-binding, which is presumably the
reason why its value of $\rho_{\rm eq}$ is too high by a few percent in most of the simulations.

The PBE approximation has rivalled BLYP in popularity as a starting point for adding dispersion, with a wide variety
of techniques used to represent the dispersion. Simulations of liquid water have been published using the representations
of dispersion due to Grimme~\cite{schmidt2009a,delben2013a,forster-tonigold2014a,gaiduk2015a} and 
Tkatchenko-Scheffler~\cite{distasio2014a}, as well as the DCACP method~\cite{lin2012a}. Furthermore, several simulations
have been reported in which the non-local correlation functional pioneered by Lundqvist, Langreth and co-workers
is paired with PBE~\cite{wang2011a,zhang2011a,zhang2011b}. We also refer here to a simulation 
of the compressed liquid based on GAP-corrected PBE~\cite{alfe2014a}.
As expected, the addition of dispersion to PBE generally increases $\rho_{\rm eq}$, though this leads to
an overestimate by $\sim 13$~\% in the case of PBE-DRSLL (see Table~\ref{tab:liq_disp_BLYP_PBE}). Interestingly, the almost exact
GAP correction for 1- and 2-body errors gives very little change of pressure in the compressed liquid at density
$1.245$~g/cm$^3$~\cite{alfe2014a}, so that the error in the pressure given by PBE appears to be mainly a 
beyond-2-body effect.

With PBE as the starting approximation, the effect of dispersion on the liquid structure depends strongly on the method
used to represent dispersion. The simulations based on 
Grimme PBE-D~\cite{schmidt2009a,delben2013a,forster-tonigold2014a,bankura2014a,gaiduk2015a} all 
show that the addition of dispersion
does not cure the marked over-structuring given by PBE itself. In these simulations, the values of $g_{\rm O O}^{\rm max}$
and $g_{\rm O O}^{\rm min}$ are $3.2$ - $3.4$ and $0.4$ - $0.5$ respectively, compared with the experimental
values $2.57$ and $0.84$ (see Table~\ref{tab:liq_disp_BLYP_PBE}), as we illustrate for 
the case of PBE-D3 in Fig.~\ref{fig:gOO_disp_incl_expt}. The same 
lack of structural softening was found with PBE-DCACP~\cite{lin2012a}, but the PBE-TS approximation
appears to produce a modest softening~\cite{distasio2014a}, the 
dispersion-corrected values of $g_{\rm O O}^{\rm max}$ and $g_{\rm O O}^{\rm min}$ being $2.99$ and $0.54$. However, with DRSLL
the softening is much stronger~\cite{wang2011a,zhang2011a,zhang2011b,corsetti2013a,bankura2014a}, 
the corrected values of $g_{\rm O O}^{\rm max}$ 
and $g_{\rm O O}^{\rm min}$ being typically $\sim 2.6$ and $\sim 0.8$ (see Fig.~\ref{fig:gOO_disp_incl_expt}). 
Such large differences between the PBE-based methods seem less surprising if one recalls (Sec.~\ref{sec:dimer})
the strong overbinding produced by PBE-DRSLL in the range of O-O distances $3.0 - 4.0$~\AA\ compared
with PBE-D3 and PBE-TS. We note also our earlier comments about
the wisdom of adding 2-body dispersion to PBE, given that PBE already slightly overbinds ice Ih and
 is known to suffer from significant beyond-2-body errors. We return to this question later.
 
We know from our discussion of ice energetics that the success of the Lundqvist-Langreth approach depends 
very much on the underlying exchange functional with which dispersion is coupled. In addition to the PBE-DRSLL work just discussed,
there have been several other simulations of the liquid in which DRSLL-type representations of dispersion are
used in conjunction with different GGAs. The  original version of DRSLL~\cite{dion2004a}
was based on the revPBE approximation, and simulations
employing this version, referred to here as revPBE-DRSLL, have been reported by 
Wang \emph{et al.}~\cite{wang2011a}. We recall that uncorrected revPBE and RPBE
are the only GGAs discussed earlier that do not suffer from significant overstructuring,  their O-O rdfs at
a density of 1.00 g/cm$^3$ being in fair agreement with experiment, though revPBE gives an
unacceptably high pressure at this density. According to the simulations~\cite{wang2011a}, the use of DRSLL dispersion
in conjunction with revPBE corrects the large error in pressure, but seriously worsens the liquid structure. In particular, revPBE-DRSLL
produces a completely spurious peak in $g_{\rm O O} ( r )$ in the region where the first minimum should be. 
The poor performance of this functional is not
unexpected, since we know that it binds the water dimer and ice Ih structure much too weakly. 
We have also found~\cite{alfe2016a} that the 2-body energy given by revPBE-DRSLL is significantly
overbound in the important region $3.0 < R_{\rm O O} < 4.0$~\AA, and this presumably
worsens the liquid structure still further.
The effect of adding Grimme D2 or D3 dispersion to revPBE has been studied by Lin \emph{et al.}~\cite{lin2012a}
and Bankura \emph{et al.}~\cite{bankura2014a}, who both found moderate agreement 
with experiment (see Fig.~\ref{fig:gOO_disp_incl_expt}). Strangely, revPBE-DCACP appears to be somewhat more structured than 
revPBE~\cite{lin2012a}, though the authors caution that the DCACP
parameters may be unphysical in the case of revPBE~\cite{lin2012a}. 

The known underbinding of revPBE-DRSLL and overbinding of PBE-DRSLL make it interesting
to investigate versions of DRSLL that give better H-bond energies, examples of which are
the optB88-DRSLL and optPBE-DRSLL functionals proposed by Klime\v{s} \emph{et al.}~\cite{klimes2010a}, and
the rPW86-DF2 approximation~\cite{lee2010a}. Simulations of the liquid
based on optB88-DRSLL and rPW86-DF2 have been described by Zhang \emph{ et al.}~\cite{zhang2011a}, who found
that the former gave a $g_{\rm O O} (r)$ in rather close agreement with experiment, while the latter
produced a somewhat understructured liquid. The rPW86-DF2 approximation was also employed by 
M\o{}gelh\o{}j \emph{et al.}~\cite{mogelhoj2011a}, who found a more severe understructuring, with the second shell of
$g_{\rm O O} (r)$ almost completely washed out. The same authors also performed simulations
based on optPBE-DRSLL, which also gave a seriously understructured liquid, with $g_{\rm O O} ( r )$
in poor agreement with experiment. 
It should be noted that the simulations of M\o{}gelh\o{}j \emph{et al.}~\cite{mogelhoj2011a} may not give
a true indication of the performance of rPW86-DF2 and optPBE-DRSLL for the liquid, since
they were performed with the
intramolecular O-H bond length held fixed at a value close to the gas-phase value. It is known that
fixing the O-H distance in this way can substantially soften the liquid structure and increase the diffusivity~\cite{allesch2004a,leung2006a}.
On the other hand, the simulations of Zhang \emph{et al.} were performed on systems of only $32$ monomers,
which may have the effect of hardening the liquid structure, according to Ref.~\cite{kuehne2009a}.
More recently, simulations employing optB88-DRSLL with a 64-molecule cell, reported by Bankura \emph{et al.}~\cite{bankura2014a},
confirm that this functional gives a $g_{\rm O O} ( r )$ in quite close agreement with experiment (Fig.~\ref{fig:gOO_disp_incl_expt}).

It has been argued in recent papers~\cite{delben2013a,distasio2014a,gaiduk2015a} that 
the most accurate XC functionals for water will need to include both dispersion
and a fraction of exact exchange. We noted above the evidence that the over-structuring of the liquid predicted by PBE
is reduced both by adding the TS form of dispersion and by replacing PBE by the hybrid functional PBE0,
and DiStasio \emph{et al.}~\cite{distasio2014a} have shown that PBE0-TS produces an O-O rdf that agrees 
fairly closely with diffraction data, if allowance is made for QNE.
However, one should bear in mind the deficiencies of PBE0-TS for ice energetics, and particularly its overbinding
of ice Ih and its substantial overestimate of the energy difference between ice~VIII and ice~Ih. There is evidence that
PBE0-D3 gives a satisfactory equilibrium density of the liquid~\cite{delben2013a,gaiduk2015a}.

Several things are clear from our survey of DFT simulations of the liquid. First, no
GGA can simultaneously describe its structure, dynamics and thermodynamics; second, a fraction of
exact exchange appears to help, but does not solve the essential problem; third, there are
promising signs that a fully satisfactory description may be achievable by dispersion-inclusive DFT,
if the large inconsistencies between different approaches can be understood and resolved. 
In the following Section, we try to draw together the main lessons that can be learnt from DFT work
on all the water systems discussed in this review. In the light of those lessons, we will then offer a more
detailed interpretation of DFT approximations for the liquid in Sec.~\ref{sec:liq_interpret}.


\section{Discussion and outlook}
\label{sec:discussion}

\subsection{Overview}
\label{sec:overview}

The extensive DFT work on water systems reviewed here contains a wealth of information
about the strengths and weaknesses of a wide variety of XC approximations. Encouragingly,
there are now dispersion-inclusive methods among them that describe fairly accurately the properties
of water clusters, ice structures and the liquid. One of our aims here has been to 
aid further progress by elucidating the main
reasons why some approximations are more successful than others. We have approached our
task by attempting to analyze the performance of XC functionals in describing the main components of the
energy, namely first-order electrostatics, polarization (induction), dispersion, and
weak covalency (charge transfer), all acting in concert with exchange-overlap interactions. In addition, we
have noted that the energetics of intra-molecular deformation can be important. We try to summarize
here what we have learnt.

\subsection{Components of the energy}
\label{sec:energy_components}

Since first-order electrostatics depends only on the charge distributions of unperturbed mole\-cules, it is accurately
reproduced by a functional that describes the monomer correctly. Likewise, the polarizability part of the energy is correct if the
functional describes the response of the monomer to appropriate electric fields. We saw in Sec.~\ref{sec:monomer}
that GGAs generally describe the dipole and quadrupole moments of the monomer
quite well. However, they somewhat overestimate polarizabilities, the predicted dipolar
polarizability being typically $10$~\% too large. Hybrid functionals such as PBE0 and B3LYP give
much more accurate polarizabilities. As noted in Sec.~\ref{sec:coop}, the GGA errors of polarizability would be
expected to exaggerate the cooperative enhancement of H-bond energies, which should be better described
by hybrids, and this is indeed the case for small clusters. In line with this, we saw in Sec.~\ref{sec:ice} that
the PBE binding energy of ice Ih is stronger than the PBE0 binding by nearly $40$~meV/monomer, which
is a significant difference. We comment below on the effect of GGA polarizability errors
in the liquid.  We saw in Sec.~\ref{sec:monomer} that the intramolecular 
vibrational frequencies provide an important measure of the deformability of the monomer from its gas-phase geometry. 
The O-H stretch frequencies are appreciably underestimated by GGAs, so that the monomer is too easily deformable,
but hybrid functionals are more accurate. The effect of deformability errors on the liquid will be noted later.

We now turn to dispersion, which has been the focus of much of the more recent work reviewed here. We have shown
that semi-local functionals are generally incapable of accounting for the energy balance between extended and compact
structures of water systems. They incorrectly favor the ring and book isomers of the hexamer over the 
prism and cage (Sec.~\ref{sec:hexamer}), and they
grossly exaggerate the energy differences between compressed ice structures such ice VIII and the 
ambient Ih structure (Sec.~\ref{sec:ice}). We have reviewed work based on a variety of
dispersion-inclusive approaches, including those based on true non-local 
functionals~\cite{dion2004a,lee2010a,klimes2010a,vydrov2010a}, those that add heuristic dispersion
potentials to chosen semi-local functionals~\cite{grimme2004a,grimme2006a,grimme2010a,tkatchenko2009a}, 
and those employing parameterized electron-ion potentials to mimic dispersion effects~\cite{vonlilienfeld2004a}.
Our survey has shown that all these approaches achieve greatly improved energy differences between extended
and compact structures in clusters and ice structures. The crucial role of dispersion is thus beyond doubt.
However, not all the dispersion-inclusive methods are equally good, because some of them achieve their
improvements at the expense of worse H-bond energies in the dimer, the ring-hexamer and ice Ih. These problems are
connected with the description of exchange-overlap interactions. We have also seen (Sec.~\ref{sec:dimer}) that
there can be substantial differences in the distance dependence of the 2-body energy predicted by different
dispersion-inclusive methods, even when these are based on exactly the same semi-local functional.

We have shown that H-bond energies predicted by GGAs and hybrids in the dimer, the ring hexamer and ice Ih vary
over a wide range, their strength being underestimated by $\sim 25$~\% by revPBE and overestimated by $\sim 30$~\%
by PBEsol. (LDA overestimates by $\sim 80$~\%.) We noted that this wide variation is mainly due to the very different
behavior of the exchange-enhancement factor at large reduced density gradient, and we recalled that this 
mechanism was first recognized many years ago as a general effect for dimers of closed-shell atoms and molecules. Some
approximations such as PBE give an attractive interaction between closed-shell systems that mimics dispersion, but is in fact a
mis-description of the exchange-overlap interaction. More strongly repulsive GGAs such as BLYP give no such attraction.
This means that in order to reproduce the H-bond energy while also obtaining the correct extended-compact balance,
dispersion should be added to a semi-local functional whose overlap interaction is given by an appropriately designed
exchange-enhancement factor. This idea underlies the design of improved versions of the dispersion-inclusive
functionals~\cite{klimes2010a,lee2010a} based on the approach of Dion \emph{et al.}~\cite{dion2004a},
and the general importance of the choice of semi-local functional in constructing dispersion-inclusive
approximations has been stressed by a number of authors~\cite{silvestrelli2009a,lee2010a,wang2011a,hamada2014a}.
An interesting example of this is that the good accuracy of the PBE approximation for the H-bond energies of the dimer, 
the ring hexamer and ice Ih is due to its spurious exchange attraction mimicking the missing dispersion.

If dispersion could be mimicked by spurious exchange attraction without other errors, the failure of PBE to describe the 
extended-compact energy balance would be paradoxical. However, we saw in our discussion of the hexamer, the larger thermal
clusters and the ice structures that the paradox is resolved by many-body errors. Exchange-enhancement factors that create a spurious
2-body exchange attraction also give a spurious beyond-2-body repulsion, and this explains why PBE
incorrectly destabilizes the prism and cage isomers of the hexamer; the same mechanism contributes to its destabilization
of compressed ice structures. The same beyond-2-body errors are seen in the large thermal clusters (Sec.~\ref{sec:larger_clusters}).
By contrast, the BLYP approximation exaggerates 2-body
exchange repulsion and beyond-2-body exchange attraction, so that correction for its 2-body errors gives a moderate
degree of overbinding, as we saw for the ice structures. The implication is that to obtain both correct binding energies of extended
structures and correct relative energies of compact and extended structures, it is essential to use functionals that give both
correct dispersion and correct exchange-overlap interactions. This is the message that emerges from the work on clusters and
ice structures.

The final energy component needing comment is weak covalency, sometimes referred to as ``charge transfer'' or
``delocalization'' energy. We noted in the Introduction that the contribution of this mechanism to H-bonding in water
systems has been much debated. It has been claimed~\cite{isaacs1999a} that Compton scattering can be used to
characterize the strength of partial covalency in ice Ih, but subsequent analysis~\cite{ghanty2000a,romero2001a}
suggests that this is not the case. However, later experimental work based on photoelectron and x-ray absorption
spectroscopy~\cite{nilsson2005a} has been interpreted to indicate that electron transfer between monomers plays a significant part
in determining charge redistribution in ice, and by implication in liquid water.
It seems clear that quantitative statements about partial covalency depend
significantly on the definition adopted. However, whatever definition is used, it appears to us that the self-interaction
error committed by most practical XC approximations is likely to exaggerate the degree of partial covalency, 
as has been noted in the general context of H-bonding in Ref.~\cite{piquemal2005a}. This may
be the reason why GGAs, and to a lesser extend hybrids, systematically overestimate the energy difference between
non-H-bonded geometries of the water dimer and its global minimum. We noted in Sec.~\ref{sec:ice} that this error may 
contribute to the destabilization of compressed ice structures. This possibility merits further study.

To summarize our analysis of DFT errors in the various energy components: The most important
deficiencies of conventional GGAs and hybrids lie in their poor description of dispersion and exchange-overlap interactions;
dispersion-inclusive methods must describe both correctly if binding energies and the extended-compact energy balance
are to come out right. Accurate polarizabilities and monomer deformation energies are also not unimportant, so
the incorporation of an appropriate fraction of exact exchange is desirable. Errors in describing partial covalency
may not be entirely negligible. 

\subsection{Interpreting DFT simulations of the liquid}
\label{sec:liq_interpret}

The strengths and weaknesses of XC approximations for clusters and ice structures that we have just outlined
help to explain the characteristic features of DFT simulations of the liquid.
We noted in Sec.~\ref{sec:liquid} that a key structural feature of the 
ambient liquid is the degree of cross-shell penetration (CSP), i.e. the probability of finding molecules at O-O
separations in the range between $3.0$ and $4.0$~\AA\ that is unoccupied in ice~Ih. We
can characterize CSP crudely by the value of $g_{\rm O O}^{\rm min}$, the value of $g_{\rm O O} ( r )$
at its first minimum. We saw that $g_{\rm O O}^{\rm min}$ depends strongly
on XC approximation, ranging from $\sim 0.4$ for uncorrected PBE to $\sim 0.9$ for 
dispersion-inclusive revPBE-DRSLL, compared with the experimental value of $0.84$ at ambient conditions.
The diffusion coefficient $D$ correlates with $g_{\rm O O}^{\rm min}$, being too small by
a factor of $\sim 10$ when $g_{\rm O O}^{\rm min} \simeq 0.4$ and being in reasonable agreement with experiment
when $g_{\rm O O}^{\rm min} \simeq 0.8$.

Two important energies would be expected to govern the degree of CSP. The first is the H-bond energy,
characterized roughly by the sublimation energy $E_{\rm sub}^{\rm Ih}$ of ice Ih, which is closely correlated
with the binding energy $E_{\rm b}^{\rm dim}$ of the dimer (Fig.~\ref{fig:H-bond_energy}). If the H-bond energy is too strong,
it will be too difficult to break H-bonds in the liquid, which will become too ordered, with too little CSP.
But we must also consider the energy change associated with the close approach of molecules that
are not H-bonded to each other, which can be crudely characterized by the difference of sublimation
energy $\Delta E_{\rm sub}^{\rm Ih-VIII}$  between ice VIII and ice Ih. Even if an XC approximation
gives $E_{\rm sub}^{\rm Ih}$ correctly, the liquid can still be over-structured because 
$\Delta E_{\rm sub}^{\rm Ih-VIII}$ is too large, as is the case with uncorrected PBE. 
In addition to these two key energies, the distance dependence of the 2-body energy
is also important.

By varying the semi-local functional and the dispersion that is added to it, we control both $E_{\rm sub}^{\rm Ih}$
and $\Delta E_{\rm sub}^{\rm Ih-VIII}$. We saw in Sec.~\ref{sec:liquid} that as we pass through the series of GGAs
in order of increasing $E_{\rm sub}^{\rm Ih}$, the values of $g_{\rm O O}^{\rm min}$ and $D$ systematically
decrease, while the equilibrium density $\rho_{\rm eq}$ increases. However, good values of all three quantities
cannot be obtained simultaneously with any GGA, and in addition $\Delta E_{\rm sub}^{\rm Ih-VIII}$ is always seriously in error.
If we add suitable dispersion to an appropriately chosen GGA, we can ensure that both $E_{\rm sub}^{\rm Ih}$ and
$\Delta E_{\rm sub}^{\rm Ih-VIII}$ are correct, as we know from the work on ice structures (Sec.~\ref{sec:ice}).
We saw in Sec.~\ref{sec:liquid} that there are functionals that are not far from satisfying these conditions,
while also giving a good description of the liquid: optB88-DRSLL is one example.
This encourages optimism that satisfactory practical
XC functionals for water systems are achievable.

We consider that correct values for $E_{\rm sub}^{\rm Ih}$ and $\Delta E_{\rm sub}^{\rm Ih-VIII}$ are necessary for an
XC functional to be satisfactory, but they may still not be sufficient to guarantee good liquid properties. We
saw in Sec.~\ref{sec:dimer} that the PBE-D3, PBE-TS and PBE-DRSLL functionals give almost the same dimer
binding energy, but the distance dependences of their 2-body energies differ greatly, and this appears
to explain their very different liquid structures. This may be an example of a general problem concerning
the distance dependence of dispersion, which deserves closer scrutiny. For the water systems we have reviewed, 
correctness of the 2-body energy over the range of O-O distances out to $\sim 4.0$~\AA\ appears to be essential.

Compared with the large effects of dispersion and exchange-overlap errors, the errors of polarizability and
monomer deformability on the properties of the liquid appear to be less important. Nevertheless,
we have noted that the overestimated polarizability found with GGAs will tend to enhance the H-bond strength,
and we reviewed the evidence for a small resulting suppression of CSP. A similar overstructuring
may well be caused by the overestimated monomer deformability given by GGAs, though as far we know
this possible effect has never been quantified. The errors of both polarizability and deformability
are much smaller with hybrid functionals, which should ideally be used as the basis for dispersion-inclusive
simulations, if the computer budget permits. 

\subsection{A scoring scheme}
\label{sec:scoring}

In order to characterize the errors of XC functionals for water systems, we have devised a scoring scheme,
which assigns a percentage score to any chosen approximation, according to its performance
for the properties of the water monomer, the dimer, the hexamer, and ice structures. 
The physical quantities employed in the present form of our scoring scheme are as follows. We score the binding
energy of the dimer $E_{\rm b}^{\rm dim}$, the binding energy per monomer of the ring-hexamer
$E_{\rm b}^{\rm ring}$ and the sublimation energy $E_{\rm sub}^{\rm Ih}$ of ice Ih,
because these characterize the H-bond energy. The difference per monomer 
$\Delta E_{\rm b}^{\rm prism-ring} \equiv E_{\rm b}^{\rm prism} -  E_{\rm b}^{\rm ring}$ between 
the binding energies of the prism and ring isomers of the hexamer, and the difference 
$\Delta E_{\rm sub}^{\rm Ih-VIII} \equiv E_{\rm sub}^{\rm Ih} - E_{\rm sub}^{\rm VIII}$ of the sublimation
energies of ice Ih and ice VIII are also crucial, as we have seen. In addition, we regard the equilibrium
O-O distance $R_{\rm OO}^{\rm dim}$ of the dimer, and the equilibrium volumes per monomer
$V_{\rm eq}^{\rm Ih}$ and $V_{\rm eq}^{\rm VIII}$ of ice phases Ih and VIII as important. Since
the O-H bond-stretch energetics of the H$_2$O monomer may be important, we
characterize this by the symmetric stretch frequency $f_{\rm ss}^{\rm mono}$ of the monomer. 
We have seen that errors in the distance dependence of the 2-body energy of the dimer can also be
troublesome, and these should clearly be scored, but we prefer not to attempt this until we have
been able to report on a more detailed examination of dimer energetics with dispersion-inclusive functionals~\cite{alfe2016a}.

For each of the scored quantities, the chosen XC functional gives some value $x$, which generally deviates from the 
benchmark value $x_{\rm bench}$. We assign a score of $100$~\% if the magnitude of the deviation
is less than a chosen tolerance $\delta x_{\rm tol}$. Otherwise, we deduct $10$~\% for each successive increment
$\delta x_{\rm tol}$ in $| x - x_{\rm bench} |$. If $| x - x_{\rm bench} | > 11 \delta x_{\rm tol}$,  a zero score is given.
For the binding energies $E_{\rm b}^{\rm dim}$, $E_{\rm b}^{\rm ring}$ and $E_{\rm sub}^{\rm Ih}$, 
and for the compact-extensive differences $\Delta E_{\rm b}^{\rm prism-ring}$ and $E_{\rm sub}^{\rm Ih-VIII}$
we adopt a tolerance of $10$~meV, except that in the case of $E_{\rm b}^{\rm prism-ring}$ prediction of the wrong sign
gets a score of zero. For the equilibrium O-O distance of the dimer, the tolerance is $0.01$~\AA\ and for the equilibrium
volumes of the ice Ih and VIII phases our tolerance is a $1$~\% error. We allow a tolerance of $20$~cm$^{-1}$
for the monomer frequency $f_{\rm ss}^{\rm mono}$.
Finally, the overall percentage score for a given functional is simply the average of all the individual scores.

We present in Table~\ref{tab:scoring} the individual and overall scores for the local and semi-local functionals LDA,
PBE, BLYP and PBE0 and for a selection of dispersion-inclusive functionals. 
The XC functionals scored here are those for which data is readily available, but in the future it would be useful
to score a wider range of functionals. We see from the Table that the local, semi-local and hybrid functionals
all score rather poorly because they completely fail to reproduce the energy
differences $\Delta E_{\rm b}^{\rm prism-ring}$ and $\Delta E_{\rm sub}^{\rm Ih-VIII}$, and they also
struggle with the volume per monomer of ice VIII.
Some of the dispersion-inclusive functionals score much better, but others, such as revPBE-DRSLL, PBE-TS
and BLYP-D3 do poorly, because of problems with ice energetics and volumes. Overall, 
the highest scoring functionals are optPBE-DRSLL,
optB88-DRSLL, PBE0-TS and rPW86-DF2, although even they have notable deficiencies. It is striking that
some functionals that score well for the binding energies of the dimer and the ring hexamer and for the relative
energy of the ring and prism isomers of the hexamer can still do poorly  for the energies and volumes of ice
structures. This may be due to errors in the distance dependence of dispersion.

Any recommendation of the ``best'' functional must be tentative, since our scoring table shows that all the functionals
assessed here have faults. The optB88-DRSLL functional appears to be the most satisfactory, since it has the highest score
and also gives a good liquid structure, according to Refs.~\cite{zhang2011a,bankura2014a} (see also Table~\ref{tab:liq_disp_revPBE_etc}).
Its strong overbinding of ice Ih is a cause for concern, but this may be due to the excessive long-range attraction of the 2-body
energy, which seems to be a feature of DRSLL non-local correlation. Of the next highest scorers, rPW86-DF2 appears to give
an understructured liquid~\cite{zhang2011a}, and optPBE-DRSLL may do the same~\cite{mogelhoj2011a}. Next in the
scoring comes the hybrid-based PBE0-TS, which gives a good liquid structure, but would be expensive for general use.

It may be worth commenting that our
scheme has something in common with the methods developed by Vega \emph{et al.}~\cite{vega2009a,vega2011a}
for characterizing the ability of force fields to describe water systems, but differs from them in important ways.
Most significantly, our scheme works only with simple and readily computable energies and structures that characterize
clusters and ice structures, focusing on those quantities highlighted in the present review. By contrast, the scheme of
Vega \emph{et al.} employs quantities such as the temperature of the maximum density and the surface tension of the liquid,
whose calculation by DFT methods would be very demanding. Our approach to scoring is based on the idea that an
XC functional should be required to work well for clusters and ice structures before being seriously considered for the liquid.
However, we stress that a high score in our scheme may not be either necessary or sufficient to ensure success
in describing the liquid at ambient conditions, since even simple, unpolarizable force fields can succeed for 
the ambient liquid~\cite{vega2009a} in spite of their poor description of clusters, 
and our scheme may not yet capture all the key features of XC functionals. On the other hand, a low score alerts one to
the failure of a functional to describe at least \emph{some} parts of the energy adequately. We should also note that our
scheme is not meant to be definitive, and it can be extended and modified to suit different purposes.

\subsection{Outlook}
\label{sec:outlook}

Kohn Sham DFT is now 50 years old, and is one of the great success stories of molecular simulations.
It has been extremely productive in furthering our understanding of water. However, here 
we have deliberately dwelt on the difficulties faced by current functionals in describing water systems, because we think
that an analysis of these difficulties is needed for future progress. The past 10 years have been immensely
fruitful in highlighting the crucial role of dispersion. However, some representations of dispersion are clearly
not under satisfactory control, since different dispersion-inclusive methods can produce very different results.
We have also emphasized here the sensitivity of predictions to the choice of the semi-local part of the
XC functional.

The erratic differences beween the various representations of dispersion already manifest themselves in the
energetics of the dimer, and we think that satisfactory functionals should be required to reproduce this
energetics accurately. In fact, the parameterization of some types of dispersion-inclusive functional,
including Grimme DFT-D and the TS functionals, is explicitly based on benchmark data for molecular
dimers such as the S22 set~\cite{jurecka2006a}. But for water the distance dependence of the dispersion damping is vitally
important, and this has sometimes not been considered. The recent development of ``extended'' benchmark
datasets which include a range of inter-monomer distances and/or monomer orientations~\cite{rezac2011a,rezac2011b} is therefore
very welcome. We plan to report elsewhere on a more detailed study of the energetics of the water dimer
than has been possible here.

Good dimer energetics is, of course, not enough for water, because some functionals suffer from substantial
many-body errors, which are unrelated to polarizability or many-body dispersion, and which depend strongly 
on the choice of semi-local functional. Until fairly recently, such many-body errors for molecular systems have
attracted rather little attention. However, the situation is now changing, with publications appearing that explicitly
investigate these errors~\cite{tkatchenko2008a,gillan2014a,rezac2015a,huang2015a}. The 
study of molecular trimers is already enough to quantify some beyond-2-body errors,
and in the case of water the energetics of the hexamer is very informative. We believe that the question of
how to achieve small errors in both the 2-body and beyond-2-body energies of molecular systems in general
and water in particular will repay deeper study.

This review has focused on the strengths and weaknesses of XC functionals 
for a fairly limited set of properties: mostly energies, structures and to some extent 
dynamics. Getting these basic issues right is a prerequisite to more detailed 
study of e.g. the electronic properties of water in the condensed phases, dissociation 
and proton transfer, the anomalies mentioned in the Introduction, and, of  course, a 
mapping of the entire phase diagram of water with DFT. Although some important work 
has been done with DFT in some of these areas recently (see 
e.g. Refs.~\cite{murray2012a,kuehne2013a,bonnet2014a,schoenherr2014a,santra2015a}), 
much more work is needed with a broad range of functionals
before we can provide a full answer to the question ``How good is DFT for 
water?''. The computational cost of DFT is the main factor that slows progress when it 
comes to establishing all the thermodynamic properties needed to map out the phase 
diagram of water. With this in mind, the development of DFT-based machine learning 
potentials such as the GAP approach discussed here or the neural networks from Behler and 
co-workers~\cite{morawietz2016a} allow 
a wider range of water properties to be explored more rapidly with DFT-level accuracy. 

As a final comment, it is worth noting that benchmark electronic-structure calculations based on MP2, CCSD(T)
and quantum Monte Carlo are playing an increasingly important role in assessing DFT methods for clusters and ice structures, but until now
experiment has been the only source of accurate data for the liquid. The recent publication of simulations of
liquid water based both on MP2~\cite{delben2013a} and on variational quantum Monte Carlo~\cite{zen2015a} suggests 
the exciting future possibility of obtaining benchmark
liquid-state data for quantities that cannot be measured experimentally.


\begin{acknowledgments}
A.M.'s work is supported by the European Research Council under the European Union's 
Seventh Framework Programme (FP/2007-2013)/ERC Grant Agreement No. 616121 (HeteroIce project) and 
the Royal Society through a Wolfson Research merit Award. We thank M. L. Klein, A. Gross and A. E. Mattsson
for providing numerical data of rdfs from their simulations. This research used resources of the Oak Ridge 
Leadership Computing Facility, which is a DOE Office of Science User Facility supported under 
Contract DE-AC05-00OR22725. We are grateful to a number of people for reading and commenting on
an earlier version of this manuscript, including M. Tuckerman, B. Santra, B. Slater, J. VandeVondele,
G. Sosso and A. Zen.
\end{acknowledgments}


\section*{Appendix: Overview of exchange-correlation functionals}
\label{appendix}

While this is primarily a review on DFT for water, we include here a brief summary of the main 
features of some of the key XC functionals discussed in the review. Our aim here is not to provide a comprehensive overview of XC functionals,
but rather to provide some useful background information for those less well versed in the details of the functionals we refer to. 
We assume here a spin unpolarized system, whose distribution of electron number density is $\rho ( {\bf r } )$.

\subsection*{Local functionals}

The local  density approximation (LDA)~\cite{kohn1965a} was the original approximation used in all the earliest applications of DFT. Its
exchange-correlation (XC) energy $E_{\rm xc}^{\rm LDA} [ \rho ]$ is the sum of the exchange energy $E_{\rm x}^{\rm LDA}$ and
the correlation energy $E_{\rm c}^{\rm LDA}$, which are expressed in terms of the exchange energy and the correlation energy per electron
$\epsilon_{\rm x}^0 ( \rho )$ and $\epsilon_{\rm c}^0 ( \rho )$ in the uniform electron gas (sometimes called ``jellium'') 
of density $\rho$. The approximation takes the form:
\begin{equation}
E_{\rm xc}^{\rm LDA} = 
E_{\rm x}^{\rm LDA} + E_{\rm c}^{\rm LDA} =
\int d {\bf r} \, \rho ( {\bf r} ) \epsilon_{\rm x}^0 ( \rho ( {\bf r} ) ) +
\int d {\bf r} \, \rho ( {\bf r} )  \epsilon_{\rm c}^0 ( \rho ( {\bf r} ) ) \; .
\end{equation}
The exchange energy $\epsilon_x^0 ( \rho )$ is given by the Hartree-Fock formula (atomic units):
$\epsilon_x^0 ( \rho ) = - \frac{3}{4} ( 3 / \pi )^{1/3} \rho^{1/3}$. Approximations for $\epsilon_{\rm c}^0 ( \rho )$
typically employ parameterized fits to quantum Monte Carlo data for the uniform electron gas, 
these fits being constrained to reproduce the leading terms
of the high-density expansion. For dimers of closed-shell atoms and molecules (e.g. rare gases), the LDA gives an
attraction that resembles the dispersion interaction but is in fact an artefact due to the approximation made for exchange~\cite{harris1985a}.

\subsection*{Generalized gradient approximations}

In generalized gradient approximations (GGAs)~\cite{becke1986a}, the XC energy depends locally on the gradient of the density $\nabla \rho$ as well
as the density $\rho$. The magnitude of the local gradient can be specified by the so-called ``reduced gradient'', defined as the dimensionless
quantity $x \equiv | \nabla \rho | / \rho^{4/3}$ or sometimes as $s \equiv x / ( 2 ( 3 \pi^2 )^{1/3} )$, 
so that GGAs have the general form:
\begin{equation}
E_{\rm xc}^{\rm GGA} = E_{\rm x}^{\rm GGA} + E_{\rm c}^{\rm GGA} =
\int d {\bf r} \, f_{\rm x} ( \rho ( {\bf r} ) , s ( {\bf r} ) ) + \int d {\bf r} \, f_{\rm c} ( \rho ( {\bf r} ) , s ( {\bf r } ) ) \; ,
\end{equation}
where $f_{\rm x}$ and $f_{\rm c}$ specify the local exchange and correlation parts. However, the exchange part can be simplified,
using an exact condition that it must obey~\cite{levy1985a} under uniform scaling of the density, as a result of which the exchange part can be expressed as:
\begin{equation}
E_{\rm x}^{\rm GGA} [ \rho ( {\bf r} ) ] = 
\int d {\bf r} \, \epsilon_{\rm x}^0 ( \rho ( {\bf r} ) ) F_{\rm x} ( s ( {\bf r} ) ) \; ,
\end{equation}
where $F_{\rm x}$, the so-called exchange-enhancement factor, depends only on $s$. Many different GGAs have been proposed, differing
in their exchange-enhancement factors and the approximations for $f_{\rm c} ( \rho , s )$. We note that the form of $F_{\rm x} ( s )$ at large
reduced gradient is particularly important for non-covalent interactions in molecular systems such as water, because
exchange-overlap interactions depend strongly on the behavior of $F_{\rm x} ( s )$ in the regions where electron densities overlap,
which is where $s$ has large values~\cite{lacks1993a,zhang1997a,wu2001a,kannemann2009a,murray2009a,kanai2009a}. 
We note next the main features of the GGAs relevant for this review.

\noindent
\textbf{BLYP:} This functional has been extensively used for water systems, and consists of the B88 and LYP approximations for exchange and
correlation respectively~\cite{becke1988a,lee1988a}. The B88 approximation of Becke~\cite{becke1988a} has 
an exchange enhancement factor designed to reproduce
exactly the exchange energy density in the limit of long distance from an isolated atom. It depends on a single parameter, which
is adjusted to minimize the errors of the exchange energies of rare-gas atoms. For molecular systems, this exchange
functional has the desirable feature~\cite{kannemann2009a} that it eliminates the spurious exchange attraction between rare-gas atoms exhibited by the LDA.
The LYP correlation functional, so named for the paper by Lee, Yang and Parr where it was introduced~\cite{lee1988a}, is derived from an older
formula for correlation due to Colle and Salvetti~\cite{colle1975a}, who approximated it in terms of the electron density and the non-interacting
kinetic-energy density. The LYP functional is a pure GGA constructed by replacing the kinetic-energy density by its
second-order gradient expansion.

\noindent
\textbf{BP86:} This consists of exactly the same B88 functional~\cite{becke1988a} for exchange as the BLYP functional, but 
replaces LYP correlation by the correlation functional known
as P86. The latter functional was constructed by Perdew~\cite{perdew1986a} so as to reproduce accurately the correlation energy of both the uniform electron
gas and the electron gas with small gradients of the reduced density, and it also known to give accurate correlation energies for
free atoms and ions.

\noindent
\textbf{PBE:} This widely used approximation~\cite{perdew1996a} adopts a simple functional form for the exchange enhancement factor $F_{\rm x} ( s )$
proposed earlier by Becke~\cite{becke1986a} to ensure satisfactory behavior in the long-distance tails of free atoms. However,
the parameters appearing in this functional form are not regarded as empirically adjustable, as was done by Becke,
but are determined by requiring that exact conditions be satisfied. The form of the correlation energy density
$f_{\rm c} ( \rho , x )$ is also chosen so as to satisfy exact conditions on its behavior for small and large
reduced gradients, and the parameters appearing in this form are likewise determined by the requirement that they
satisfy exact conditions. PBE is therefore one of the few XC functionals whose parameters are all (or almost all) fundamental
constants, rather than being adjusted to fit empirical data.  However, the form of the exchange enhancement factor at large
reduced gradient produces a spurious exchange binding of closed-shell atoms and molecules~\cite{murray2009a}.

\noindent
\textbf{PW91:} This approximation~\cite{perdew1992a,perdew1993a} is a forerunner of the better known PBE functional. It differs from PBE in employing more complicated
functional forms for the exchange and correlation energies. These functional forms and the parameters that enter them
attempt to satisfy a wider range of exact conditions. The energy predictions of PW91 are often close to those of PBE. However, the
forms of their exchange enhancement factors at large reduced gradient are substantially different~\cite{kannemann2009a}, and PW91 produces a 
spurious exchange binding of closed-shell atoms and molecules that is larger than that of PBE.

\noindent
\textbf{revPBE:} This is a modification of the PBE functional. The modification was proposed by Zhang and Yang~\cite{zhang1998a}, and consists solely
of changing the value of one of the constants  appearing in the exchange enhancement factor. 
This change achieves considerably improved values of atomic total energies and molecular
atomization energies. For molecular systems such as water, the change is important because it greatly increases the
overlap-repulsion between pairs of molecules~\cite{kannemann2009a}. 

\noindent
\textbf{RPBE:} This functional~\cite{hammer1999a} resembles revPBE in being a modification of PBE formed by changing the exchange enhancement
factor, while leaving the correlation functional unchanged. However, in the case of RPBE, the functional form of $F_{\rm x} ( s )$ is modified.
The resulting $F_{\rm x} ( s )$ is almost identical to that of revPBE for small and moderate values of $s$, but increases more slowly at large
$s$ values~\cite{hammer1999a}. Physical properties predicted with RPBE and revPBE generally differ rather little.

\noindent
\textbf{PBEsol:} This is a modification of the PBE functional designed to improve predictions for the energetics of densely packed
solids and their surfaces~\cite{perdew2008a}. It achieves this by keeping the functional forms used for the exchange and correlation energies of
PBE, while modifying the two coefficients specifying the expansion of these energies to second order in the
density gradient.

\noindent
\textbf{AM05:} This functional, proposed by Armiento and Mattsson~\cite{armiento2005a}, may be classified as a GGA, but the principles on which
it is based differ from those used to develop conventional GGAs. Its aim is to improve predictions for surface properties, which are
often poorly predicted by functionals such as PBE. It has the unique feature that it is constructed to reproduce exactly
the exchange-correlation energies of \emph{two} types of model system: uniform jellium and the jellium surface.

\subsection*{Hybrids}

A hybrid XC functional is one formed by addition of a fraction of exact Hartree-Fock exchange to a chosen linear combination
of local or semi-local functionals. For this purpose, the Hartree-Fock exchange energy is defined to be:
\begin{equation}
E_{\rm x}^{\rm HF} = - \frac{1}{2} \sum_{i, j} \int \int d {\bf r} \, d {\bf r}^\prime \,
\psi_i^\star ( {\bf r} ) \psi_j ( {\bf r} ) \frac{1}{| {\bf r} - {\bf r}^\prime |} \psi_i ( {\bf r}^\prime ) \psi_j^\star ( {\bf r}^\prime ) \; ,
\end{equation}
where $\psi_i ( {\bf r} )$ are the Kohn-Sham orbitals of DFT, and the double sum goes over all pairs of occupied orbitals. Some justification~\cite{becke1993a}
for mixing in a fraction of $E_{\bf x}^{\rm HF}$ is supplied by the adiabatic connection 
theorem~\cite{harris1974a,gunnarsson1976a}, which underpins DFT methods. We
provide here brief notes about the hybrid methods that are relevant to this review.

\noindent
\textbf{PBE0:} This approximation is a hybrid version of the semi-local PBE functional, defined by~\cite{perdew1996b,adamo1999a}:
\begin{equation}
E_{\rm xc}^{\rm PBE0} = \frac{1}{4} E_{\rm x}^{\rm HF} + \frac{3}{4} E_{\rm x}^{\rm PBE} + E_{\rm c}^{\rm PBE} \; ,
\end{equation}
so that the exchange part of the functional is a linear combination of $E_{\rm x}^{\rm HF}$ and the usual PBE for exchange, $E_{\rm x}^{\rm PBE}$,
while the correlation part of the functional has the normal PBE form.

\noindent
\textbf{B3LYP:} This widely used hybrid is somewhat more complicated than PBE0, since the semi-local parts of the XC functional
themselves involve a linear combination of GGAs and LDA. It is defined as~\cite{becke1993a,stephens1994a}:
\begin{equation}
E_{\rm xc}^{\rm B3LYP} = a_0 E_{\rm x}^{\rm HF} + a_{\rm x} E_{\rm x}^{\rm B88} + ( 1 - a_0 - a_{\rm x} ) E_{\rm x}^{\rm LDA} +
a_{\rm c} E_{\rm c}^{\rm LYP} + ( 1 - a_{\rm c} ) E_{\rm c}^{\rm LDA} \; .
\end{equation}
Here, $E_{\rm x}^{\rm B88}$ and $E_{\rm c}^{\rm LYP}$ are the exchange and correlation parts of the BLYP functional (see above), and
the numerical values of the mixing coefficients are: $a_0 = 0.20$, $a_{\rm x} = 0.72$, and $a_{\rm c} = 0.81$.

\subsection*{Dispersion-inclusive methods}

Dispersion refers to the attraction between atoms or molecules due to the Coulomb interaction between correlated quantum fluctuations of their
electron densities. These are also known as London forces and in the DFT functional community often referred to as van der Waals forces. At long distances, where overlap of these densities is negligible, the dispersion energy
falls off as the inverse 6$^{th}$ power of the intermolecular distance $R$. At distances where the overlap between the electron
distributions vanishes, local, semi-local and hybrid functionals account only for interactions due to
electrostatics and polarization, and are incapable of describing the $1 / R^6$ behavior of dispersion. Several
approaches have been developed for overcoming this problem, as described in recent reviews~\cite{klimes2012a,dilabio2014a}.

One possible approach is simply to ignore the ``asymptotic'' region where dispersion falls off as $1 / R^6$, and to treat
only the intermediate and short-range region where non-local correlation is expected to contribute most to bonding.
This is the approach of the so-called Minnesota functionals~\cite{zhao2007a}, in which meta-GGAs and their hybrids
are carefully parameterized to reproduce both covalent and non-covalent binding energies, reaction-barrier
energies, etc. An alternative approach, which also aims to treat only intermediate and short distances, involves
the addition of artificial nucleus-electron potentials designed to mimic the energetics of non-local correlation, as in
the DCACP and DCP methods. The Minnesota functionals, as well as the 
DCACP~\cite{vonlilienfeld2004a} and DCP~\cite{torres2012a} methods, have enjoyed
some success for molecular systems, and have been used to a limited extent for water systems.

The simplest methods capable of accounting for dispersion in the asymptotic $1 / R^6$ region involve the
introduction of atom-atom pair potentials of the form $- C_{\rm A B} / R^6$ between every pair of atoms A, B,
which are added to chosen semi-local or hybrid functionals, referred to here as ``base'' functionals. This idea
goes back nearly $20$ years~\cite{meijer1996a,gianturco2000a,wu2001a}, and 
has been extensively developed by Grimme and co-workers~\cite{grimme2004a,grimme2006a,grimme2010a}.
The $1 / R^6$ form must be damped at intermediate and short distances, and Grimme proposed a damping
scheme, as well as a procedure for assigning values to the coeffficients $C_{\rm A B}$ for several choices
of base functional. This approach, known as DFT-D or DFT-D2~\cite{grimme2006a}, has seen widespread use and achieves major improvements in the
description of non-covalent interactions in molecular systems.

One deficiency of DFT-D2 is that it does not account for the dependence of atom-atom dispersion interactions
on the chemical state and environment of the atoms. For example, the electron distribution on an atom
may be compressed by its neighbors, so that its polarizability and hence its dispersion interaction with other atoms
are reduced. The Tkatchenko-Scheffler (TS) scheme~\cite{tkatchenko2009a} is in the same general spirit as Grimme's DFT-D2 method, but
it accounts for the dependence of dispersion interactions on changes of electron density, using the Hirshfeld
partitioning scheme~\cite{hirshfeld1977a,johnson2005a} to assign a density distribution to each atom, which can vary as the atoms move. The dependence
of dispersion on the chemical state of the atoms is also accounted for in a later development of the Grimme scheme,
known as DFT-D3~\cite{grimme2010a}. In principle, both the Grimme and the TS schemes can be used in conjunction with any base functional.
We have referred to the use of the DFT-D3 and the TS methods for water systems throughout the present review, and in relation to ice we also 
referred to a modified version of the TS scheme that accounts for many-body dispersion \cite{tkatchenko2012a}.

A rather different approach to the representation of dispersion is to include in the XC functional a non-local correlation
term $E_{\rm c}^{\rm nl}$ depending explicitly on electron densities at spatially 
separated positions~\cite{rapcewicz1991a,andersson1996a,rydberg2003a,dion2004a}. 
When paired with a conventional GGA representation of semi-local exchange-correlation, the total XC
functional then becomes:
\begin{equation}
E_{\rm xc} = E_{\rm x}^{\rm GGA} + E_{\rm c}^{\rm LDA} + E_{\rm c}^{\rm nl} \; .
\end{equation}
The general form assumed for the non-local correlation term has usually been:
\begin{equation}
E_{\rm c}^{\rm nl} = \int d {\bf r}_1 \, d {\bf r}_2 \, n ( {\bf r}_1 ) \phi ( {\bf r}_1 , {\bf r}_2 ) n ( {\bf r}_2 ) \; ,
\end{equation}
where the kernel $\phi$ is itself a functional of the density, but ensures the correct asymptotic behavior of the dispersion
interaction by falling off as $1 / | {\bf r}_1 - {\bf r}_2 |^6$ at large separations $| {\bf r}_1 - {\bf r}_2 |$. The earliest forms of
this approach~\cite{rapcewicz1991a,andersson1996a,rydberg2003a} were restricted to interactions 
between atoms or molecules whose densities do not overlap, but an important
advance came with the proposal by Dion \emph{et al.}~\cite{dion2004a} of a functional form valid for overlapping molecules in arbitrary
geometries. The base functional proposed in that paper consisted of revPBE exchange and LDA correlation, and we refer
to the entire functional here as revPBE-DRSLL. 

It was shown later that the particular choice of base functional made by Dion \emph{et al.} substantially underbinds molecular
dimers~\cite{gulans2009a,klimes2010a}, and that the performance can be greatly improved by choosing other base functionals. If PBE is paired with the DRSLL
form of non-local correlation, giving the functional referred to here as PBE-DRSLL, molecular dimers are significantly overbound~\cite{klimes2010a}.
However, if the exchange parts of PBE of BLYP are appropriately modified, giving base functionals known as optPBE
and optB88~\cite{klimes2010a}, then the resulting dispersion-inclusive functionals, here called optPBE-DRSLL and optB88-DRSLL, give much
improved binding energies for dimers. Similar arguments underlie the modified version of the DRSLL approach known
as rPW86-DF2~\cite{lee2010a}, in which a revised form of PW86 exchange is paired with a new form of the non-local correlation
functional $E_{\rm c}^{\rm nl}$.

A simplified version of the non-local correlation kernel $\phi ( {\bf r} , {\bf r}^\prime )$ known as VV10 has been developed
by Vydrov and van Voorhis~\cite{vydrov2010a}. This contains a free parameter $b$, which allows the short-range damping of dispersion to be adapted to
the base functional. Technical developments due to Sabatini \emph{et al.}~\cite{sabatini2013a} allow this approach to be implemented at a cost not much
greater than that of standard GGAs. The base functional originally recommended for VV10 consisted of revised PW86 exchange
and PBE correlation, but other base functionals could also be used.

\subsection*{And finally: computational cost}

Although it is difficult to be precise, we comment on the relative approximate computational cost of the various functionals discussed. 
In brief, there is no significant difference between the cost of LDA and the GGAs. Dispersion correction schemes such as the 
various generations of the DFT-D approach or the TS method also do not incur any major computational overhead compared to the base GGA they are added to. 
Methods based on non-local correlation functionals can lead to a slow-down of no more than $ca.$~50\% compared to a GGA calculation, when implemented 
efficiently~\cite{roman-perez2009a}.
The hybrid functionals, when employed in periodic boundary conditions, as done for liquid water simulations, can 
be significantly (roughly an order of magnitude) more expensive than a GGA.



\clearpage

\begin{table}[!htb]
\centering
\begin{tabular}{lcc|lcc}
\hline
\multicolumn{3}{c}{semi-local} & \multicolumn{3}{c}{disp-inclusive} \\
\hline
method & $E_{\rm b}^{\rm dim}$ & $R_{\rm O O}^{\rm dim}$ & method & $E_{\rm b}^{\rm dim}$ & $R_{\rm O O}^{\rm dim}$ \\
\hline
LDA & $380^c$ & $2.72^c$& & & \\
PBEsol & $265^c$ & $2.81^c$ & & & \\
PBE & $220^a$ & $2.90^c$ & PBE-D3 & $239^c$ & $2.89^c$ \\
         & & & PBE-DRSLL & $245^c$ & $2.94^c$ \\
         & & & PBE-TS & $241^c$ & $2.89^c$ \\
BLYP & $181^a$ & $2.95^c$ & BLYP-D3 & $219^c$ & $2.94^c$ \\
revPBE & $156^c$ & $3.01^c$ & revPBE-DRSLL & $183^c$ & $3.03^c$ \\
PBE0 & $215^a$ & $2.89^b$ & PBE0-TS & $234^c$ & $2.89^c$ \\
           & & & optPBE-DRSLL & $215^c$ & $2.95^c$ \\
           & & & optB88-DRSLL & $212^c$ & $2.96^c$ \\
           & & & rPW86-DF2 & $217^c$ & $2.97^c$ \\
\hline
\end{tabular}
\caption{Binding energies $E_{\rm b}^{\rm dim}$ (meV units) and equilibrium O-O distances $R_{\rm O O}^{\rm dim}$
(\AA\ units) of the H$_2$O dimer in its global minimum configuration computed with a variety
of semi-local, hybrid and dispersion-inclusive XC approximations. Benchmark values from CCSD(T)
calculations are $E_{\rm b}^{\rm dim} = 217.6$~meV, $R_{\rm O O}^{\rm dim} = 2.909$~\AA.
References to semi-local functionals: LDA~\cite{ceperley1980a,perdew1981a}, PBE~\cite{perdew1996a},
revPBE~\cite{zhang1998a}, PBEsol~\cite{perdew2008a}, BLYP~\cite{becke1988a,lee1988a},
PBE0~\cite{perdew1996a,adamo1999a}.
References to numerical values: $a$:~\cite{santra2007a}, $b$:~\cite{zhang2011a}, $c$: this work.}
\label{tab:dimer_gm}
\end{table}


\clearpage

\begin{table}[!htb]
\centering
\begin{tabular}{c|cccccccc}
\hline
s.p. & bench & PBE & BLYP & PBE0 & PBE-TS & PBE0-TS & revPBE-DRSLL & optPBE-DRSLL \\
\hline
1 & $0$ & $0$ & $0$ & $0$ & $0$ & $0$ & $0$ & $0$ \\
2 & $21$ & $25$ & $23$ & $24$ & $24$ & $23$ & $17$ & $20$ \\
3 & $25$ & $35$ & $32$ & $30$ & $35$ &$29$ & $24$ & $28$ \\
4 & $30$ & $46$ & $49$ & $44$ & $48$ & $47$ & $35$ & $36$ \\
5 & $41$ & $65$ & $65$ & $60$ & $68$ & $63$ & $45$ & $49$ \\
6 & $44$ & $74$ & $73$ & $65$ & $78$ & $70$ & $49$ & $54$ \\
7 & $79$ & $96$ & $96$ & $92$ & $93$ & $90$ & $66$ & $79$ \\
8 & $154$ & $160$ & $155$ & $160$ & $149$ & $151$ & $121$ & $143$ \\
9 & $77$ & $95$ & $95$ & $90$ & $92$ & $88$ & $63$ & $77$ \\
10 & $117$ & $132$ & $125$ & $127$ & $130$ & $125$ & $92$ & $112$ \\
\hline
\end{tabular}
\caption{Energies (meV units) of the Smith stationary points of the water dimer relative to the energy of the global-minimum
geometry given by benchmark CCSD(T) calculations (bench) and
by a selection of semi-local and dispersion-inclusive exchange-correlation functionals. 
Benchmark values and values computed with PBE, BLYP and PBE0 are taken from Ref.~\cite{gillan2012a};
dispersion-inclusive values were obtained within the present work. Geometries
of the stationary points are shown in Fig.~\ref{fig:Smith_SP}.}
\label{tab:Smith_energies}
\end{table}


\clearpage

\begin{table}[!htb]
\centering
\begin{tabular}{c|cccc}
\hline
Method &  Prism &     Cage &    Book &    Ring \\
\hline
\hline
CCSD(T) \cite{bates2009a}
 & \textbf{337}   & 335  (2) & 329 (8) & 324 (13) \\
DMC \cite{santra2008a} &\textbf{332} & 330 (2) & 328 (4) & 321 (11)  \\
\hline
PBE   \cite{santra2008a}   &336 & 339  (-3) &  \textbf{346} (-10)  & 344 (-8) \\
BLYP   \cite{santra2008a}  &274   & 277  (-3) & 288  (-14) & \textbf{290} (-16) \\
PBE0  \cite{santra2008a}   &323  & 325 (-2) &  \textbf{331} (-8) & \textbf{331} (-8) \\
B3LYP \cite{santra2008a}   &294  & 297 (-3) &  305 (-11) & \textbf{307} (-13) \\
\hline
PBE-D \cite{santra2008a} & 378 & \textbf{380}  (-2) & 378 (0) & 367  (11) \\
BLYP-D \cite{santra2008a} & \textbf{360} & \textbf{360} (0) & 356 (4) & 345 (15)\\
BLYP-D3 \cite{pruitt2013a} & \textbf{353} & 347 (6) & 344 (9) & 339 (24)\\
%
PBE-TS \cite{santra2008a} & 370 & \textbf{373} (-3) & 371 (-1) & 361 (9) \\
PBE0-TS$^z$ & 353 & \textbf{356} (-3) & 354 (-1) & 347 (6) \\
PBE0-D \cite{santra2008a} & 361 & \textbf{362}  (-1) & 359 (2) & 351 (10) \\
revPBE-DRSLL$^z$ & 275  & 275 (0) & \textbf{276} (1) & 272 (3)  \\
rPW86-DF2$^z$  & \textbf{329}  & 328 (1) & 325 (4) & 316 (13) \\
optPBE-DRSLL  \cite{klimes2010a} & \textbf{335} & 334 (1) & 332 (3) & 323 (13) \\
optB88-DRSLL \cite{klimes2010a}  & \textbf{347} & \textbf{347} (0) & 344 (3) & 334 (13)  \\
\end{tabular}
\caption{Binding energies (meV/monomer) of the prism, cage, book and ring isomers of the H$_2$O hexamer
according to CCSD(T) and DMC benchmark calculations and a selection of semi-local, hybrid and dispersion-inclusive
exchange-correlation functionals. The lowest energy structure(s) for each approach is indicated in bold. Energies
relative to the prism are given in parenthesis. $z$: this work. Geometries of the isomers are shown in Fig.~\ref{fig:hexamers}.}
\label{tab:hexamer_binding}
\end{table}


\clearpage

\begin{table}[!htb]
\centering
\begin{tabular}{c|cccc}
\hline
Method &  $E_{\rm sub}^{\rm Ih}$ & $E_{\rm sub}^{\rm VIII}$   &  $\Delta E^{\rm Ih-VIII}$  \\
\hline
\hline
Expt. \cite{whalley1984a} & 610 &  577 &    33 \\
DMC \cite{santra2011a} & 605 & 575 & 30   \\
\hline
LDA    \cite{fang2013a}  & 943 (333) & 813  (233) & 130 (97)   \\
PBE    \cite{santra2013a}  & 636 (26) & 459  (118) & 177 (143)   \\
BLYP  \cite{brandenburg2015a} & 555   (-55) & 347  (-230) & 208 (175)   \\
revPBE   \cite{brandenburg2015a} & 499 (-111) & 291(-286) & 208 (175)   \\
PBE0 \cite{santra2013a}   & 598 (-12) & 450 (-127)& 148 (115) \\
\hline
PBE-TS \cite{santra2013a} & 714 (114) & 619 (42) & 95 (62) \\
PBE0-TS \cite{santra2013a} & 672 (62) & 596 (19)& 76 (43) \\
PBE-D3   \cite{brandenburg2015a} & 755 (145) & 624 (47) & 131 (98) \\
BLYP-D3  \cite{brandenburg2015a} & 690 (80) & 594 (17) & 96 (63) \\
revPBE-D3   \cite{brandenburg2015a} & 659 (49) & 555 (-22) & 104 (71) \\
revPBE-DRSLL  \cite{santra2013a} & 559 (-51) & 517 (-60) & 42 (9)\\
rPW86-DF2 \cite{santra2013a} & 619 (9) & 586 (9) & 33 (0) \\
optPBE-DRSLL  \cite{santra2013a} & 668 (58) & 630 (53) & 38  (5) \\
optB88-DRSLL  \cite{fang2013a} & 696 (86) & 670 (93) & 26  (7) \\
\end{tabular}
\caption{
Sublimation energies (meV units) $E_{\rm sub}^{\rm Ih}$ and $E_{\rm sub}^{\rm VIII}$ of ice Ih and VIII 
(zero-point vibrational energies omitted) from experiment, DMC and a 
number of XC functionals. Also reported is the difference of sublimation energies 
$\Delta E_{\rm sub}^{\rm Ih-VIII}$ between ice Ih and VIII. The deviation
of each computed value from experiment is given in parenthesis. 
}
\label{tab:ice_energies}
\end{table}


\clearpage

\begin{table}[!htb]
\centering
\begin{tabular}{c|cc}
\hline
Method &  Ih    &   VIII    \\
\hline
\hline
Expt. \cite{whalley1984a} & 32.05 &  19.1 $\lbrack$20.09$\rbrack$ \\
DMC \cite{santra2011a} & 31.69 & 19.46 \\
\hline
PBE    \cite{santra2013a}  & 30.79 (3.9) & 20.74  (8.6) \\
BLYP  \cite{brandenburg2015a} & 32.2 (0.5) & 22.0 (15.2)  \\
revPBE   \cite{brandenburg2015a} & 33.1 (3.3) & 24.5  (28.3)  \\
PBE0 \cite{santra2013a}   & 30.98 (3.3) &  20.27 (6.1) \\
\hline
PBE-TS \cite{santra2013a} & 29.67 (7.4) & 20.13 (5.4)  \\
PBE0-TS \cite{santra2013a} & 29.88 (6.8) & 19.70  (3.1) \\
PBE-D3   \cite{brandenburg2015a} & 29.1 (9.2) & 19.1 (0.0) \\
BLYP-D3  \cite{brandenburg2015a} & 30.4 (5.1) & 19.4 (1.6) \\
revPBE-D3   \cite{brandenburg2015a} &  30.4 (5.1) & 19.7  (3.1) \\
revPBE-DRSLL  \cite{santra2013a} & 34.38 (7.3) & 22.96 (20.2)\\
rPW86-DF2 \cite{santra2013a} & 33.69 (5.1) & 21.27 (11.4) \\
optPBE-DRSLL  \cite{santra2013a} & 31.63 (1.3) & 20.55  (7.6) \\
optB88-DRSLL \cite{santra2015a} & 30.2 (5.8) & 19.1 (0.0) \\
%
%
\end{tabular}
\caption{
Comparisons of the calculated and experimental equilibrium volumes (\AA$^3$/H$_2$O) of ice Ih and VIII. Note 
that calculations with a range of XC functionals \cite{murray2012a,santra2013a,brandenburg2015a} suggest that 
zero-point energy effects are \emph{ca.}~5\% in ice VIII. The experimental 
value has therefore been corrected by removing this 5\% zero-point expansion, with the uncorrected 
value given in square brackets.  In ice Ih, the zero-point expansion is much less and varies by several percent from one functional to the next \cite{murray2012a,santra2013a,brandenburg2015a}, so we have not corrected the volume of this phase. The percentage deviation
of each computed value from experiment is given in parenthesis. 
}
\label{tab:ice_volumes}
\end{table}


\clearpage

\renewcommand\tabcolsep{2pt}
\renewcommand{\arraystretch}{0.8}

\begingroup
\squeezetable
\begin{table}[!htb]
\centering
\begin{tabular}{ccccccccccc}
\hline
ref & sim-alg & $N$ &  $T$ & $\rho_{\rm eq}$ & $t_{\rm pr}$ & $r_{\rm O O}^{\rm max}$ & $g_{\rm O O}^{\rm max}$ & $r_{\rm O O}^{\rm min}$ 
& $g_{\rm O O}^{\rm min}$ & $D$ \\
\hline
\multicolumn{11}{c}{BLYP functional} \\
\hline
\cite{grossman2004a} & $NVE$/CP & $32$ & $293$ & - & $20$ & 2.73 & 3.60 & 3.33 & 0.39 & 0.1   \\
\cite{kuo2004a} & $NVT$/CP/slab$^a$ & $256$ & $300$ & $0.85$ & $5$ & - & - & - & - & -   \\
\cite{kuo2004b} & $NVE$/$NVT$/CP/BO/MC & $64$ & $323$ & - & $10$ & $2.76$ & $2.98$ & $3.32$ & $0.60$ & $0.4$   \\
\cite{fernandez-serra2004a} & $NVE$/BO & $64$ & $300$ & - & $11$-$32$ & $2.78$ & $3.20$ & $3.33$ & $0.60$ & $0.2$   \\
\cite{fernandez-serra2005a} & $NVE$/BO & $32$ & $300$ & - & $17$-$32$ & $2.78$ & $3.30$ & $3.30$ & $0.55$ & -   \\
\cite{vandevondele2005a} & $NVE$/CP,BO & $32$ & $324$ & - & $20$ & $2.75$ & $3.03$ & $3.28$ & $0.55$ & $0.5$   \\
\cite{mcgrath2005a} & $NPT$/MC & $64$ & $298$ & $0.79$ & - & $2.80$ & $\sim 2.70$ & $3.40$ & $\sim 0.65$ & -   \\
\cite{mcgrath2006a} & Gibbs/MC$^b$ & $64$ & $323$ & $0.84$ & - & $2.80$ & $3.10$ & $3.45$ & $0.50$ & -   \\
\cite{lee2006a} & $NVT$/CP & $32$ & $300$ & - & $30$ & $2.80$ & $2.90$ & $3.65$ & $0.60$ & -   \\
\cite{todorova2006a} & $NVT$/CP & $32$ & $350$ & - & $10$ & $2.79$ & $3.00$ & $3.31$ & $0.48$ & $0.48$   \\
\cite{kuo2006a} & $NVE$/CP$^c$ & $64$ & $423$ & - & $30$ & $2.80$ & $2.30$ & $3.6$ & $0.75$ & $12$   \\
\cite{lee2007a} & $NVE$/CP & $32$ & $309$ & - & $60$ & $2.80$ & $2.88$ & $3.35$ & $0.62$ & $0.6$   \\
\cite{morrone2008a} & $NVT$/CP(PI)$^d$ & $64$ & $300$ & - & $13$ & $2.75$ & $3.20 (2.83)$ & $3.32$ & $0.42 (0.60)$ & -   \\
\cite{mattsson2009a} & $NVT$/BO & $64$ & $300$ & - & $15$-$50$ & $2.73$ & $3.40$ & $3.25$ & $0.43$ & -   \\
\cite{kuehne2009a} & $NVT$/BO & $64$ & $300$ & - & $<250$ & $2.79$ & $2.92$ & $3.33$ & $0.57$ & -   \\
\cite{schmidt2009a} & $NPT$/BO & $64$ & $330$ & $0.77$ & $25$-$45$ & $2.83$ & $3.18$ & $3.50$ & $0.35$ & -   \\
\cite{lin2009a} & $NVE$/CP & $64$ & $316$ & - & $20$ & $2.77$ & $2.94$ & $3.30$ & $0.60$ & -   \\
\cite{jonchiere2011a} & $NVT$/BO & $128$ & $317$ & - & $57$ & $2.80$ & $3.30$ & $3.40$ & $0.40$ & $0.20$   \\
\cite{baer2011a} & $NVT$/BO/slab$^e$ & $216$ & $300$ & $\sim 0.8$ & $15$ & $2.82$ & $3.17$ & $3.53$ & $0.23$ & -   \\
\cite{yoo2011a} & $NVT$/BO & $125$ & $436$ & - & ? & $2.82$ & $2.25$ & $3.50$ & $0.95$ & -   \\
\cite{lin2012a} & $NVE$/CP & $64$ & $319$ & - & $50$-$117$ & $2.77$ & $2.86$ & $3.31$ & $0.66$ & $1.0$   \\
\cite{ma2012a} & $NPT$/CP & $64$ & $300$ & $0.92$ & $20$ & $2.77$ & $3.16$ & $3.40$ & $0.45$ & $-$   \\
\cite{bartok2013a} & $NVE$/BO & $64$ & $308$ & - & $25$ & $2.77$ & $3.20$ & $3.30$ & $0.57$ & -   \\
\cite{alfe2013a} & $NVT$/BO$^f$ & $64$ & $350$ & $0.78$ & $20$ & $2.82$ & $2.90$ & $3.50$ & $0.48$ & -  \\
\cite{delben2013a} & $NPT$/MC & $64$ & $295$ & $0.78$ & - & $2.83$ & $2.44$ & $3.46$ & $0.35$ & -   \\
\cite{bankura2014a} & $NVE$/BO & $64$ & $353$ & - & $40$ & $2.77$ & $2.99$ & $3.27$ & $0.62$ & $1.14$   \\
\cite{delben2015a} & $NPT$/MC & $64$ & $295$ & $0.80$ & - & $2.83$ & $3.04$ & $3.46$ & $0.44$ & -   \\
\hline
\end{tabular}
\caption{
Summary of DFT simulations of liquid water performed with BLYP functional. Columns show: literature
reference, simulation-algorithm (ensemble and sampling technique: BO = Born-Oppenheimer, CP = Car-Parrinello, 
MC = Monte Carlo), number $N$ of monomers in cell, temperature $T$~(K), 
equilibrium density $\rho_{\rm eq}$ (g/cm$^3$), duration of production run
$t_{\rm pr}$ (ps) (MD simulations only),
distance to first maximum $r_{\rm O O}^{\rm max}$ (\AA), height of first maximum $g_{\rm O O}^{\rm max}$, 
distance to first minimum $r_{\rm O O}^{\rm min}$ (\AA) and height of first minimum $g_{\rm O O}^{\rm min}$ in O-O rdf,
diffusion coefficient $D$~($10^{-9}$~m$^2$/s).
All constant-volume simulations performed at density close to $1.0$~g/cm$^3$, unless otherwise noted.
a: slab with free surfaces; 
b: Gibbs ensemble~\cite{panagiotopoulos1987a}, $\rho_{\rm eq}$ density of liquid in coexistence with vapor;
c: density of simulated system $0.71$~g/cm$^3$;
d: performed with and without path-integral (PI) representation of quantum nuclear effects, values with PI in parentheses;
e: slab with free surfaces;
f: density of simulated system $0.78$~g/cm$^3$ gives pressure close to $0$.
}
\label{tab:liq_BLYP}
\end{table}
\endgroup


\clearpage

\renewcommand\tabcolsep{4pt}
\renewcommand{\arraystretch}{0.6}

\begingroup
\squeezetable
\begin{table}[!htb]
\centering
\begin{tabular}{ccccccccccc}
\hline
ref & sim-alg & $N$ &  $T$ & $\rho_{\rm eq}$ & $t_{\rm pr}$ & $r_{\rm O O}^{\rm max}$ & $g_{\rm O O}^{\rm max}$ & $r_{\rm O O}^{\rm min}$ 
& $g_{\rm O O}^{\rm min}$ & $D$  \\
\hline
\multicolumn{11}{c}{PBE functional} \\
\hline
\cite{allesch2004a} & $NVE$/CP & $54$ & $291$ & - & $20$ & $2.70$ & $3.35$ & $3.27$ & $0.45$ & $0.3$   \\
\cite{grossman2004a} & $NVE$/CP & $54$ & $295$ & - & $20$ & $2.71$ & $3.27$ & $3.32$ & $0.46$ & 0.1   \\
\cite{schwegler2004a} & $NVT$/BO & $64$ & $306$ & - & $20$ & $2.72$ & $3.83$ & $3.25$ & $0.33$ & $0.1$  \\
\cite{sit2005a} & $NVT$/CP & $32$ & $325$ & - & $20$ & $2.70$ & $3.32$ & $3.30$ & $0.27$ & $0.12$   \\
\cite{vandevondele2005a} & $NVE$/CP,BO & $32$ & $337$ & - & $20$ & $2.71$ & $3.18$ & $3.25$ & $0.42$ & $0.6$   \\
\cite{todorova2006a} & $NVT$/CP & $32$ & $350$ & - & $10$ & $2.70$ & $2.99$ & $3.29$ & $0.47$ & $0.47$   \\
\cite{leung2006a} & $NVT$/BO & $54$ & $320$ & - & $20$ & $2.78$ & $3.0$ & $3.30$ & $0.60$ & -   \\
\cite{rempe2008a} & $NVT$/BO & $64$ & $300$ & - & $15$ & $2.72$ & $3.57$ & $3.22$ & $0.30$ & -   \\
\cite{guidon2008a} & $NVE$/BO & $64$ & $313$ & - & $13$ & $2.70$ & $3.40$ & $3.28$ & $0.38$ & -   \\
\cite{mattsson2009a} & $NVT$/BO & $64$ & $300$ & - & $15$-$50$ & $2.70$ & $3.50$ & $3.25$ & $0.40$ & -   \\
\cite{kuehne2009a} & $NVT$/BO$^a$ & $64$ & $300$ & - & $< 250$ & $2.73$ & $3.25$ & $3.28$ & $0.44$ & -   \\
\cite{schmidt2009a} & $NPT$/BO & $64$ & $330$ & $0.87$ & $25$-$45$ & $2.76$ & $3.54$ & $3.40$ & $0.30$ & -   \\
\cite{guidon2010a} & $NVT$/BO & $64$ & $330$ & - & $30$ & $2.70$ & $3.55$ & $3.30$ & $0.35$ & -   \\
\cite{zhang2011a} & $NVE$/BO & $32$ & $297$ & - & $20$ & $2.71$ & $3.67$ & $3.24$ & $0.28$ & -   \\
\cite{zhang2011c} & $NVE$/BO & $32$ & $367$ & - & $17$ & $2.70$ & $3.35$ & $3.33$ & $0.40$ & -   \\
\cite{mogelhoj2011a} & $NVE$/BO$^b$ & $64$ & $299$ & - & $10$ & $2.75$ & $2.93$ & $3.38$ & $0.63$ & -   \\
\cite{lin2012a} & $NVE$/CP & $64$ & $314$ & - & $50$-$117$ & $2.72$ & $3.19$ & $3.27$ & $0.43$ & $0.3$   \\
\cite{forster-tonigold2014a} & $NVE$/BO & $64$ & $334$ & - & $17$ & $2.72$ & $3.30$ & $3.27$ & $0.37$ & -   \\
\cite{distasio2014a} & $NVT$/CP & $64$ & $300$ & - & $20$ & $2.69$ & $3.28$ & $3.28$ & $0.37$ & $0.20$   \\
\cite{miceli2015a} & $NPH$/BO & $64$ & $349$ & $0.87$ & $20$ & $2.75$ & $3.40$ & $3.40$ & $0.30$ & -   \\
\cite{gaiduk2015a} & $NVT$/BO & $64$ & $400$ & $0.81$ & $8 \times 20$ & - & - & - & - & -   \\
\hline
\multicolumn{11}{c}{revPBE functional} \\
\hline
\cite{todorova2006a} & $NVT$/CP & $32$ & $350$ & - & $10$ & $2.81$ & $2.29$ & $3.34$ & $0.80$ & $1.8$   \\
\cite{kuehne2009a} & $NVT$/BO$^a$ & $64$ & $300$ & - & $<250$ & $2.77$ & $3.01$ & $3.31$ & $0.50$ & -   \\
\cite{wang2011a} & $NVE$/BO$^c$ & $64$ & $341$ & - & $20$ & $2.83$ & $2.35$ & $3.45$ & $0.85$ & $2.7$  \\
\cite{lin2012a} & $NVE$/CP & $64$ & $323$ & - & $50$-$117$ & $2.80$ & $2.38$ & $3.34$ & $0.90$ & $2.1$   \\
\hline
\multicolumn{11}{c}{RPBE functional} \\
\hline
\cite{fernandez-serra2005a} & $NVE$/BO & $32$ & $300$ & - & $17$-$32$ & $2.82$ & $2.50$ & $3.40$ & $0.95$ & -   \\
\cite{mattsson2009a} & $NVT$/BO & $64$ & $300$ & - & $15$-$50$ & $2.78$ & $2.60$ & $3.27$ & $0.70$ & -   \\
\cite{kuehne2009a} & $NVT$/BO$^a$ & $64$ & $300$ & - & $<250$ & $2.75$ & $3.19$ & $3.32$ & $0.42$ & -   \\
\cite{forster-tonigold2014a} & $NVE$/BO & $64$ & $295$ & - & $17$ & $2.76$ & $2.80$ & $3.30$ & $0.68$ & -   \\
\hline
\multicolumn{11}{c}{PBE0 functional} \\
\hline
\cite{todorova2006a} & $NVT$/CP & $32$ & $350$ & - & $10$ & $2.74$ & $2.58$ & $3.35$ & $0.73$ & $2.8$   \\
\cite{guidon2008a} & $NVE$/BO & $64$ & $325$ & - & $13$ & $2.70$ & $3.30$ & $3.30$ & $0.40$ & -   \\
\cite{guidon2010a} & $NVT$/BO & $64$ & $330$ & - & $30$ & $2.70$ & $3.45$ & $3.30$ & $0.40$ & -   \\
\cite{zhang2011a} & $NVE$/BO & $32$ & $330$ & - & $20$ & $2.68$ & $3.01$ & $3.31$ & $0.58$ & -   \\
\cite{zhang2011c} & $NVE$/BO & $32$ & $374$ & - & $17$ & $2.73$ & $2.95$ & $3.37$ & $0.60$ & -   \\
\cite{distasio2014a} & $NVT$/CP & $64$ & $300$ & - & $20$ & $2.71$ & $2.96$ & $3.30$ & $0.53$ & $0.67$   \\
\cite{gaiduk2015a} & $NVT$/BO & $64$ & $400$ & $0.71$ & $8 \times 20$ & - & - & - & - & -   \\
\cite{delben2015a} & $NPT$/MC & $64$ & $295$ & $0.83$ & - & $2.78$ & $3.28$ & $3.42$ & $0.35$ & -   \\
\hline
\end{tabular}
\caption{
Summary of simulations of liquid water performed with PBE, revPBE, RPBE and PBE0 functionals.
Columns as in Table~\ref{tab:liq_BLYP}.
All constant-volume simulations performed at density close to $1.0$~g/cm$^3$, unless otherwise noted.
a: employs a form of coupled electron-ion dynamics that maintains the system close to the BO surface;
b: intramolecular O-H bond-lengths fixed at gas-phase value;
c: density of simulated system $0.95$~g/cm$^3$.
}
\label{tab:liq_PBE}
\end{table}
\endgroup


\clearpage

\renewcommand\tabcolsep{4pt}
\renewcommand{\arraystretch}{0.8}

\begingroup
\squeezetable
\begin{table}[!htb]
\centering
\begin{tabular}{cccccccccccc}
\hline
ref & disp & sim-alg & $N$ &  $T$ & $\rho_{\rm eq}$ & $t_{\rm pr}$ & $r_{\rm O O}^{\rm max}$ & $g_{\rm O O}^{\rm max}$ & $r_{\rm O O}^{\rm min}$ 
& $g_{\rm O O}^{\rm min}$ & $D$  \\
\hline
\multicolumn{11}{c}{based on BLYP functional} \\
\hline
\cite{schmidt2009a} & D1 & $NPT$/BO & $64$ & $330$ & $0.99$ & $25$-$45$ & $2.80$ & $2.78$ & $3.45$ & $0.80$ & -   \\
\cite{ma2012a} & D2 & $NPT$/CP & $64$ & $300$ & $1.07$ & $20$ & $2.73$ & $2.84$ & $3.50$ & $1.00$ & -   \\
\cite{baer2011a} & D2 & $NVT$/BO/slab$^a$ & $216$ & $330$ & $\sim 1.0$ & $15$ & $2.77$ & $2.87$ & $3.35$ & $0.52$ & -   \\
\cite{lin2012a} & D2 & $NVE$/CP & $64$ & $321$ & - & $50$-$117$ & $2.78$ & $2.83$ & $3.44$ & $0.77$ & $1.6$   \\
\cite{bankura2014a} & D2 & $NVE$/BO & $64$ & $328$ & - & $40$ & $2.76$ & $3.17$ & $3.40$ & $0.63$ & $0.71$   \\
\cite{jonchiere2011a} & D3 & $NVT$/BO & $128$ & $322$ & - & $122$ & $2.85$ & $2.76$ & $3.50$ & $0.80$ & $1.6$   \\
\cite{yoo2011a} & D3 & $NVT$/BO & $125$ & $385$ & - & $13$ & $2.79$ & $2.40$ & $3.65$ & $0.90$ & -   \\
\cite{delben2013a} & D3 & $NPT$/MC & $64$ & $295$ & $1.07$ & - & $2.78$ & $3.01$ & $3.51$ & $1.00$ & -   \\
\cite{bankura2014a} & D3 & $NVE$/BO & $64$ & $330$ & - & $40$ & $2.79$ & $2.80$ & $3.47$ & $0.80$ & $1.81$   \\
\cite{delben2015a} & D3 & $NPT$/MC & $64$ & $295$ & $1.07$ & - & $2.78$ & $2.78$ & $3.51$ & $0.92$ & -   \\
\cite{lin2009a} & DCACP & $NVE$/CP & $64$ & $325$ & - & $20$ & $2.79$ & $2.67$ & $3.38$ & $0.85$ & -   \\
\cite{lin2012a} & DCACP & $NVE$/CP & $64$ & $308$ & - & $50$-$117$ & $2.79$ & $2.72$ & $3.36$ & $0.85$ & $1.7$   \\
\cite{bartok2013a} & GAP & $NVE$/BO & $64$ & $308$ & - & $25$ & $2.79$ & $2.85$ & $3.35$ & $0.76$ & $1.7$   \\
\cite{alfe2013a} & GAP & $NVT$/BO$^b$ & $64$ & $350$ & $1.05$ & $20$ & $2.75$ & $2.60$ & $3.50$ & $0.90$ & -   \\
\hline
\multicolumn{11}{c}{based on PBE functional} \\
\hline
\cite{schmidt2009a} & D1 & $NPT$/BO & $64$ & $330$ & $0.94$ & $25$-$45$ & $2.76$ & $3.35$ & $3.35$ & $0.45$ & -   \\
\cite{lin2012a} & D2 & $NVE$/CP & $64$ & $324$ & - & $50$-$117$ & $2.72$ & $3.23$ & $3.30$ & $0.47$ & $0.6$   \\
\cite{delben2013a} & D3 & $NPT$/MC & $64$ & $295$ & $1.06$ & - & $2.73$ & $3.24$ & $3.15$ & $0.73$ & -   \\
\cite{forster-tonigold2014a} & D3 & $NVE$/BO & $64$ & $348$ & - & $17$ & $2.72$ & $3.30$ & $3.27$ & $0.42$ & -   \\
\cite{bankura2014a} & D3 & $NVE$/BO & $64$ & $324$ & - & $40$ & $2.72$ & $3.38$ & $3.29$ & $0.42$ & $0.39$   \\
\cite{gaiduk2015a} & D3 & $NVT$/BO & $64$ & $400$ & $1.02$ & $8 \times 20$ & - & - & - & - & -   \\
\cite{delben2015a} & D3 & $NPT$/MC & $64$ & $295$ & $1.06$ & - & $2.73$ & $3.07$ & $3.25$ & $0.69$ & -   \\
\cite{lin2012a} & DCACP & $NVE$/CP & $64$ & $323$ & - & $50$-$117$ & $2.71$ & $3.27$ & $3.28$ & $0.40$ & $0.5$   \\
\cite{distasio2014a} & TS & $NVT$/CP & $64$ & $300$ & - & $20$ & $2.71$ & $2.99$ & $3.27$ & $0.54$ & $0.44$   \\
\cite{wang2011a} & DRSLL & $NVE$/BO & $64$ & $304$ & $1.13$ & $20$ & $2.83$ & $2.17$ & $3.38$ & $0.92$ & $2.08$   \\
\cite{zhang2011a} & DRSLL & $NVE$/BO & $32$ & $295$ & - & $20$ & $2.80$ & $2.77$ & $3.34$ & $0.71$ & -  \\
\cite{corsetti2013a} & DRSLL & $NVE$/BO & $128$ & $301$ & $1.18$ & $20$ & $2.80$ & $2.51$ & $3.40$ & $1.00$ & $1.7$   \\
\cite{bankura2014a} & DRSLL & $NVE$/BO & $64$ & $328$ & - & $40$ & $2.82$ & $2.41$ & $3.58$ & $1.03$ & $2.57$   \\
\hline
\end{tabular}
\caption{
Summary of DFT simulations of liquid water performed with dispersion-inclusive functionals based on BLYP and PBE.
Columns as in Tables~\ref{tab:liq_BLYP} and \ref{tab:liq_PBE}, but with column 2 showing dispersion type.
All constant-volume simulations performed at density close to $1.0$~g/cm$^3$, unless otherwise noted.
Dispersion types: D1, D2, D3 are Grimme dispersion of Refs.~\cite{grimme2004a,grimme2006a,grimme2010a} respectively;
DCACP (dispersion-correcting atomic-centered potentials) is technique of Ref.~\cite{vonlilienfeld2004a};
TS is Tkatchenko-Scheffler technique~\cite{tkatchenko2009a};
DRSLL is non-local correlation technique of Dion \emph{et al.}~\cite{dion2004a};
GAP (Gaussian approximation potential) is technique of Refs.~\cite{bartok2010a,bartok2013a}.
a: slab with free surfaces;
b: density of simulated system $1.05$~g/cm$^3$ gives pressure close to $0$.
}
\label{tab:liq_disp_BLYP_PBE}
\end{table}
\endgroup


\clearpage

\renewcommand\tabcolsep{4pt}
\renewcommand{\arraystretch}{0.8}

\begingroup
\squeezetable
\begin{table}[!htb]
\centering
\begin{tabular}{cccccccccccc}
\hline
ref & disp & sim-alg & $N$ &  $T$ & $\rho_{\rm eq}$ & $t_{\rm pr}$ & $r_{\rm O O}^{\rm max}$ & $g_{\rm O O}^{\rm max}$ & $r_{\rm O O}^{\rm min}$ 
& $g_{\rm O O}^{\rm min}$ & $D$  \\
\hline
\multicolumn{11}{c}{based on revPBE functional} \\
\hline
\cite{lin2012a} & D2 & $NVE$/CP & $64$ & $322$ & - & $50$-$117$ & $2.80$ & $2.34$ & $3.55$ & $0.95$ & $3.4$   \\
\cite{bankura2014a} & D3 & $NVE$/BO & $64$ & $316$ & - & $40$ & $2.81$ & $2.54$ & $3.51$ & $0.83$ & $1.85$   \\
\cite{lin2012a} & DCACP & $NVE$/CP & $64$ & $331$ & - & $50$-$117$ & $2.74$ & $2.94$ & $3.35$ & $0.76$ & $1.6$   \\
\cite{wang2011a} & DRSLL & $NVE$/BO & $64$ & $300$ & $1.02$ & $20$ & $2.92$ & $2.35$ & $3.46$ & $1.10$ & $2.63$   \\
\hline
\multicolumn{11}{c}{based on RPBE functional} \\
\hline
\cite{forster-tonigold2014a} & D3 & $NVE$/BO & $64$ & $314$ & - & $17$ & $2.80$ & $2.58$ & $3.40$ & $0.83$ & -   \\
\hline
\multicolumn{11}{c}{based on optB88 functional} \\
\hline
\cite{zhang2011a} & DRSLL & $NVE$/BO & $32$ & $326$ & - & $20$ & $2.78$ & $2.83$ & $3.33$ & $0.73$ & -   \\
\cite{bankura2014a} & DRSLL & $NVE$/BO & $64$ & $341$ & - & $40$ & $2.76$ & $2.75$ & $3.33$ & $0.75$ & $1.57$   \\
\cite{delben2015a} & DRSLL & $NPT$/MC & $64$ & $295$ & $1.08$ & - & $2.74$ & $2.94$ & $3.34$ & $0.80$ & -   \\
\hline
\multicolumn{11}{c}{based on optPBE functional} \\
\hline
\cite{mogelhoj2011a} & DRSLL & $NVE$/BO & $64$ & $276$ & - & $10$ & $2.88$ & $2.32$ & - & - & -   \\
\hline
\multicolumn{11}{c}{based on rPW86 functional} \\
\hline
\cite{zhang2011a} & rPW86-DF2 & $NVE$/BO & $32$ & $291$ & - & $20$ & $2.84$ & $2.54$ & $3.56$ & $0.94$ & -   \\
\cite{mogelhoj2011a} & rPW86-DF2 & $NVE$/BO & $64$ & $283$ & - & $10$ & $2.90$ & $2.50$ & - & - & -   \\
\cite{corsetti2013a} & VV10 & $NVE$/BO & $200$ & $300$ & $1.19$ & $20$ & $2.75$ & $3.1$ & $3.3$ & $0.60$ & -   \\
\cite{miceli2015a} & rVV10-b9.3 & $NPH$/BO & $64$ & $342$ & $0.99$& $20$ & $2.75$ & $2.95$ & $3.30$ & $0.63$ & $1.5$   \\
\cite{delben2015a} & rVV10 & $NPT$/MC & $64$ & $295$ & $1.08$& - & $2.73$ & $3.22$ & $3.32$ & $0.79$ & -   \\
\hline
\multicolumn{11}{c}{based on PBE0 functional} \\
\hline
\cite{distasio2014a} & TS & $NVT$/CP & $64$ & $300$ & - & $20$ & $2.72$ & $2.76$ & $3.31$ & $0.70$ & $0.98$   \\
\cite{delben2013a} & D3 & $NPT$/MC & $64$ & $295$ & $1.02$ & - & $2.74$ & $3.23$ & $3.30$ & $0.67$ & -   \\
\cite{gaiduk2015a} & D3 & $NVT$/BO & $64$ & $400$ & $0.96$ & $8 \times 20$ & - & - & - & - & -   \\
\cite{delben2015a} & D3 & $NPT$/MC & $64$ & $295$ & $1.05$ & - & $2.74$ & $2.88$ & $3.29$ & $0.79$ & -   \\
\hline
\end{tabular}
\caption{
Summary of DFT simulations of liquid water performed with dispersion-inclusive functionals based on revPBE,
RPBE, optB88, optPBE, rPW86 and PBE0. Columns as in Table~\ref{tab:liq_disp_BLYP_PBE}.
All constant-volume simulations performed at density close to $1.0$~g/cm$^3$, unless otherwise noted.
Dispersion types: D2, D3 are Grimme dispersion of Refs.~\cite{grimme2006a,grimme2010a} respectively;
DCACP (dispersion-correcting atomic-centered potentials) is technique of Ref.~\cite{vonlilienfeld2004a};
TS is Tkatchenko-Scheffler technique~\cite{tkatchenko2009a};
DRSLL is non-local correlation technique of Dion \emph{et al.}~\cite{dion2004a};
rPW86-DF2 is revised form of DRSLL due to Lee \emph{et al.}~\cite{lee2010a};
VV10 is non-correlation technique of Vydrov and Van Voorhis~\cite{vydrov2010a};
rVV10-b9.3 is revised form of VV10 due to Sabatini \emph{et al.}~\cite{sabatini2013a}.
}
\label{tab:liq_disp_revPBE_etc}
\end{table}
\endgroup


\clearpage

\begin{table}[!htb]
\centering
\begin{tabular}{l|ccccccccc|c}
\hline
method & $f_{\rm ss}^{\rm mono} $ & $E_{\rm b}^{\rm dim}$ & $E_{\rm b}^{\rm ring}$ & $E_{\rm sub}^{\rm Ih}$ &
$\Delta E_{\rm b}^{\rm prism-ring}$ & $E_{\rm sub}^{\rm Ih-VIII}$ & $R_{\rm O O}^{\rm dim}$ &
$V_{\rm eq}^{\rm Ih}$ & $V_{\rm eq}^{\rm VIII}$ & Total \\
\hline
LDA   & 60 & 0     & --  & -- & 0   & 10 & 0 & 0 & -- &  \textbf{8} \\
PBE   & 50 & 100 & 90 & 80 & 0 & 0 & 100 & 70 & 20 &  \textbf{57} \\
BLYP   & 20 & 70 & 70 & 50 & 0 & 0 & 60 & 100 & 0 &  \textbf{41} \\
PBE0   & 80 & 100 & 100 & 90 & 0 & 0 & 90 & 70 & 40 &  \textbf{63} \\
revPBE-DRSLL  & 30 & 70 & 50 & 50 & 100 & 100 & 0 & 30 & 0 & \textbf{48}  \\
optPBE-DRSLL  & 40 & 100 & 100 & 50 & 100 & 100 & 60 & 90 & 30 & \textbf{74}  \\
optB88-DRSLL   & 60 & 100 & 100 & 20 & 100 & 100 & 50 & 50 & 100 &   \textbf{76} \\
rPW86-DF2   & 20  & 100 & 100 & 100 & 100 & 100 & 70 & 50 & 0 &   \textbf{71} \\
PBE-TS   & 50 & 80 & 70 & 0 & 100 & 40 & 90 & 30 & 50 &  \textbf{57} \\
PBE0-TS   & 80 & 90 & 80 & 40 & 100 & 60 & 90 & 40 & 70 &  \textbf{72} \\
BLYP-D3  & 20 & 100 & 90 & 30 & 100 & 40 & 70 & 50 & 90 &  \textbf{66} \\
\hline
\end{tabular}
\caption{
Percentage scores of selected XC functionals computed using the
scheme explained in Sec.~\ref{sec:scoring}. Physical quantities scored are:
monomer symmetric stretch frequency
$f_{\rm ss}^{\rm mono}$,
dimer binding energy $E_{\rm b}^{\rm dim}$,
ring-hexamer binding energy per monomer $E_{\rm b}^{\rm ring}$,
ice Ih sublimation energy $E_{\rm sub}^{\rm Ih}$,
difference $\Delta E_{\rm b}^{\rm prism-ring}$ of binding energies per monomer of prism and ring isomers of the hexamer,
difference $\Delta E_{\rm sub}^{\rm Ih-VIII}$ of sublimation energies of ice Ih and VIII,
equilibrium O-O distance $R_{\rm O O}^{\rm dim}$ in dimer, and
equilibrium volumes per monomer $V_{\rm eq}^{\rm Ih}$, $V_{\rm eq}^{\rm VIII}$ of ice Ih and VIII.
The total score in the final column is the average of the nine individual scores. Numerical values leading
to the individual scores are taken from earlier Tables in this paper, except for the stretch frequencies $f_{\rm ss}^{\rm mono}$.
These frequencies were computed as part of this work, apart from those predicted by LDA, BLYP and BLYP-D3 functionals.
For LDA the value is taken from Ref.~ \cite{xu2004a}, for BLYP the value is taken from Ref.~\cite{santra2009a}
and the BLYP-D3 and BLYP values are assumed to be the same. Note that data is not available for LDA for all criteria
but from the limited data available it is clear that it is abysmal.
}
\label{tab:scoring}
\end{table}


\clearpage

\clearpage

\clearpage

\clearpage

\clearpage

\clearpage

\clearpage

\clearpage

\clearpage

\clearpage

\end{document}